\documentclass[12pt]{article}
\usepackage{amsmath}
\usepackage{graphicx,epsf}
\usepackage{enumerate}
\usepackage{natbib}
\usepackage{url} 

\newcommand{\blind}{0}

\usepackage{graphicx}
\RequirePackage{amsthm,amsmath}
\RequirePackage{natbib,graphicx}
\usepackage{float}

\usepackage{sectsty}
\sectionfont{\centering}

\usepackage{booktabs,caption}
\usepackage{threeparttable}

\usepackage{colortbl}
\usepackage{graphicx}
\RequirePackage{amsthm,amsmath}
\RequirePackage{natbib,graphicx}

\usepackage{subfig}
\usepackage{graphicx}

\usepackage{booktabs,caption}
\usepackage{threeparttable}
\usepackage{hyperref}

\usepackage{amsmath}
\usepackage[table]{xcolor}

\begin{document}

\def\spacingset#1{\renewcommand{\baselinestretch}%
{#1}\small\normalsize} \spacingset{1}


\if0\blind
{
  \title{\bf Regression Modeling and File Matching Using Possibly Erroneous Matching Variables}
  \author{Nicole M. Dalzell
   \hspace{.2cm}\\
    Department of Mathematics and Statistics, Wake Forest University \\
    and \\
    Jerome P. Reiter \\
    Department of Statistical Science, Duke University}
  \maketitle
} \fi

\if1\blind
{
  \bigskip
  \bigskip
  \bigskip
  \begin{center}
    {\LARGE\bf Title}
\end{center}
  \medskip
} \fi

\bigskip
\begin{abstract}
Many analyses require linking records from two databases comprising overlapping sets of individuals. In the absence of unique identifiers, the linkage procedure often involves matching on a set of categorical variables, such as demographics, common to both files. Typically, however, the resulting matches are inexact: some cross-classifications of the matching variables do not generate unique links across files. Further, the variables used for matching can be subject to reporting errors, which introduce additional uncertainty in analyses. We present a Bayesian file matching methodology designed to estimate regression models and match records simultaneously when categorical variables used for matching are subject to errors. The method relies on a hierarchical model that includes (1) the regression of interest involving variables from the two files given a vector indicating the links, (2) a model for the linking vector given the true values of the variables used for matching, (3) a model for reported values of the variables used for matching given their true values, and (4)  a model for the true values of the  variables used for matching. We describe algorithms for sampling from the posterior distribution of the model.  We illustrate the methodology using artificial data and data from education records in the state of North Carolina.
\end{abstract}

\noindent%
{\it Keywords:}  Fusion; imputation; mixture; record linkage.
\vfill

\newpage
\spacingset{1.45} 

\section{Introduction} 

Increasingly, analysts seek to integrate data from two files
comprising overlapping sets of individuals, for example, by linking
subjects in planned studies to administrative databases like
electronic health records or tax records.  Often analysts are unable
to base linkages on unique identifiers, such as Medicare
identification numbers or social security numbers, because they are unavailable on at least one of the  files, e.g., due
to privacy requirements.  Analysts instead may have to base
linkages on combinations of categorical variables available in both files, such as demographic  
variables (e.g., age, race, sex) or group membership variables (e.g., county of residence, hospital, school attended). 
Depending on the richness and quality of these categorical 
variables, this file matching may not result in unique links for every record
across the files.  

In this article, we consider file matching settings where
the ultimate goal is to regress some outcome $Y_1$ available in file $F_1$ on a set
of predictors $Y_2$ available in the other file $F_2$ (and possibly
other predictors available in $F_1$). In this setting, the most
obvious approach is to base regression inferences on only the
linked data from the unique matches.  However, this discards
information from the other records, resulting in inefficient and
possibly even biased estimates.  An alternative is the approach of
\citet{Gutman}, which simultaneously estimates the regression and
performs linkages using the information from all records in
the files; we describe this model, which we call the GAZM, in detail in
Section \ref{Paper::sec::Methods}. The GAZM not only takes advantage of information from
non-unique matches, it also leverages relationships among $Y_1$ and $Y_2$ to improve the linking process. 

The GAZM utilizes blocking to reduce the number of paired comparisons considered in the matching; that is, possible matches are restricted to records with identical values of some set of categorical variables available in both files. This works best when the values of these variables for any individual present in both files are the same across the files. Otherwise, the records corresponding to the same individual are placed in different blocks in the two files and never given the chance to be correctly linked.
In some settings, however, variables that analysts might like to use in blocking might differ across the files for the same individual.  This can happen, for example, because of typographical  errors \citep{Jaro}, discrepancies in the coding systems used to
record information \citep{Winkler2004}, and legitimate changes in
field values, such as a student who moves to a new school. Such differences 
can generate incorrect matches, which in turn 
adversely impacts the results of regression analyses based on linked files \citep[e.g.,][]{NeterEtAl, ScheurenWinkler1993, ScheurenWinkler1997,
  LahiriLarsen2005, Chambers2, Chambers1, MMDLCh4}.
  
In this article, we extend the GAZM to handle scenarios where some categorical variables used for file matching may be faulty, i.e., subject to reporting error. To do so, we classify these categorical variables into two types, blocking variables (BVs) and matching variables (MVs). For BVs, the reported values agree across files for all true matches.  We require any plausible link to have the same values of all BVs across the files.  For MVs, the reported values can differ across files  for true matches.  However, the latent, true values of the MVs are identical across files for true matches.  We would like to use these true values when linking the files, but with MVs we do not know {\em a priori} which reported values equal the true values. Thus, in extending the GAZM, we seek to allow analysts to use faulty MVs in linking while accounting for the uncertainty when making inferences with the linked data that results from not having the true MV values. 

To do so, we use a Bayesian hierarchical model that includes four layers.
First, we specify a model for the regression of interest conditional
on a vector $C$ that indicates which records are linked
across files. Second, within each pool of possible matches formed by the cross-classifications of the reported BVs and the latent, true values of the MVs, we specify
a model for the corresponding elements of $C$.  
Third, we specify a 
model for the observed MV values subject to error.  
This comprises a model for the
probability that an observed MV field is faulty coupled
with a model that describes the relationship
between the observed MV value and its corresponding true value. Finally, we specify
a model for the true MVs using a mixture of
products of multinomial distributions \citep{DunsonXing2009}, which is 
a variant of latent class models.  
We call the full hierarchical model BLASE, which stands for Bayesian linkage and simultaneous
estimation. BLASE can be used to obtain posterior inferences for the
regression parameters.  Alternatively, the MCMC sampler for BLASE generates posterior draws of $C$ and imputed values of $(Y_1, Y_2)$ for non-matches as a byproduct.  These draws can be used as completed datasets to facilitate other analysis questions.

We present the model for the case where the analyst treats all the variables in $F_1$ as a ground truth and allows the MVs in $F_2$ to be subject to reporting errors. 
The modeling strategy can accommodate cases where the MVs in both files are subject to reporting errors, as discussed in Section \ref{Paper:concl}.  
We assume that reporting errors are made completely at random. One can adapt BLASE for informative reporting error models, 
as we discuss in Section \ref{Paper:concl}. 

The remainder of this article is organized as follows. 
In Section \ref{Paper::sec::Methods}, we introduce
BLASE as an extension of the GAZM. In Section \ref{Paper::sec::PostComp}, we describe algorithms
for sampling from the posterior distribution of BLASE. In
Section \ref{Paper::sec::Sim}, we use simulation studies to investigate the performance of BLASE, using a range of illustrative scenarios with different assumptions about the nature of errors in the MVs.
In Section \ref{Paper::sec::NCERDC}, we apply
BLASE to data on students in public schools in North
Carolina. In Section \ref{Paper:concl}, we conclude with a summary of findings, and we suggest extensions of the model as well as future research directions. 

\section{The BLASE Model} \label{Paper::sec::Methods}

After defining notation and key concepts in Section \ref{notation}, we
describe the sub-models in BLASE in Sections \ref{model:regression} to \ref{model:DP}, followed by some intuition about the model in Section \ref{model:intuition}. In Section \ref{gazm}, we present the GAZM to clarify how BLASE extends that model.  

\subsection{Notation and Key Concepts}\label{notation}

Let the two files to be integrated, $F_1$ and $F_2$, comprise $n_1$ and $n_2$ records,
respectively. We assume each record corresponds to a single true
individual; see \cite{Steorts2014} for linking data with replicates. 

Let $J$ be the total number of BVs and MVs. We refer to the combined set of BVs and MVs as the in-common variables.  Any in-common variable for which reporting errors are deemed possible is designated as an MV. Let $\hat{B}_{fij}$ denote the reported value for record $i$ in file $f$ on field $j$, where $j=1, \dots, J$,  $i=1, \dots, n_f$, and $f = 1, 2$.  Let ${B}_{fij}$ be the fault-free value of $\hat{B}_{fij}$. By definition, for all BVs, $B_{fij} = \hat{B}_{fij}$ for all $i,j$. For MVs, we say record $i^{'}$ is faulty if $(i,i^{'})$ is a match and $\hat{B}_{2i^{'}j}$ does not equal $\hat{B}_{1ij}$ for some $j = 1, \dots, J$.  For all $(f,i)$, let $\hat{B}_{fi}=(\hat{B}_{fi1}, \dots, \hat{B}_{fiJ})$, and let $\boldsymbol{\hat{B}}_f
= \{\hat{B}_{fi}: i=1, \dots, n_f\}$. We define  ${B}_{fi}$ and
$\boldsymbol{B}_f$ analogously. Finally, let $\boldsymbol{\hat{B}} =
(\boldsymbol{\hat{B}}_1, \boldsymbol{\hat{B}}_2)$ and $\boldsymbol{B} =
(\boldsymbol{B}_1, \boldsymbol{B}_2)$. 

In the model presented here, we allow faulty MVs only in $F_2$, and we set ${B}_{1ij} = \hat{B}_{1ij}$ for all $i,j$.   Effectively, this treats the MVs in $F_1$ as the ground truth for matching purposes.  Of course, the MVs in $F_1$ could be reported with error as well.  We discuss how to modify BLASE to incorporate such errors in Section \ref{Paper:concl}. 

Through validation, data cleaning, or other techniques, it is
typically possible to determine with high certainty that a subset of
the records are not faulty on some or all of the $J$ fields. A record
is a \textit{seed} on field $j$ if it is known \textit{a priori} that
$\hat{B}_{fij} = B_{fij}$. For any BV, all records are seeds on that BV field. If the relationship between $\hat{B}_{fij}$ and $B_{fij}$ is unknown \textit{a priori}, then $i$ is a {\em non-seed} on field $j$. 
When $i$ is a seed for all $J$ fields, we call $i$ a Type 2 ($T_2$) seed.
We also define Type 1 ($T_1$) seeds to be records where it is known
that records $i \in F_1$  and $i' \in F_2$ belong to the same individual,
and that ${\hat{B}}_{fi} = {B}_{fi}$ for $f=1,2$. 
When it is known that
an $(i,i^{'})$ pair belong to the same individual yet $\hat{B}_{1ij} \neq
\hat{B}_{2i^{'}j}$ for some $j$, the analyst can select
one of $(\hat{B}_{1ij}, \hat{B}_{2i^{'}j})$ to be correct. As we shall discuss, the information in the $T_1$ and $T_2$ seeds is crucial
for helping BLASE estimate model parameters reliably.

For any record $i$ in $F_1$ and $i^{'}$ in $F_2$, we say the records are a possible match if and only if $B_{1i} = B_{2i^{'}}$.  Otherwise, the records cannot be a match.  Thus, for $i$ and $i^{'}$ to be a match, we require their reported BVs to be the same and their latent, true MVs to be the same.  Any $(i, i^{'})$ pair that does not agree on the BVs is automatically ruled out as a match. We call each unique combination of values in $\boldsymbol{B}$ a matching pool, which is akin to a block but defined on both the BVs and latent MVs.

Let $K$ be the number of unique pools in a given 
$\boldsymbol{B}$. For $k=1, \dots, K$, let ${n}_{1(k)}$ and ${n}_{2(k)}$ denote
the number of records in pool $k$ from $F_1$ and $F_2$,
respectively, that are not $T_1$ seeds.  In general,  ${n}_{1(k)}$ need not equal ${n}_{2(k)}$.  Following
\citet{Gutman}, in each pool $k$ we match all
records from the smaller sized file for that pool to records from the
larger sized file for that pool; we treat the unmatched records as if
they were missing at random from the smaller file.  Thus, for the
given $\boldsymbol{B}$, the total number of unique individuals in any
pool $k$ is $n_{(k)} = c_{(k)}+ u_{(k)}$, where  $c_{(k)} =
\max({n}_{1(k)}, n_{2(k)})$ and $u_{(k)}$ is the number
of $T_1$ seeds in pool $k$. The total number of unique individuals in
$F_1 \cup F_2$ is $N_B = \sum_{k=1}^K n_{(k)}$.  The sizes of the  pools can change with different plausible values of $\boldsymbol{B}$.

For a given $\boldsymbol{B}$ and for $k=1, \dots, K$, let $F_{1k}$ and $F_{2k}$ represent the records in pool $k$ from $F_1$ and $F_2$, respectively. 
Without loss of generality, within each pool $k$ fix the ordering of
the records from $F_{1k}$.  For each pool $k$, we define a $n_{(k)}$
element permutation vector $C_k$ that specifies how records 
from $F_{2k}$ must be re-organized to match their corresponding records
in $F_{1k}$.  
To illustrate, suppose that  ${n}_{1(k)} = {n}_{2(k)} = 2$ and we link
the first record in $F_{1k}$ to the second record in $F_{2k}$.  Then,
$C_k = (2,1)$.  
 
When ${n}_{1(k)} > {n}_{2(k)}$, we create ${n}_{1(k)} - {n}_{2(k)}$
dummy records and add them to $F_{2k}$. We then match 
${n}_{2(k)}$ records from $F_{1k}$ to actual records from $F_{2k}$,
and match the remaining ${n}_{1(k)} - {n}_{2(k)}$ records from $F_{1k}$ to the dummy records.  
Likewise, when ${n}_{1(k)} < {n}_{2(k)}$, we add ${n}_{2(k)} - {n}_{1(k)}$
dummy records to $F_{1k}$. When ${n}_{1(k)} = {n}_{2(k)}$, no dummy records are needed.
Thus, for a given $\boldsymbol{B}$, the linkage in BLASE is one-to-one and complete, in the sense that every record from $F_1$ is matched to an actual or dummy record from $F_2$, and vice versa. We refer to pools with complete linkage as balanced. 

For any given $\boldsymbol{B}$, we define $C = \{C_1, \dots, C_K\}$ as the collection of all $K$ permutation vectors. We emphasize that in BLASE $C$ derives from $\boldsymbol{B}$; hence, it can change with different plausible values of $\boldsymbol{B}$.
Similar linkage structures are used in other Bayesian record linkage approaches \citep[e.g.,][]{Fortini2001,LarsenRubin2001,Larsen2004,TancrediLiseo2011, Steorts2014,sadinle16}. 

Finally, for $i \in F_1$, let $\boldsymbol{Y}_{1i} = (Y_{1i1}, \dots, Y_{1ip})$ comprise
the $p$ variables of interest for the analysis that are exclusive to $F_1$.  For $i \in F_2$, let $\boldsymbol{Y}_{2i} = (Y_{2i1}, \dots, Y_{2iq})$ comprise the $q$ variables of interest for the analysis that are exclusive to
$F_2$.  Let $\boldsymbol{Y}_1 = \{\boldsymbol{Y}_{1i}: i \in F_1 \}$ and $\boldsymbol{Y}_2 = \{\boldsymbol{Y}_{2i}: i \in F_2 \}$.


\subsection{Analysis Model}\label{model:regression}


We specify a joint distribution, $f(\boldsymbol{Y}_{1}, \boldsymbol{Y}_{2} \mid \Theta, C, \boldsymbol{B})$, where $\Theta$ represents the parameters of the distribution.    For any record $i' \in F_2$ linked to a dummy record, we treat 
its value of $\boldsymbol{Y}_{1i'}$ as missing at random \citep{Rubin1976}; similarly, we treat $\boldsymbol{Y}_{2i}$ as missing at random for records $i \in F_1$ linked to dummy records in $F_2$. Thus, within any pool $k$ in the given $\boldsymbol{B}$, the contribution to the likelihood function from the linking analysis is the product of $f(y_{1i}, y_{2i})$ for all $(i, i')$ in the pool that are linked, and 
$f(y_{1i})$ and $f(y_{2i'})$  for $i$ and $i'$ in the pool that are linked to  dummy records. 

In the examples in Sections \ref{Paper::sec::Sim} and \ref{Paper::sec::NCERDC}, we make $Y_1$ univariate with a 
normal linear regression on $(Y_2, \boldsymbol{B})$, with parameters  $(\boldsymbol{\beta}, \sigma_1)$.  We make $Y_2$ univariate with a normal linear regression on $\boldsymbol{B}$, 
with parameters  $(\boldsymbol{\eta}, \sigma_2)$. As a prior distribution for $\Theta = (\boldsymbol{\beta}, \sigma_1, \boldsymbol{\eta}, \sigma_2)$, we use $p(\Theta) \propto (1/\sigma_1^2) (1/\sigma_2^2)$.


\subsection{Model for \textit{C}}\label{model:C}


In any pool $k$, by definition the $T_1$ seeds never change links,
whereas the links for the non-$T_1$ seeds are unknown.  Hence, there are $c_{(k)}!$ possible values of $C_k$, which we denote as $C_{kl}$, where $l \in \lbrace 1, \dots ,c_{(k)}!\rbrace$. We assume each permutation is {\em a priori} equally likely given a realization of $\boldsymbol{B}$, so that
\begin{align} \label{Ch2::eq::BCPrior} 
p(C_k = C_{kl} \mid \boldsymbol{B}) & = \frac{1}{ {c_{(k)}}!}.
\end{align}
We model each $C_k$ independently, given $\boldsymbol{B}$.


\subsection{Reporting model}\label{model:errors}


The distinction between MVs and BVs necessitates a model for the errors in the reported MVs in $F_2$. 
For each $i^{'}$, we define  $E_{2i} = \lbrace E_{2i^{'}j}
: j \in (1, \dots, J) \rbrace$, where  each 
  \begin{equation} \label{Ch2::eq::ErrorVector}
   E_{2i^{'}j} = \left\{
     \begin{array}{ll}
     1  &  : \text{ if } \hat{B}_{2i^{'}j} \neq B_{2i^{'}j}  \\
     0  & :  \text{ otherwise.}
     \end{array}
   \right.
\end{equation}
For $T_1$ and $T_2$ seeds, as well as all BVs, $E_{2i^{'}j} = 0$ for all $j$. For non-seeds, each $E_{2i^{'}j}$ corresponding to a MV is a random variable.  We let $\boldsymbol{E}_2 = \{E_{2i^{'}}: i \in F_2\}$. We use independent beta-Bernoulli models for each $E_{2i^{'}j}$. We have 
\begin{align}  
  E_{2i^{'}j}| \gamma_{2j} & \sim Bernoulli(\gamma_{2j}) \label{eq::Ch2::ErrorVector::e}\\ 
 \gamma_{2j} & \sim Beta( a_{\gamma_{2j}}, b_{\gamma_{2j}}), \label{eq::Ch2::ErrorVector::b}
\end{align}
where $E(\gamma_{2j}) = a_{\gamma_{2j}}/(a_{\gamma_{2j}} + b_{\gamma_{2j}})$ is the
expected prior probability of observing a faulty value on non-seed
field $j$ in $F_2$. We discuss selection of $(a_{\gamma_{2j}}, b_{\gamma_{2j}})$ in
Section \ref{Paper::sec::Sim}.  

This effectively assumes that
MVs are faulty at random \citep{kim:cox:karr:reiter:2015} at potentially different
rates, which is a reasonable default model in the absence of specific
information about the mechanisms that generate faulty values.  If available, 
such information could be encoded  in the model for
$E_{2i^{'}j}$. For example, suppose that gender is a BV and that men are more likely than women to have errors in some MV $j$. We can use separate values of $\gamma_{2j}$ for men and women in \eqref{eq::Ch2::ErrorVector::e} and \eqref{eq::Ch2::ErrorVector::b}. For more complicated violations of faulty at random, we can use logistic regression on some function of $\boldsymbol{B}$. 

For $\boldsymbol{\hat{B}}_2$, we use a model similar to that used in
\citet{TancrediLiseo2011}, \citet{Steorts2014}, and \citet{manriqueReiterUpdated}.  We
have 
\begin{equation} \label{Ch2::eq::MeasurementError} 
   f \left( \hat{B}_{2i^{'}j}| E_{2i^{'}j}, B_{2i^{'}j} \right) = \left\{
     \begin{array}{ll}
      B_{2i^{'}j}  &  : \text{if } E_{2i^{'}j} = 0 \\
      \text{Uniform discrete on all but } B_{2i^{'}j} & : \text{if } E_{2i^{'}j} = 1. 
     \end{array}
   \right.
\end{equation}
This model implies that reported values for faulty
items are generated completely randomly. This also is a default
model. If information is available on the mechanisms that introduce
faults in the MVs, one could use other reporting error models as appropriate. We describe an example in Section \ref{Paper:concl}.


\subsection{Model for \textit{B}}\label{model:DP}


BLASE can use any model that adequately describes the distribution of
$\boldsymbol{B}$.  Here, we use the highly flexible and
computationally efficient latent class model, which has been shown to
be a useful tool for modeling and imputation of categorical data \citep[e.g.,][]{Vermunt, ReiterSi2011}. 

In this article, we assume that $B_{fi}$ follows the same distribution for all unique individuals $i$ in the target population for $F_1 \cup F_2$. 
 This is often a reasonable assumption in linkage settings, where we believe that many of the same individuals are present in both files. Assuming a common distribution, when reasonable, can improve the quality of the matches and estimates from BLASE compared to estimating separate latent class models for $\boldsymbol{B}_1$ and $\boldsymbol{B}_2$. The information in $\boldsymbol{B}_1$ encourages the latent class model to estimate the joint distribution of the BVs and latent MVs accurately. With accurate estimation of this distribution, BLASE can determine which reported values of the MVs are unlikely under the model, and hence possibly faulty. In turn, this can improve the ability of BLASE to move cases into pools that correspond to more plausible matches.  

When the distribution of the in-common variables varies by file, disregarding this difference can worsen the performance of BLASE.  In particular, the latent class model estimates will be a compromise between the distributions of $\boldsymbol{B}_1$ and $\boldsymbol{B}_2$.  Hence, the distribution of plausible values of the MVs in $\boldsymbol{B}_2$ will be pulled away from their actual distribution and toward the distribution of $\boldsymbol{B}_1$.  This could cause imputations of the MVs in $\boldsymbol{B}_2$ to be inaccurate, thereby putting some cases in $F_2$ in unreasonable pools.  The impacts of these effects on the performance of BLASE is complicated to characterize, as it depends on the nature of the differences in the distributions, the fault rates, and the strength of the association between $\boldsymbol{Y}_1$ and $\boldsymbol{Y}_2$ for true matches.  When analysts suspect that the distributions for $\boldsymbol{B}_1$ and $\boldsymbol{B}_2$ differ, they can specify separate models for  $\boldsymbol{B}_1$ and $\boldsymbol{B}_2$. 
We leave evaluating the performance of BLASE with separate models, as well as the impacts of assuming a common $\boldsymbol{B}$ model when not warranted, to future research. In Section \ref{Paper:concl}, we suggest some diagnostic checks of this modeling assumption.

For each in-common variable $j$, where $j=1, \dots, J$, without loss of
generality assume its values take on elements in $\{1, \dots, d_j\}$.
As is usual in latent class models, assume each individual  
belongs to one of $H$ latent classes. Let $z_{i} \in \lbrace 1, \dots,  H\rbrace$ represent the latent class assignment for record $i$.
We assume that  $Pr(z_i = h) = \pi_h$ for all $i$ and $h$. Let $\pi =
(\pi_1, \dots, \pi_{H})$. Within each latent class $h$, all variables follow independent multinomial distributions.  
For $h = 1, \dots, H$ and $j=1, \dots, J$, let $\phi_{h j b } =
Pr(B_{ij} = b \mid z_i = h)$, where $b \in \{1, \dots, d_j\}$.  Let
$\phi_{hj} = \left( \phi_{hj1}, \dots, \phi_{hjd_{j}} \right)$.
Mathematically, we have
\begin{align}
B_{ij} | \phi_{hj}, z_{i} & \sim Multinomial \left( 1; \lbrace 1, \dots, d_j \rbrace, \phi_{z_{i} j} \right) \label{eq:DP:a}\\
z_{i}|\pi & \sim Multinomial( \pi_{1}, \dots,\pi_{H} ). \label{eq:DP:b}
\end{align}
As prior distributions on $\pi$ and each $\phi_{hj}$, we use a
truncated Dirichlet process mixture model representation
\citep{DunsonXing2009}. We have 
\begin{align}
\pi_{h} & = V_{h} \prod_{g < h} (1-V_g) , h = 1, \dots, H \label{eq:DP:c}\\
V_{h} & \sim Beta( 1, \alpha ), \textrm{for } h = 1, \dots, H-1; \,\, V_{H} = 1 \label{eq:DP:d}\\ 
\alpha & \sim Gamma( a_{\alpha} , b_{\alpha} ) \label{eq:DP:e}\\
\left( \phi_{hj1}, \dots, \phi_{hjd_{j}} \right) & \sim Dirichlet( a_{j1}, \dots, a_{j d_j} ). \label{eq:DP:f}
\end{align}

Following \citet{DunsonXing2009} and \citet{ReiterSi2011}, we set
$a_{\alpha} = b_{\alpha} = 0.25$ to represent low prior sample size, and
{$a_j = (a_{j1},\dots,a_{jd_j}) = (1,\dots,1)$ for all $j$.}
For $H$, we recommend starting with a large value, such as $H = 30$. When
posterior runs indicate that all $H$ classes are consistently
occupied, we increase $H$ and restart the sampling.
We represent the collection of parameters as $\Psi=\left( \phi,
  \pi, \alpha, z \right)$. 

\subsection{Intuition about block moves and roles of seeds}\label{model:intuition}

The distributions of $(\boldsymbol{\hat{B}}_2,\boldsymbol{E}_2)$ and 
$\boldsymbol{B}$ define models for the 
MVs that incorporate the possibility of faulty MVs. At each
iteration $s$ of the MCMC for BLASE, a posterior sample for
$\boldsymbol{B}^{(s)}$ is drawn, and records are shifted into the matching pools defined by $\boldsymbol{B}^{(s)}$. Thus, BLASE allows records initially in a separate pool from their true match potentially to shift into the pool in which their true match lies. 

Moves into new pools are influenced by multiple levels of the hierarchical model. BLASE favors shifting records from a current pool into one that represents a higher value of the likelihood function for the regression models of interest. Thus, the accuracy of pool moves is reliant upon the strength of the relationships between $\boldsymbol{B}$ and $\boldsymbol{Y}_{1}$, and between $\boldsymbol{B}$ and $\boldsymbol{Y}_{2}$. When $\boldsymbol{B}$ is not important in the regression models, posterior samples for $\boldsymbol{B}^{(s)}$ tend not to move from $\boldsymbol{\hat{B}}$, resulting in few records shifting pools. Moves also are reliant on the model for the faulty MVs, particularly when \eqref{eq::Ch2::ErrorVector::b} is highly informative. Finally, values of the MVs with high posterior probability under the latent class model are more likely to be sampled for $\boldsymbol{B}^{(s)}$, whereas unlikely combinations are sampled less often.

The $T_1$ and $T_2$ seeds inform multiple parts of the model.
Because their true MV values are known, $T_1$ and $T_2$ seeds provide information on the joint distribution of the MVs.
The $T_1$ seeds additionally provide essential 
information about $\Theta$; effectively, they identify regions of support with non-trivial mass.
At each posterior update, $T_1$ seeds are used along with
estimated links to update $C$ and the parameters of the
linking analysis. In this way, the influence of the $T_1$ seed
information persists throughout posterior estimation. Thus,
it is crucial that the $T_1$ seeds be representative of
the underlying relationships between $\boldsymbol{Y}_{1}$ and $\boldsymbol{Y}_{2}$.  When this is
not the case, strong $T_1$ seed information could encourage
BLASE to favor choices of $C$ that reflect the relationships among the seeds 
rather than the true $\Theta$. 

\subsection{The GAZM}\label{gazm}

We conclude this section by showing how BLASE extends and differs from the GAZM. The key difference is that the GAZM requires all in-common variables to be BVs and not MVs.  Hence, for the GAZM we set $\boldsymbol{B} = \boldsymbol{\hat{B}}$ and eschew the sub-models in Section \ref{model:errors} and \ref{model:DP}.  The joint distribution in Section \ref{model:regression} conditions on $\boldsymbol{\hat{B}}$ rather than the latent $\boldsymbol{B}$ in  $f(\boldsymbol{Y}_{1}, \boldsymbol{Y}_{2} \mid \Theta, C, \boldsymbol{\hat{B}})$. The prior distribution for $C_k$ uses independent uniform distributions,  $p(C_k = C_{kl} \mid \boldsymbol{\hat{B}})  = 1/c_{(k)}!$, within each block defined by $\boldsymbol{\hat{B}}$.
Unlike in BLASE,  in the GAZM each $n_{(k)}$ and $N_B$ are fixed since $\boldsymbol{B}$ is constant over all iterations of the MCMC sampler. 

\section{Posterior Sampling Algorithm for BLASE} \label{Paper::sec::PostComp}

To estimate the posterior distribution of the parameters in BLASE, we
leverage the MCMC algorithms described in \citet{Gutman}. 
As BLASE considers $\boldsymbol{\hat{B}}$
to be subject to errors and allows the  pools to change, we need to augment the sampler in 
 \citet{Gutman} with additional steps.  

We initialize the MCMC sampler using information from the $T_1$ and $T_2$ seeds. Let $\mathbf{T}_1$ be the subset of $(\boldsymbol{Y}_1,\boldsymbol{Y}_2,\boldsymbol{B})$ corresponding to the $T_1$ seeds. We initialize $\Theta^{(0)}$ by taking a random draw from  $f(\Theta \mid \mathbf{T}_1)$. 
We initialize $\boldsymbol{B}^{(0)} = \boldsymbol{\hat{B}}$, and put all records in $F_1 \cup F_2$ into the matching pools defined by $\boldsymbol{B}^{(0)}$. We create dummy records as needed to balance all initial pools. We initialize $C_{k}^{(0)}$ for each $k$ using the step described in S2 below. 
Let $\boldsymbol{B}_{12}$ contain the rows of $\boldsymbol{B}$ corresponding to 
the $T_1$ and $T_2$ seeds. Conditioning on $\boldsymbol{B}_{12}$, we initialize $\Psi^{(0)}$ using one draw from the block Gibbs sampler \citep{IshwaranJames2001} described in the supplementary material.

The posterior sampling for BLASE includes the following six steps to update from iteration $(s)$ to iteration $(s+1)$. The supplementary material includes additional details on the steps and sampler.
\begin{enumerate}
\item[S1.] For each non-seed record in  $F_2$, propose $({E}_{2i^{'}}^{*}, {B}^{*}_{2i^{'}})$ using a Metropolis Hastings (MH) step, resulting in the updated draw
  $(\boldsymbol{E}_2^{(s+1)},\boldsymbol{B}_2^{(s+1)})$. Once $\boldsymbol{E}_2$ and $\boldsymbol{B}_2$ have been updated, shift records into pools defined by $\boldsymbol{B}^{(s+1)}$.  Add or delete dummy records as needed to
  ensure all pools are balanced. See Section \ref{Paper::sec::BLASEMH}
  for details.
\item[S2.] Sample a new $C^{(s+1)}_k$ for each pool $k$; see
  Section \ref{Paper::sec::SampleC} for details. 
\item[S3.] Impute missing values of $\boldsymbol{Y}_1$ or $\boldsymbol{Y}_2$ for records linked to dummy
  records, using the appropriate conditional distribution derived from
  $f(\boldsymbol{Y}_1, \boldsymbol{Y}_2 \mid  C^{(s+1)},\boldsymbol{B}^{(s+1)},
  \Theta^{(s)})$.  As explained in \citet{Gutman}, imputation helps us avoid working with the marginal distributions of $\boldsymbol{Y}_1$ and $\boldsymbol{Y}_2$, which could be complicated for general specifications of $f(\boldsymbol{Y}_{1}, \boldsymbol{Y}_{2} \mid \Theta, C, \boldsymbol{B})$, especially when some 
items are missing within $\boldsymbol{Y}_1$ or $\boldsymbol{Y}_2$ themselves. Let $\boldsymbol{Y}_1^{(s+1)}$ comprise $\boldsymbol{Y}_1$ and the imputed values of $\boldsymbol{Y}_1$,  and let $\boldsymbol{Y}_2^{(s+1)}$ comprise $\boldsymbol{Y}_2$ and the imputed values of $\boldsymbol{Y}_2$.
\item[S4.] Sample $(\Theta^{(s+1)} \mid C^{(s+1)},\boldsymbol{B}^{(s+1)},
\boldsymbol{Y}_1^{(s+1)}, \boldsymbol{Y}_2^{(s+1)})$ from its full conditional.
\item[S5.] 
Sample $\gamma^{(s+1)}$ from 
\begin{align} \label{Ch2::eq::PostGamma} 
 \gamma_{2j}\mid \boldsymbol{E}_2^{(s+1)}   & \sim Beta\left( a_{\gamma_{2j}} + \sum_{i} E_{2ij}^{(s+1)},  b_{\gamma_{2j}} + \sum_{i}(1-E_{2ij}^{(s+1)}) \right). 
\end{align} 
\item[S6.]
Sample $\Psi^{(s+1)}$ using the block Gibbs sampler in the supplementary material. 
\end{enumerate}

In addition to facilitating computations, using imputed values in S3 ensures that completed datasets generated as byproducts of the MCMC---which could be further analyzed or disseminated as public use files---do not have missing values of  $(\boldsymbol{Y}_1, \boldsymbol{Y}_2)$.

\subsection{Further details on sampling \textit{C}} \label{Paper::sec::SampleC}

Given $\boldsymbol{B}^{(s+1)}$, to sample $C^{(s+1)}$ we use the
strategies in \citet{Gutman}, which we summarize here due to their
importance in the algorithm. Let $\boldsymbol{Y}_{1(k)}^{(s)}$ and $\boldsymbol{Y}_{2(k)}^{(s)}$ be 
the values of $\boldsymbol{Y}_{1}^{(s)}$ and $\boldsymbol{Y}_{2}^{(s)}$ in pool $k$ defined by $\boldsymbol{B}^{(s+1)}$.
We sample each $C_k$ independently from  
\begin{align}  \label{Ch2::eqn:multBern}
p(C_k = C_{k\ell} \mid \boldsymbol{Y}_1^{(s)}, \boldsymbol{Y}_2^{(s)}, \boldsymbol{B}^{(s+1)}, \Theta^{(s)})& = 
\frac{ f( \boldsymbol{Y}_{1(k)}^{(s)}, \boldsymbol{Y}_{2(k)}^{(s)} \mid \Theta^{(s)}, C_{k\ell}, \boldsymbol{B}^{(s+1)})  }{ \sum_{l'=1}^{c_{(k)}!} f(  \boldsymbol{Y}_{1(k)}^{(s)}, \boldsymbol{Y}_{2(k)}^{(s)} \mid \Theta^{(s)}, C_{k\ell^{'}}, \boldsymbol{B}^{(s+1)}) }.
\end{align} 
We refer to sampling $C_k$ from \eqref{Ch2::eqn:multBern} as the Exact Step. 
It requires enumerating all possible permutations within pool $k$,
and computing the likelihood for the analysis model within the pool for each possible permutation. 

Evaluating all possible permutations becomes computationally expensive 
when the number of records within a pool is even modestly large. 
To mitigate this for large pools, we adapt the MH step in the algorithm
for the GAZM, which itself is adapted from 
\cite{Wu1995Proceed}, as an alternative sampling technique for $C_k$. We refer to this 
alternative as a Switch Step. For any $k$ in which $c_{(k)}$ is larger than a user specified threshold, we do the following.
\begin{enumerate} \label{Ch2::step::Switch} 
 \item Propose a new permutation $C_{k\ell^{*}}$ by selecting with uniform probability two within-pool records from $F_{2}$, and swapping their positions. 
\item Compute the ratio of the likelihoods under the current and proposed $C_{k\ell}$ and $C_{k\ell^{*}}$,
\begin{align}
r = \frac{ f( \boldsymbol{Y}_{1(k)}^{(s)}, \boldsymbol{Y}_{2(k)}^{(s)} \mid \Theta^{(s)}, {C_{k\ell}}^{*}, \boldsymbol{B}^{(s+1)})}{  f(  \boldsymbol{Y}_{1(k)}^{(s)}, \boldsymbol{Y}_{2(k)}^{(s)} | \Theta^{(s)}, C_{k\ell}, \boldsymbol{B}^{(s+1)}) }.
\end{align} 
\item Set $C_k = {C_{k\ell}}^{*}$ when $log(u) < log(r), u \sim Unif(0,1)$. Otherwise, set $C_k = C_{k\ell}$. 
\item Repeat these three steps $R>1$ times to encourage sampling high probability permutations.  The constant $R$ is user specified ($R=30$ in our applications).  The value of $C_{k}$ in the $R$th
iteration is taken for $C^{(s+1)}_k$. This also reduces correlation between parameters in iterations $(s)$ and $(s+1)$.
\end{enumerate}
The Switch Step avoids computing the likelihood for every possible $C_{kl}$, as only two 
likelihoods need to be computed. 
Following the GAZM, we sample $C_k$ for all pools $k$ with $c_{(k)} \geq
5$ using the Switch Step and for all smaller pools 
using the Exact Step. 

The collection $\lbrace C_k^{(s+1)}: k = 1, \dots, K \rbrace$ defines
the sampled value for $C^{(s+1)}$.  We link $F_1$ and $F_2$ according to this
permutation.  We note that for BLASE, since  $\boldsymbol{B}^{(s+1)}$ potentially can change at each MCMC iteration, to make pools balanced we must introduce dummy variables  prior to updating $C^{(s)}$.

\subsection{Further details on sampling \textit{E} and \textit{B}} \label{Paper::sec::BLASEMH}

The most complicated step of BLASE involves proposing values for
$(\boldsymbol{E}_2, \boldsymbol{B}_2)$. The basic strategy is as
follows. For each record $i \in F_2$ that is not a $T_1$ or $T_2$
seed, we propose a new value of $({(E_{2i})}^*, {(B_{2i})}^{*})$.  We accept or reject that proposal using a MH step, with an acceptance ratio defined in the supplementary material. We repeat these two steps independently for all non-seed records in
$F_2$, resulting in $(\boldsymbol{E}_2^{(s+1)},
\boldsymbol{B}_2^{(s+1)})$.  Using independent, record by record
updates allows for the updates to be done in parallel using multiple
processors (although we do not utilize this feature in the examples).

We now describe the updating process for any record $i \in F_2$ that is
not a $T_1$ or $T_2$ seed.
For any eligible record $i$,  we randomly select  one non-seed MV field $j^{*}$ uniformly from all non-seed MV fields for that record.
Thus, at any iteration in the MCMC, $E_{2i}^{(s)}$ and $E_{2i}^{(s+1)}$ differ on at
most one field, and record $i$ may move only to some pool
$k^{*}$ that disagrees with its current pool $k$ on exactly 
one field. This 
simplifies the computation of the MH acceptance ratio. 

We propose a value for $E_{2ij^{*}}$ from 
\begin{align}
    ({E_{2i{j}^{*}}})^{*}|\gamma_{2}^{(s)},j^{*} \sim Bernoulli( \gamma_{2j^{*}}^{(s)} ).
\end{align}
  When $({E_{2i{j^{*}}} })^{*} = E_{2ij^{*}}^{(s)} = 0$, we set
$E_{2i}^{(s+1)} = E_{2i}^{(s)}$ and $B_{2i}^{(s+1)} =
B_{2i}^{(s)}$. 
Otherwise, we set 
${(E_{2ij})}^{*} = E_{2ij}^{(s)}$ for all $j \neq j^{*}$, and  
propose a new $({B_{2ij^{*}}})^{*}$ via one of three general patterns. 
\begin{subequations}
\begin{align}
\text{If } {(E_{2i{j^{*}}})}^{*} = 0 \text{ and } E_{2ij^{*}}^{(s)} = 1 & \Rightarrow \text{Propose } {(B_{2ij^{*}})}^{*} = \hat{B}_{2ij^{*}}.  \label{eq::Bprop::line1} \\ 
\text{If } {(E_{2i{j^{*}}})}^{*} = 1 \text{ and } E_{2ij^{*}}^{(s)} = 0 & \Rightarrow \text{Propose } {(B_{2ij^{*}})}^{*} \neq \hat{B}_{2ij^{*}}. \label{eq::Bprop::line2} \\ 
\text{If } {(E_{2i{j^{*}}})}^{*} = 1 \text{ and } E_{2ij^{*}}^{(s)} = 1 & \Rightarrow \text{Propose } {(B_{2ij^{*}})}^{*} \not\in \lbrace \hat{B}_{2ij^{*}} \cup {B}_{2ij^{*}}^{(s)} \rbrace. \label{eq::Bprop::line3}
\end{align}
\end{subequations}
To aid understanding, call the pool defined by $\hat{B}_{2i}$ the
reported pool for record $i$, and any pool that is different
from the reported pool an alternate pool. Hence, \eqref{eq::Bprop::line1} shifts $i$ from an alternate
pool to a reported pool, \eqref{eq::Bprop::line2} 
shifts from a reported pool to an alternate
pool, and \eqref{eq::Bprop::line3} shifts $i$ from one
alternate pool to another. 

In the cases of \eqref{eq::Bprop::line2} and \eqref{eq::Bprop::line3}, we propose ${(B_{2ij^{*}})}^{*}$ by drawing from the mixture model described 
in Section \ref{model:DP}. However, we restrict ${(B_{2ij^{*}})}^{*}$ to ensure the conditions in \eqref{eq::Bprop::line2} and \eqref{eq::Bprop::line3} are not violated.  We also find that requiring any proposed value of ${(B_{2ij^{*}})}^{*}$ to be observed in $\boldsymbol{\hat{B}}_1$ can improve efficiency of the sampler; see supplementary material for details.

Based on the sampled value of ${ (B_{2 i j^{*}} )}^{*}$, we shift record $i$ from the pool $k$ defined by ${B_{2i}}^{(s)}$ into the pool $k^{*}$ defined
 by ${ ( B_{2 i } )  }^{*}$. Pool $k$ loses a record from $F_2$ while pool $k^{*}$ gains one. We then add or remove dummy records to rebalance both affected pools. 
When pool $k^{*}$ contains at least one record from $F_{1k^*}$ matched to a dummy record, we temporarily replace one of 
the dummy records with the record $i$ shifted into the pool. {When pool $k^{*}$ does not contain a dummy record in $F_{2k}$, we impute the missing $\boldsymbol{Y}_1$ for record $i$ from
 $f(\boldsymbol{Y}_1| \boldsymbol{Y}_2^{(s)}, C^{(s)},\boldsymbol{B}^{*}, \Theta^{(s)})$. For pool $k$, if record $i$ was matched to a dummy record when in $F_{2k}$, 
we temporarily remove the dummy record from pool $k$. Otherwise, we impute the missing $\boldsymbol{Y}_2$ value for record $i$ from $f(\boldsymbol{Y}_2| \boldsymbol{Y}_1^{(s)}, C^{(s)},\boldsymbol{B}^{*}, \Theta^{(s)})$.
 
Shifting records in and out of pools requires two new permutations, $(C_{k})^{*}$ and $(C_{k^{*}})^{*}$. For small pools, we propose these permutations using the Exact Step. {For large pools, we use a variation of the Switch Step; see the supplementary material for details.}
We accept or reject $(E_{2ij^{*}}^{*},B^{*}_{2ij^{*}})$ based on the MH acceptance ratio described in the supplementary material, the result of which we use for $(E_{2ij^{*}}^{(s+1)}, B_{2ij^{*}}^{(s+1)})$. 

When updating each non-seed's MVs in $F_2$ in serial, we return all values to those in \\ $(E^{(s)},B^{(s)}, C^{(s)},\boldsymbol{Y}_1^{(s)},\textbf{Y}_2^{(s)})$ before moving to the proposal for the next record. 

\section{Simulation Studies} \label{Paper::sec::Sim} 

In this section, we present results of simulation studies designed to illustrate the performance of BLASE with varying amounts of faulty values, numbers of $T_1$ seeds, and hyper-parameter settings for the prior distribution of $\gamma_2$. We use two faultiness levels, high (40\%) and low (5\%), and two $T_1$ seed counts, high (60\%) and low (20\%), to define four primary simulation groups. For each simulation, we examine three prior distributions for $\gamma_{2}$, representing diffuse (D) prior beliefs, concentrated and appropriate (CA) prior beliefs, and concentrated but poorly specified (CP) beliefs. The exact specifications of the priors are described in the supplementary material.  Broadly, with the CA prior we concentrate prior mass at the true faultiness level, e.g., in the scenario with 40\% faulty values, we specify hyperparameters that appropriately reflect a strong belief in a high degree of faultiness. The CP prior concentrates prior mass at an incorrect fault level, e.g., inappropriately reflects prior beliefs of a high degree of faultiness in the presence of low faults, and vice versa. The D prior is relatively flat with a small prior sample size.

\begin{table}[t]
\small 
\begin{center}
  \begin{threeparttable}
    \caption{Variable descriptions for the \textsf{hsbdemo} data
        set.  The BVs include \textsf{female, schtyp, ses, honors, cid}, and the MV is \textsf{prog}.}
       \label{Ch2::table::SimVars}
 		 		\begin{tabular}{lllll}
 			\hline
 			\hline 
 			{Name}          &  Type  & Description         & Levels   & In File\\ 
 			\hline 
 			\textsf{read}   &Cont    &  Reading test score &          & $F_1$ only \\ 
 			\textsf{math}   &Cont    &  Math test score    &          & $F_2$ only \\ 
 			\textsf{female} &Binary  &                     &0 = Male  & Both \\
 			                &        &                     &1 = Female&   \\
 			\textsf{schtyp} &Binary  & Type of school      &0 = Public& Both\\
 			                &        &                     &1 = Private&   \\
 			 \textsf{ses}   &  Cat  & Level of school      &1 = Lower & Both\\
 			                &       &                      &2 = Middle&   \\
 			                &       &                      &3 = High  & \\
 			 \textsf{prog}  &Cat    & Program of study     &1 = General& Both \\
 			                &       &                      &2 = Academic & \\
 			                &       &                      &3 = Vocational&  \\
 			 \textsf{honors}&Binary & Honors status        &0 = Not in Honors & Both \\
 			                &       &                      &1 = In Honors &   \\
 			 \textsf{cid}   & Cat   & Unspecified          & 1-30  & Both \\
 				\hline 
		\end{tabular}
  \end{threeparttable}
   \end{center}
\end{table}

We use variables from the \textsf{hsbdemo} dataset \citep{hsbdemo} to create simulated files $F_1$ and $F_2$. This dataset focuses on school testing, as does the NC education data we use in Section 
\ref{Paper::sec::NCERDC}. Table \ref{Ch2::table::SimVars} describes the variables. 
Because \textsf{hsbdemo} has only 200 records, we simulate 5000 record
pairs by sampling from predictive distributions estimated on the
observed data; details are presented in the supplementary material. We generate each replicated dataset ensuring that
$c_{(k)} \leq 10$ for all $k$, as, in general, file linking applications perform better with small $c_{(k)}$. We use five categorical variables as BVs and one as an MV.  To make the MV, we introduce errors in the variable 
\textsf{prog}. To do so, we select the
record pairs with faults on \textsf{prog} at random, and  
sample the faulty reported values of \textsf{prog} uniformly from the available incorrect values. 
We repeat this process independently 100 times at each simulation
setting, generating new sets of  5000 records each time.
We refer to the generated data in each replication with no MV faults,
i.e., $\boldsymbol{\hat{B}} = \boldsymbol{B}$, 
as the perfectly blocked (PB) data. 

We apply BLASE in each data set. We also apply the GAZM to each data set, treating the faulty version of \textsf{prog} as a BV. Of course, the GAZM is not intended for data with MVs; it is intended for data with all BVs.  We use the GAZM with the faulty data to investigate the potential gains from accounting for errors in MVs when file matching.  
Finally, to get a sense of the best BLASE possibly could do, we apply the GAZM using the PB data.  

For each replicate and each estimation method, we run a MCMC chain for 10000 iterations, thinned every two iterations, with 500 iterations used for burnin. Based on experimentation and investigations with trace plots and autocorrelation checks, this iteration length facilitates exploration of the posterior distribution when running BLASE.  Experiments also show that it is more than adequate for the GAZM, which converges more quickly than BLASE.  When fitting BLASE, we only allow \textsf{prog} to 
change; that is, we set the other five in-common variables as BVs. 

The analysis model includes a linear regression for $Y_1=$ \textsf{read} on \textsf{math} and \textsf{prog},
\begin{align}
Y_1 
&=  \beta_0 + \beta_1 \textsf{math} + \beta_2 I(\textsf{prog} = academic) + \beta_3 I(\textsf{prog} = vocational) + \epsilon_1,
\end{align}
where $\epsilon_{1} \sim N(0,\sigma_1^2)$, and a linear regression for $Y_2=$ \textsf{math} on \textsf{female}, \textsf{prog}, and \textsf{ses}, 
\begin{equation}
\begin{split}
Y_2 = & \eta_0 + \eta_1 I(\textsf{female} = yes) + \eta_2 I(\textsf{prog} = academic) + \\ & \eta_3 I(\textsf{prog} = vocational) + \eta_4 I(\textsf{ses} = Middle) +
\eta_5 I(\textsf{ses} = High) + \epsilon_2, \\
\end{split}
\end{equation}
where $\epsilon_{2} \sim N(0,\sigma_2^2)$. As the parameter values for data generation, we use the maximum likelihood estimates from the \textsf{hsbdemo} data.  We use the same regressions when estimating BLASE and the two applications of the GAZM, estimating all parameters.

We examine the performance of BLASE across the simulation scenarios using several metrics. First, we compute the posterior match rate (PMR), which is
the number of correctly matched pairs divided by the total number of
matched pairs. Since for BLASE the number of 
matched pairs can vary at each MCMC iteration, we compute the PMR for
each post-burnin posterior sample of $C$, and average this across
posterior samples. We also evaluate posterior inferences for $\Theta$,
the coefficients in both regressions in the linking analysis
model. Finally, we compute an average root mean square error (RMSE) for
prediction on out-of-sample cases. Here, we create a single 
test data set of 500 records simulated from the
models used to create the 5000 records. In each replication, we then compute $RMSE = \sqrt{(1/500) \sum_{i=1}^{500} {(\hat{Y}_{1i} - Y_{1i})}^2 }$.
Each $\hat{Y}_{1i}$ is sampled from $f(Y_1|\hat{\Theta}, Y_{2i}, B_{2i})$, where $\hat{\Theta}$ equals the posterior mean of $\Theta$ for that replication. In each simulation setting and for each method, we average the 100 RMSE values.

\begin{figure}[t]
    \centering
    \subfloat[Posterior means for \textsf{math} coefficient.]{{\includegraphics[scale=.30]{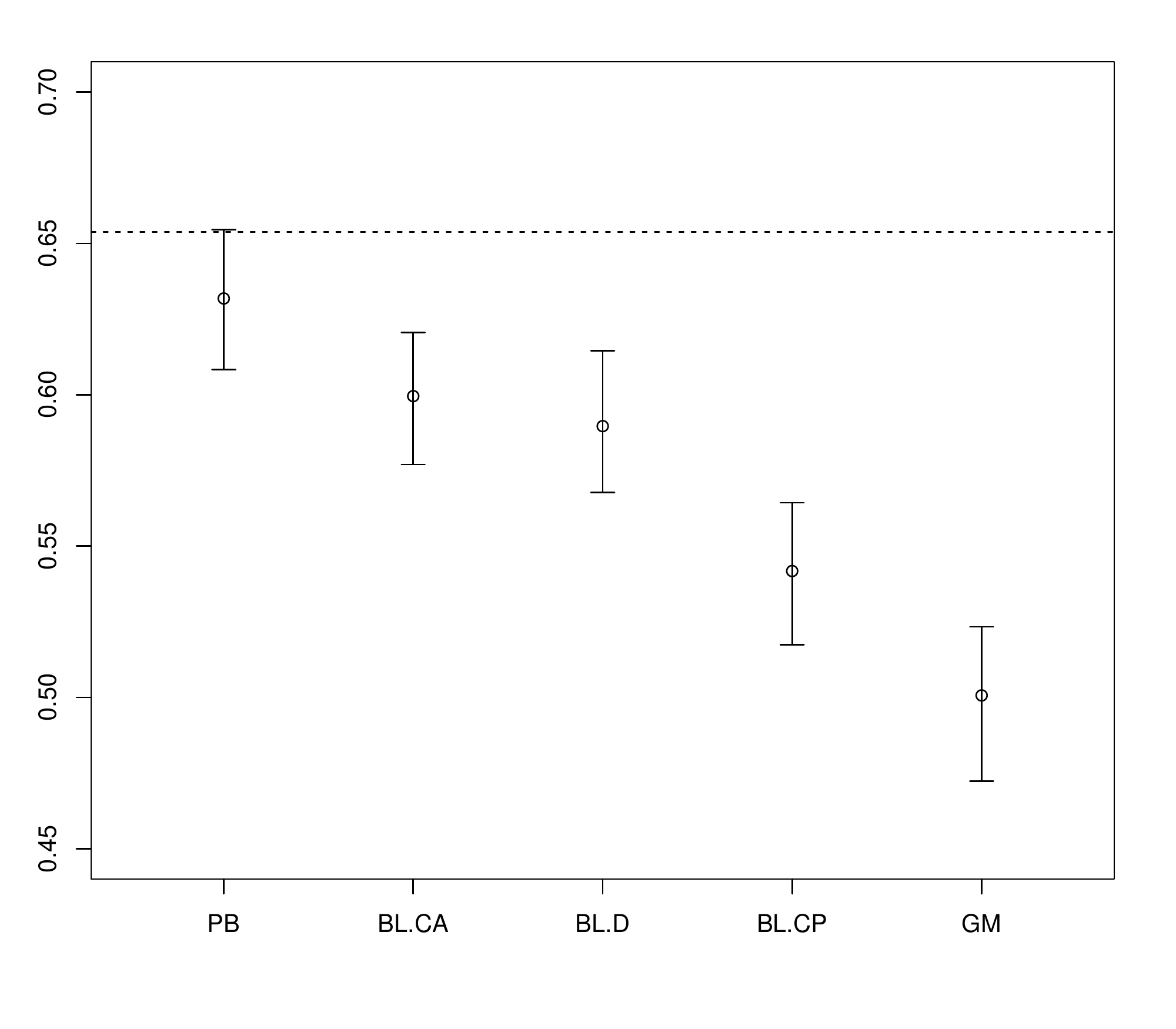} }\label{fig:testa}}%
    \qquad
     \subfloat[Non-seed PMRs.]{{\includegraphics[scale=.30]{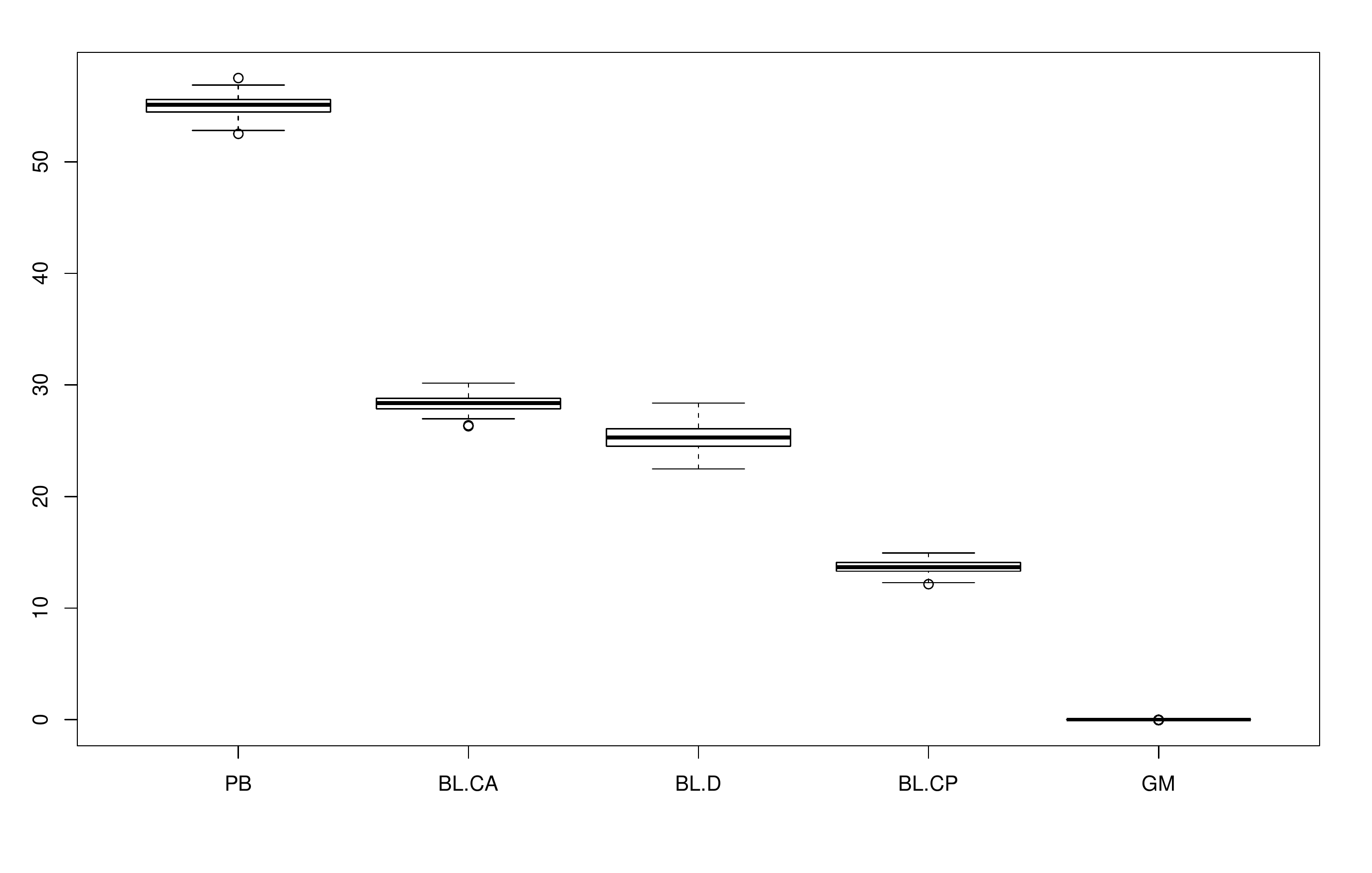}} \label{fig:testb}}%
    \caption{Results from the HSHF simulation. Left panel (a) displays 95\% interval for the 100 posterior means for \textsf{math } coefficient. Dashed horizontal line indicates true value of coefficient. Right panel (b) displays average posterior match rates from 100 simulations.  In both panels, PB stands for using the GAZM on the perfectly blocked data, BL.CA for using BLASE with the concentrated and appropriate prior, BL.D for using BLASE with the diffuse prior, BL.CP for using BLASE with the concentrated but poorly specified prior, and GM for using the GAZM on the faulty data.}
    \label{fig:test}
\end{figure}

We begin with a scenario with high levels of faulty values, which is where one would expect accounting for errors in the MVs to matter the most. Figure \ref{fig:test} displays key results for the coefficient for \textsf{math} in the regression for $Y_1$ in the high seed, high fault (HSHF) scenario.  We focus on  this coefficient because it is the most important predictor of $Y_1$ by far, as evident in  the final row of Table \ref{Ch2::table::SimResults}.
The posterior means based on BLASE with the 
 appropriately concentrated or diffuse priors tend to be similar to those based on the analysis with the PB data.  Not surprisingly, using the poorly concentrated prior distribution worsens the performance of BLASE. Figure \ref{fig:testa}, as well as Table \ref{Ch2::table::SimResults}, also suggest that accounting for potential errors in the MVs can improve inferences relative to treating the faulty MV as an error-free BV.  This is corroborated in Figure \ref{fig:testb}, which shows that allowing the file matching algorithm to change faulty MVs from their reported values can increase match rates. 

\begin{figure}[t]
    \centering
    \subfloat[Posterior means for  \textsf{math}  coefficient.]{{\includegraphics[scale=.30]{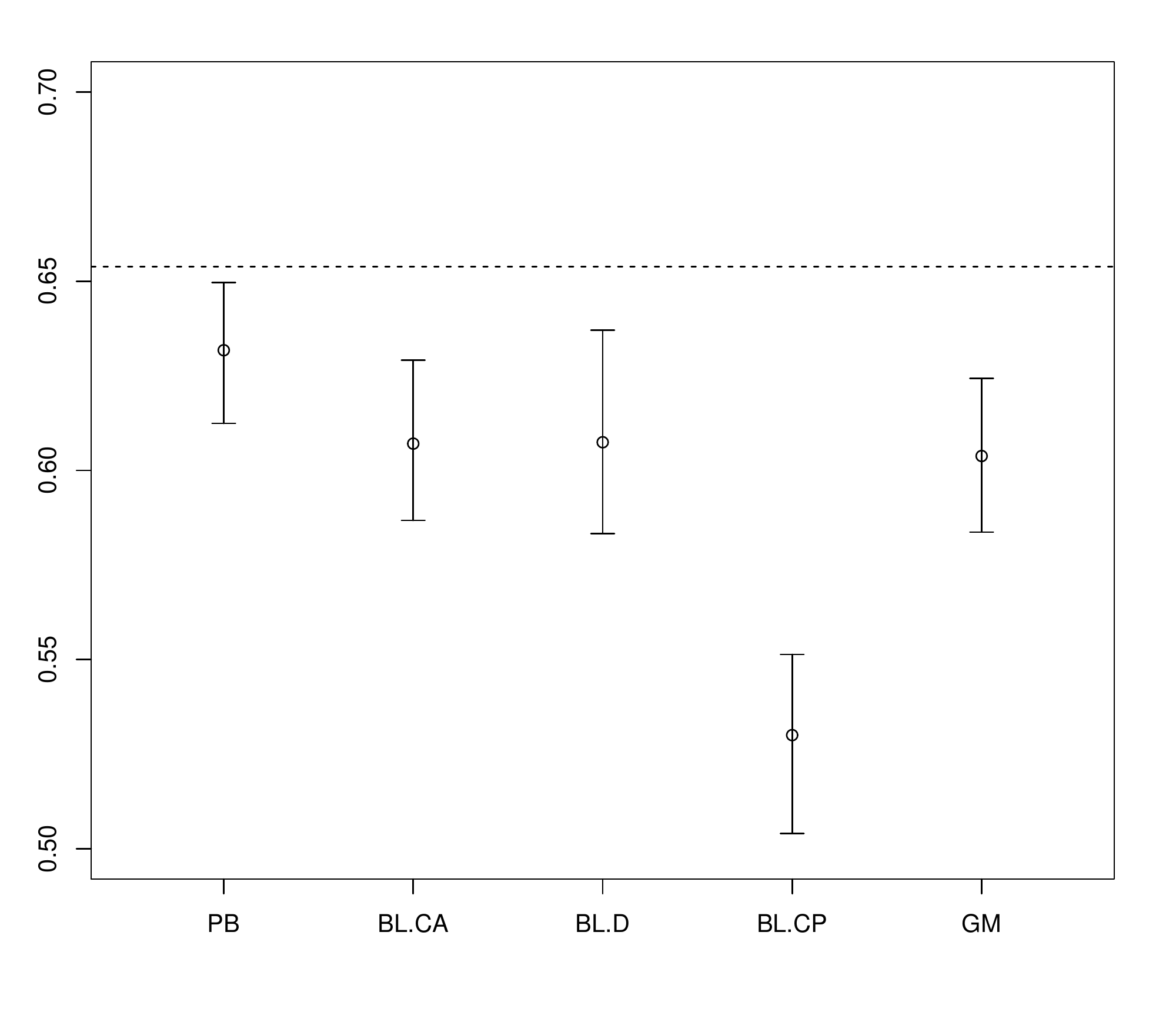} }\label{fig:HSLFa}}%
    \qquad
     \subfloat[Non-seed PMRs.]{{\includegraphics[scale=.30]{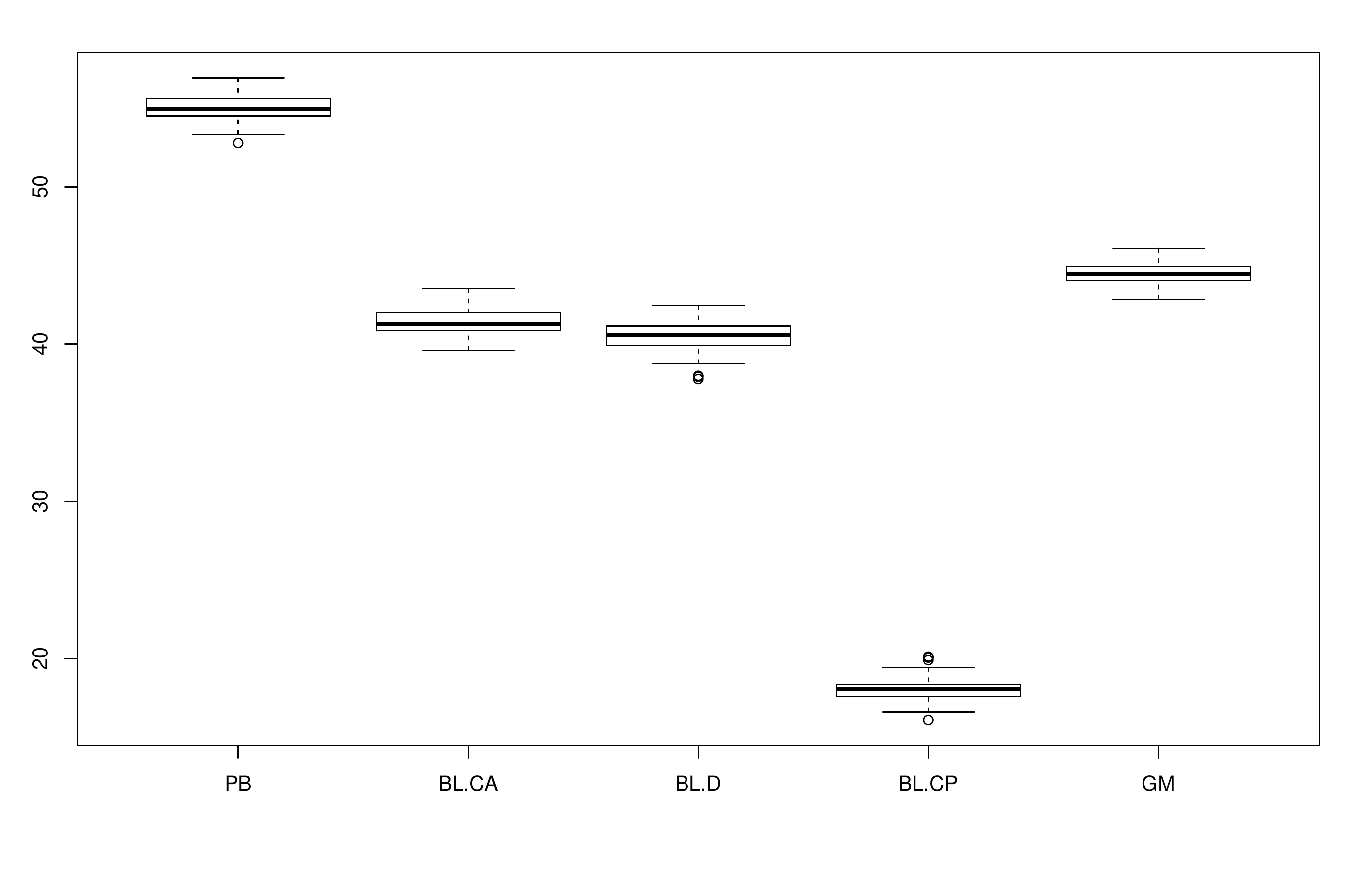}} \label{fig:HSLFb}}%
    \caption{Results from the HSLF simulation. Left panel (a) displays 95\% intervals for the 100 posterior means for \textsf{math } coefficient. Dashed horizontal line indicates true value of coefficient. Right panel (b) displays average posterior match rates from 100 simulations.  In both panels, PB stands for using the GAZM on the perfectly blocked data, BL.CA for using BLASE with the concentrated and appropriate prior, BL.D for using BLASE with the diffuse prior, BL.CP for using BLASE with the concentrated but poorly specified prior, and GM for using the GAZM on the faulty data.}
    \label{fig:HSLF}
\end{figure}

We next examine the performance of BLASE with low levels of faulty values, focusing on the high seed, low fault (HSLF) scenario. Key results are displayed in Figure \ref{fig:HSLF} and Table \ref{Ch2::table::SimResults}.
Here, BLASE with the CA or D priors continues to estimate the coefficient of \textsf{math} reasonably well. 
However, gains from accounting for the faulty MVs are not as large as in the HSLF scenario. Specifically, the RMSE and match rates for BLASE with the CA or D priors are similar to those for the model that treats the faulty MV as an error-free BV. 

\begin{table}[t]
\begin{center}
  \begin{threeparttable}
    \caption{Summary of results of simulation studies.}
    \label{Ch2::table::SimResults}
		\begin{tabular}{cccccccc}
  \hline
\hline
  &  & \multicolumn{4}{c}{$Y_1$ Regression Linking Analysis Parameters} &  & \\ \cline{3-6} 
 			{}    &     & {Intercept}    & Math  & \textsf{prog} = Acad & \textsf{prog} = Voc  & {dPMR } & {dRMSE}       \\[+5pt]
 			\hline
			        & CA & \textbf{29.5} & \textbf{15.2} & \textbf{-14.7} & \textbf{42.8}  & \textbf{11.4} & \textbf{3.3}\\
			HSHF    & D  & \textbf{25.6} & \textbf{13.3}  & \textbf{-14.0} & \textbf{41.0}   & \textbf{10.1}& \textbf{2.5}\\
 			        & CP & \textbf{11.9}& \textbf{6.3}  & -3.1 & \textbf{55.0} & \textbf{5.5}& \textbf{.50} \\
 			         \\[-5pt]

 			        & CA & \textbf{.9} & \textbf{.5} & \textbf{2.1} & \textbf{4.4}   & \textbf{-1.2} & .006\\
 			HSLF    & D   & .16 & \textbf{.15} & \textbf{1.7} & \textbf{3.2}  & \textbf{-1.6} & \textbf{-.02} \\
 			        & CP & \textbf{-21.4} & \textbf{-11.0} & \textbf{-37.6} & \textbf{-69.1}  & \textbf{-10.6}  & \textbf{-.96} \\
 			         &    &&      && &&            \\
 			         
 			     & CA & {.3} & .06 & \textbf{-21.2}& 1.2&  \textbf{.95}& \textbf{.003}\\
 			LSHF   & D  & \textbf{5.9} &\textbf{2.8} & \textbf{-9.9} & \textbf{20.3}  & \textbf{3.3} & \textbf{.008}\\
 			        & CP & \textbf{9.9}& \textbf{4.8}  & \textbf{-4.4} & \textbf{27.0}  & \textbf{4.3}  & \textbf{.011}  \\
 			        &  & &&      &                &               & \\
 			         			 
 			        & CA & \textbf{-12.6}& \textbf{-6.3}  & .6 &  \textbf{-9.3} & \textbf{-1.9}  &  \textbf{-1.3} \\
 		LSLF	         & \multicolumn{1}{l}{D}   & \textbf{-9.9} & \textbf{-5.0} & \textbf{-6.2} & \textbf{-13.0}& \textbf{-1.3}  & \textbf{-.9}\\
 			        & CP & \textbf{-40.0}  & \textbf{-20.2} & \textbf{-45.7}  & \textbf{-82.3}  & \textbf{-9.9}&  \textbf{-3.4} \\
			 &    &&      && &&            \\
		t-values & & 27.7&  56.3 & 8.9 & -5.3 & $R^2$ & .46 \\ 
		\hline 
		\end{tabular}
    	\begin{tablenotes}[flushleft]
      		\item NOTE: Positive values indicate BLASE is more accurate than the file linking model that incorrectly treats the faulty MV as a BV, and negative values indicate the opposite. Letting GM stand for the latter modeling strategy, and using subscripts to indicate the method used, the entries under each predictor are averages across 100 replications of $100\left( \left| \hat{\theta}_{GM} -\theta \right| - \left| \hat{\theta}_{BLASE} - \theta\right|\right)/\theta$; the entries under dPMR equal 100$(PMR_{BLASE} - PMR_{GM})$; and, the entries under dRMSE are averages across 100 replications of $( RMSE_{GM} - RMSE_{BLASE} )/ RMSE_{PB}$. Bold values indicate differences more extreme than would be expected under Monte Carlo error (p-values $<.05$ in appropriate t-tests) if both modeling strategies had the same properties. Final row is the average t-values and $R^2$ values of the regression linking analysis across the 100 PB data sets.
    	\end{tablenotes}
 	 \end{threeparttable}
	\end{center}
\end{table}

As is evident in Figure \ref{fig:test} and Figure \ref{fig:HSLF}, using strong but inappropriate prior
beliefs to specify $(a_{\gamma_{2j}}, b_{\gamma_{2j}})$ can degrade the 
  performance of BLASE.  This is true in most other simulation
  scenarios as well, as seen in Table \ref{Ch2::table::SimResults}. The
  most severe degradation occurs in the low seed, low fault (LSLF) scenario, where the poorly specified prior distribution is informative enough to cause many unnecessary changes in blocks. In contrast, across all scenarios, the diffuse prior distribution performs comparably to the better of the CA and CP prior
  distributions.  Thus, in the absence of reliable prior information
  about the faultiness rates, we 
  recommend using diffuse priors in which the choice of
  $(a_{\gamma_{2j}}, b_{\gamma_{2j}})$ corresponds to a small prior
  sample size.
  
Focusing on results in Table \ref{Ch2::table::SimResults} for the diffuse prior distribution, we can discern general patterns over the four simulation scenarios.  First, in both HF scenarios, the results suggest that applying BLASE tends to result in improvements in PMR, lower average RMSE, and more accurate estimates of the $Y_1$ model coefficients than a model which incorrectly assumes that faulty MVs can be used as BVs. In contrast, in low fault scenarios there is less to be gained by applying BLASE. Second, in the HSLF scenario, BLASE using the CA or D priors can offer improvements, albeit smaller than those for the HSHF scenario. Apparently, even when only a modest number of records are reported with errors, BLASE can leverage the information in the $T_1$ seeds to identify plausible moves. Third, in the LSLF scenario, BLASE is not as effective as treating all the variables as BVs.  Without adequate numbers of $T_1$ seeds, BLASE does not have enough information to determine accurate moves. 
We note that, although not apparent from Table  \ref{Ch2::table::SimResults}, in scenarios with low numbers of $T_1$ seeds, none of the file matching algorithms provide accurate estimates of the coefficients; see the supplementary material for supporting results.

\clearpage

\section{Illustrations with NC Education Data} \label{Paper::sec::NCERDC}

In this section, we illustrate the performance of BLASE using data from the North Carolina Education Research Data Center \citep{Data}. The data consist of 83,957 record pairs 
containing end-of-grade (EOG) math test scores for $7^{th}$ graders from 2011 ($F_1$) and their accompanying $8^{th}$ grade scores from 2012 ($F_2$). We work specifically with a subset of 77,998 record pairs corresponding to the three largest ethnic groups in the data. These records are known from 
extensive clerical review to be perfectly matched, so we can treat the records as a truth for 
purposes of evaluating the performance of BLASE with genuine data. 

We use six in-common variables: {birth day}, {birth month}, {birth year}, {sex},  ethnicity (White, Black, Hispanic), and an identifier for the school attended in a given year (704 schools). Let $Y_1$ be the centered math EOG score for 2011, and let $Y_2$ be the centered math EOG score for 2012. As the linking analysis for BLASE, as well as for model evaluation, we use   
\begin{eqnarray} \label{Ch2::eq::NCERDC::linking} 
Y_1 &=& \beta_0 + \beta_1 Y_2 + \beta_2 I( \text{Black} ) + \beta_3 I(\text{Hispanic}) + \beta_4 I(\text{Male}) + \epsilon_1\\ 
Y_2 &=& \eta_0 + \eta_1 I( \text{Black} ) + \eta_2 I(\text{Hispanic}) + \eta_3 I(\text{Male}) + \epsilon_2,
\end{eqnarray}
where $\epsilon_1 \sim N(0,\sigma^2_1)$ and $\epsilon_2 \sim N(0,\sigma^2_2)$. In both models, we use a baseline of a female with ethnicity equal to white. 
In the perfectly matched data, the $Y_1$ regression has $R^2= .69$, and the $Y_2$ regression has $R^2= .1$.

Students can move to new schools over the year.  Thus, for the same student, the school attended in 2012 does not have to be the same as the school attended in 2011.  This invalidates the use of school attended as a BV.  However, the large majority of students stay in the same school over the year, making school attended an appealing variable to use as an in-common variable.  We therefore treat school attended in 2012 as an error-prone measurement of school attended in 2011, allowing us to treat school attended in 2011 as a MV and take advantage of the information this variable provides for matching. 

The data contain a high number of seeds, with 62,276 pairs (roughly 80\%) of the true matches agreeing uniquely on all six in-common variables.  About 9.2\% of the true matches disagree on at least one in-common variable, with school being the most commonly different. 
For purposes of illustration, we adopt the default assumption that all variable disagreements occur completely at random.   Accordingly, we fit BLASE using the diffuse prior, treating school as a MV and the other 5 variables as BVs. We run the model for 10000 iterations, with 1000 iterations discarded for burn-in, and compute the posterior credible intervals for the coefficients in \eqref{Ch2::eq::NCERDC::linking}. The posterior estimates are comparable to the maximum likelihood estimates obtained using the true linked data, as shown in Section 6 of the supplement. 
For comparison, we also use the GAZM treating all six variables as BVs.  Based on the simulation results of Section \ref{Paper::sec::Sim}, one would expect BLASE to perform similarly to 
this other model, given the low error rates and high fraction of $T_1$ seeds. Indeed, there is 
little difference in the sets of results. Additionally, in these data the school attended is not an important predictor of the other variables.
 As a result, BLASE has little information to encourage moves to the correct blocks.   

As more stringent tests of BLASE, we generate a new type of disagreement by perturbing ethnicity. This variable is an important predictor in \eqref{Ch2::eq::NCERDC::linking}, so that reporting errors in this variable could adversely affect the estimated regression.
 For all true matches $(i,i^{'})$ which disagree on school, we set $\hat{B}_{2i^{'}j} = B_{1ij}$ for $j$ corresponding to the school attended variable, thereby turning school into a BV. We then 
make 15\% of the values of ethnicity faulty; this corresponds to 75\% of the values for the non-seeds. We sample 75\% of the non-seeds completely at random, and for selected records 
sample a new value for ethnicity uniformly from incorrect values. 
We then treat ethnicity as a MV and the other five variables as BVs in BLASE.
We call this the block faults at random (FAR) scenario.
 
We also create a scenario in which the faulty value of ethnicity is not chosen completely at random. 
Specifically, for each record  chosen to be faulty, we make Black and Hispanic students have a reported value of 
white, and we make white students have a reported value of Black or Hispanic, chosen with equal probability.  
We call this the faulty not at random (FNAR) scenario.  
The FNAR scenario assesses the performance of BLASE when its model assumptions do not match the 
error generating process for the data. 

\begin{figure}[!t]
  \centering
  \includegraphics[scale=.6]{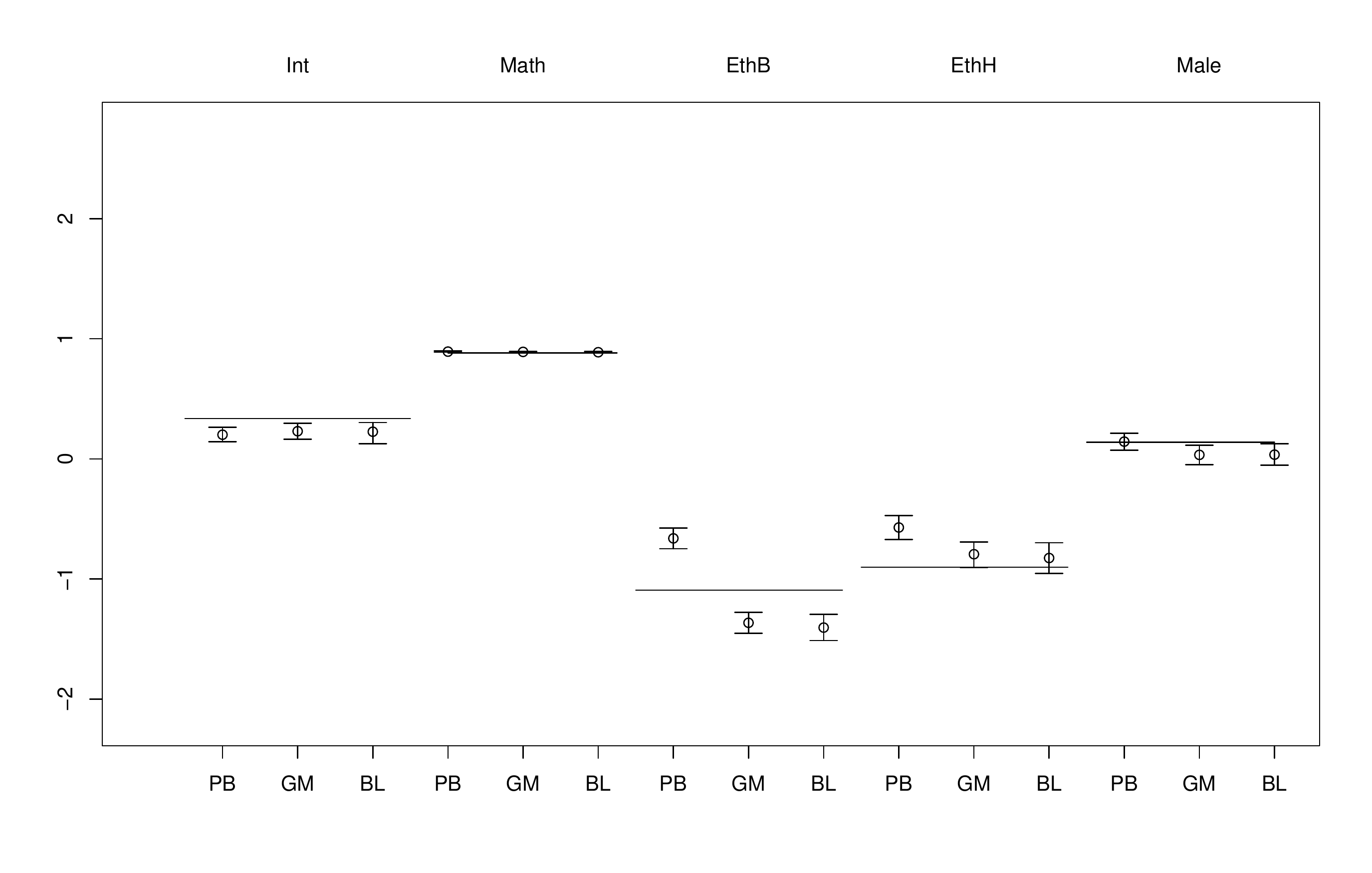}
  \caption{95\% posterior intervals for the coefficients of the linking analysis in the FAR simulation. Horizontal bars are OLS estimates from the correctly linked data.  PB stands for using the GAZM on the perfectly blocked data, BL for using BLASE with the diffuse prior, and GM for using the GAZM on the faulty data.  EthB stands for Black, and EthH stands for Hispanic.  For the math coefficient, the correctly linked estimate is .883.  The PB mean is .893 (.889,.898), GM mean is .891 (.886,.896), and BL mean is .888 (.882,.896). Results are from MCMC chain run for 10000 iterations, thinned every two draws.}
  \label{Paper::fig:FAR}
\end{figure}

\begin{figure}[!t]
  \centering
  \includegraphics[scale=.6]{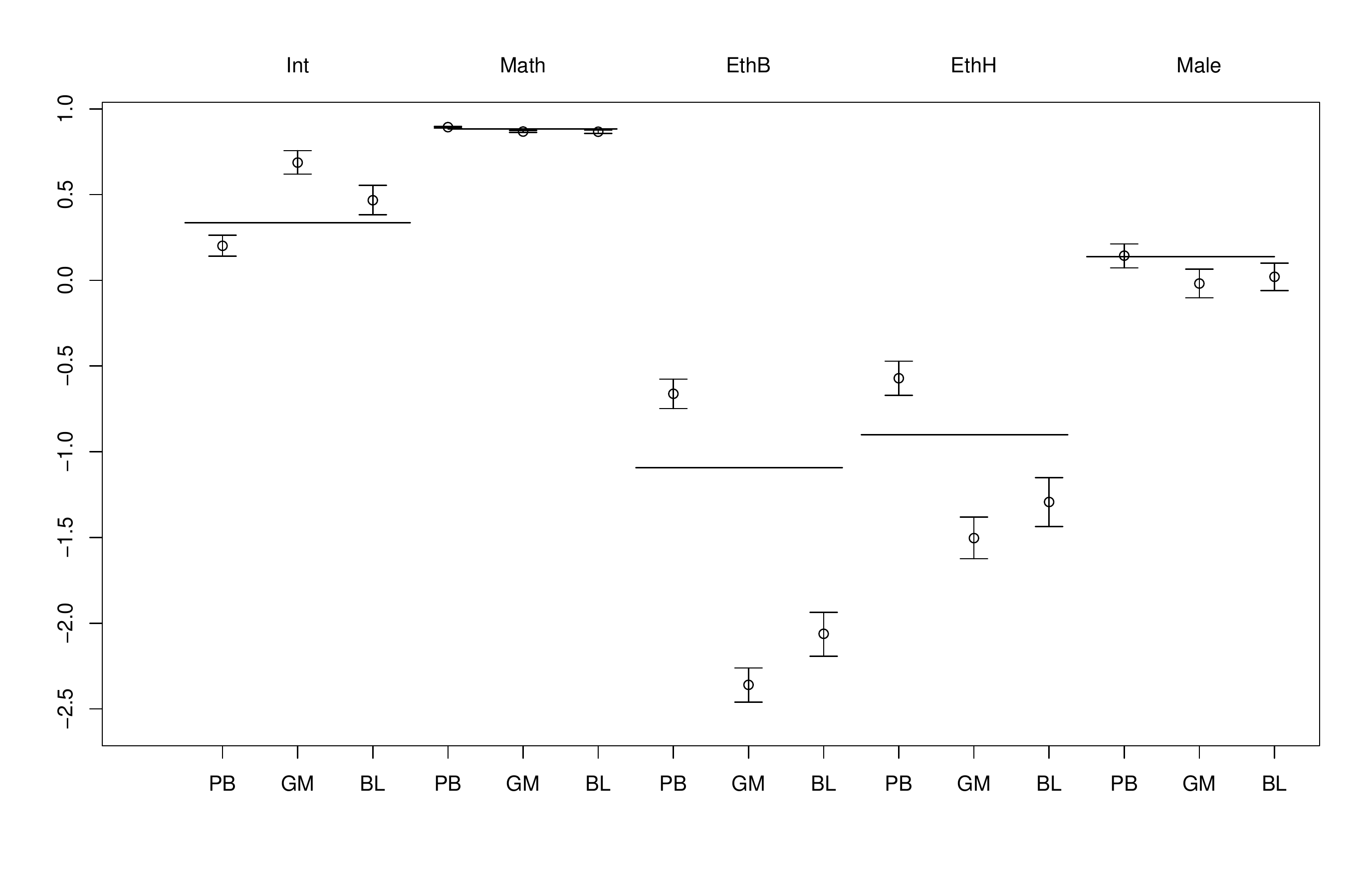}
  \caption{95\% posterior intervals for the coefficients of the linking analysis in the FNAR simulation. Horizontal bars are OLS estimates from the correctly linked data.  PB stands for using the GAZM on the perfectly blocked data, BL for using BLASE with the diffuse prior, and GM for using the GAZM on the faulty data.  EthB stands for Black, and EthH stands for Hispanic.  For the math coefficient, the correctly linked estimate is .883, the PB mean is .893 (.889,.898), GM mean is .867 (.862,.873), and BL mean is .867 (.856,.875).}
  \label{Paper::fig::FNAR}
\end{figure}

Figure \ref{Paper::fig:FAR} displays the results from the FAR simulation. This simulation mirrors the characteristics of the high seed, low fault (HSLF) simulation from Section \ref{Paper::sec::Sim}. On the whole, the inferences with BLASE are similar to those obtained with the perfectly linked data, as shown in Figure \ref{Paper::fig:FAR}. For example, for the math coefficient, the BLASE posterior mean is .893 while the estimate from the correctly linked data is .883.  Arguably, inferences from BLASE are no worse than inferences based on the perfectly blocked data. 
To see whether accounting for errors in the MV makes much practical difference in inferences,  we also estimate the GAZM treating all values as BVs.
Differences in the two sets of inferences are minor, as might be expected in a setting that mimics the HSLF simulations.  Apparently, the $T_1$ seeds are representative of the relationships in the data,
 and the signal from these complete cases dominates the aberrant records with faulty MVs. 
BLASE does result in larger posterior uncertainty about $\Theta$ than the model that disregards potential errors in the MVs, 
reflecting the fact that the true values of the MVs are unknown.  The PMRs for both models are generally around 87\%.

Figure \ref{Paper::fig::FNAR} displays the  results from the FNAR scenario. The estimated coefficients from BLASE are further from the coefficients from the perfectly linked data 
than they were in the FAR scenario. This is to be expected, since the reporting model in the default application of BLASE does not reflect the FNAR error generation process. 
Interestingly, using a BLASE model that allows for errors in MVs, even with a mis-specified error generation process, results in more accurate estimates than the model that 
treats ethnicity as a BV.  The increase in accuracy is accompanied  by an increase in PMR from 82\% for the ethnicity-as-a-BV model to 86\% for BLASE. 
Apparently, by allowing the in-common variables to move off their reported values, BLASE can use information from the
 seeds to improve the file matching, even though it uses a mis-specified reporting model.

\section{Concluding Remarks}\label{Paper:concl}

The simulation results suggest that, in scenarios with high numbers of $T_1$ seeds and a high number of faulty values in the MVs, using BLASE can result 
in more accurate estimates of the linking analysis parameters and higher match rates compared to a file matching procedure that wrongly disregards reporting errors by treating the faulty MVs as BVs.
In the simulations with high numbers of $T_1$ seeds and small rates of faulty values, accounting for errors in the MVs with BLASE offers little gain over disregarding the errors, although BLASE does 
appropriately inflate posterior uncertainty in these scenarios.    
When using BLASE, we recommend diffuse prior values, especially absent reliable information
 to think otherwise, as strongly concentrated prior distributions reflecting inaccurate prior beliefs can degrade the performance of BLASE.

When MVs are subject to errors, an alternative to BLASE is to include the MVs with errors as part of the linkage analysis models, leaving only BVs for matching, and use the GAZM.    
This alternative approach trades modeling the reporting errors for modeling the relationships between the faulty MVs and the other variables on the file.
Whether this is preferable to BLASE is particular to the problem of interest. For an illustrative example, when a faulty MV is conditionally independent of other variables, including it in the linking analysis 
is unlikely to improve the quality of matches. More generally, analysts who discard the information on the faulty MV from one of the files, say $F_1$, could sacrifice information that is useful for linking. For example, when the MV has the same value in both files for a high fraction of true matches, 
the information from the MV in $F_1$ could improve the matching for most records in the files, even if it does not do so for all records. As a final consideration, suppose that some of the linking variables subject to errors are multinomial with many levels, like school or county of residence.  There may be settings where, when one acts as if these variables are present in only $F_2$ (or $F_1$) and includes them in the linkage analysis, the block sizes without this multinomial variable are too large to be practically useful for a GAZM or other file matching method based on blocking on the BVs alone. 

A key step in BLASE, and any file matching strategy, is to decide which variables to treat as BVs and as MVs. Ideally, one can identify true matches through 
pre-processing, e.g., with clerical review of a sample of cases, so as to examine rates of error in the MVs. See \citet{herzog2007data} for general advice on selecting blocking variables.  Absent such  information, analysts can 
use BLASE with the diffuse prior, as well as the GAZM, with multiple subsets of variables to define the BVs in  sensitivity analyses.

In the variant of BLASE presented here, we assume that the joint distribution of BVs and MVs is the same in both files.  As a partial check of this assumption, 
analysts can evaluate the empirical distributions of the BVs in $F_1$ and $F_2$.  When dissimilar, assuming a common distribution may be unreasonable.  
Comparing the distributions of the MVs is not as straightforward, as they are potentially subject to errors. Of course,  checking any modeling assumption when data are incomplete or 
faulty is a difficult task.  One possibility is to   
estimate the model in Sections  \ref{model:errors} and \ref{model:DP} for $F_2$ separately disregarding information from $Y_2$, and generate posterior predictive draws of plausible values of $\boldsymbol{B}_2$.
If their empirical distributions look  dissimilar to the distributions in $\boldsymbol{B}_1$, then assuming  a common distribution may be unreasonable.
Quantifying the impact on inferences and match rates of departures from the common distribution assumption is an interesting topic for future research.

The modeling framework underpinning BLASE offers flexibility for tuning specifications to different assumptions.  For example, if 
one believes the MVs in both $F_1$ and $F_2$ are subject to errors, one can allow errors in $F_1$ using definitions akin to \eqref{Ch2::eq::ErrorVector} and similar modeling approaches.
Although we have not tried this model, based on results in \citet{manriqueReiterUpdated},
we conjecture that it should perform like BLASE but with increased posterior uncertainty, provided that the model for $(\boldsymbol{B}, \boldsymbol{E})$ is approximately correct and that there are a large number of $T_1$ seeds. Without these conditions, we conjecture that the performance of BLASE will be poorer than  
that observed here. One also can specify alternative models for the reported values and the errors, as mentioned in Section \ref{model:errors}, particularly when external information is available that indicates 
errors may be systematic.  For example, suppose that one considers school attended as a MV as we did in Section \ref{Paper::sec::NCERDC}.  If external information indicates that
students of different demographic backgrounds are more likely to switch schools, one can use a logistic regression for differences in the reported school variable as a function of those demographic 
variables.  Or, if rates of transfer from one school to another are available, one could use these rates as the reporting model.  Once again, without some validation sample from pre-processing 
it is difficult to evaluate the reasonableness of these assumptions.  Thus, another fruitful topic for future research is to determine how departures from assumptions about the models for $\boldsymbol{B}$ and $\boldsymbol{E}$, and from the linking analysis model, affect the quality of inferences from BLASE, as well as when such departures offset potential gains from accounting for errors in the MVs.

\appendix 

\section*{Acknowledgements}
This work was supported by NSF Grant SES 1131897, and by the Duke University Energy Initiative Energy Research Seed Fund, with co-funding from the Information Initiative at Duke. Data was provided by the North Carolina Education Research Data Center. 


\bibliographystyle{./jasa} 
\cleardoublepage
\normalbaselines 
\bibliography{./NicoleRef} 

\pagebreak 

\bigskip
\begin{center}
{\large\bf SUPPLEMENTARY MATERIAL}
\end{center}

The following sections comprise supplementary material to complement the main article.  In Section \ref{Supp::MHProposals}, we 
expand on the updating process for any record $i \in F_2$ that is not a $T_1$ or $T_2$ seed. In Section \ref{SuppA::ComputingA}, we derive the Metropolis-Hastings (MH) acceptance ratio 
for $( {(E_{2ij^{*}})}^{*},  {(B_{2ij^{*}})}^{*} )$. In Section \ref{SuppA::DPFCs}, we detail the full conditionals for the truncated Dirichlet process mixture of products of multinomial model. These are adapted from \cite{IshwaranJames2001}. In Section \ref{SuppA::DataGen}, we present the data generation process for the simulation studies discussed in the main text. In Section \ref{SuppA::Results}, we present extended results for these studies. In Section \ref{Supp::NCERDC}, we present a supplemental plot for the NCERDC illustration.

\section{Further details on proposing \texorpdfstring{$(\boldsymbol{E}_2^{*}, \boldsymbol{B}^{*}_2)$}{C}} \label{Supp::MHProposals} 

In this section, we describe in detail the updating process for any record $i \in F_2$ that is not a $T_1$ or $T_2$ seed. Some of this material is repeated from the main text for completeness. 

The most complicated step of BLASE involves proposing values for
$\boldsymbol{E}_2$ and $\boldsymbol{B}_2$. The basic strategy is as
follows. For each record $i \in F_2$ that is not a $T_1$ or $T_2$
seed, we propose a new value of $({(E_{2i})}^*, {(B_{2i})}^{*})$.  We accept or reject that proposal using a MH step, with an acceptance ratio defined in Section \ref{SuppA::ComputingA} of this supplement. We repeat these two steps independently for all non-seed records in $F_2$, resulting in $(\boldsymbol{E}_2^{(s+1)}, \boldsymbol{B}_2^{(s+1)})$.  Using independent, record by record updates allows for the updates to be done in parallel using multiple processors (although we do not utilize this feature in the examples).

We now describe the updating process for any record $i \in F_2$ that is
not a $T_1$ or $T_2$ seed. For any eligible record $i$, we randomly select one non-seed MV field $j^{*}$ uniformly from all non-seed MV fields for that record.
Thus, at any iteration in the MCMC, $E_{2i}^{(s)}$ and $E_{2i}^{(s+1)}$ differ on at
most one field, and record $i$ may move only to some pool
$k^{*}$ that disagrees with its current pool $k$ on exactly 
one field. This simplifies the computation of the MH acceptance ratio derived in Section \ref{SuppA::ComputingA}.

\subsection{\texorpdfstring{${(E_{2ij^{*}})}^{*}$}{E2} } \label{Supp:ssec:ProposingE}

We propose a value for $E_{2ij^{*}}$ from 
\begin{align}
    ({E_{2i{j}^{*}}})^{*}|\gamma_{2}^{(s)},j^{*} \sim Bernoulli( {\gamma_{2j^{*}}}^{(s)} ).
\end{align}

\subsection{\texorpdfstring{${(B_{2ij^{*}})}^{*}$}{B2} } \label{Supp:ssec:ProposingB}

When $({E_{2i{j^{*}}} })^{*} = E_{2ij^{*}}^{(s)} = 0$, we set $E_{2i}^{(s+1)} = E_{2i}^{(s)}$ and $B_{2i}^{(s+1)} = B_{2i}^{(s)}$. Otherwise, we set ${(E_{2ij})}^{*} = E_{2ij}^{(s)}$ for all $j \neq j^{*}$, and propose a new $({B_{2ij^{*}}})^{*}$ via one of three general patterns. 
\begin{subequations}
\begin{align}
\text{If } {(E_{2i{j^{*}}})}^{*} = 0 \text{ and } E_{2ij^{*}}^{(s)} = 1 & \Rightarrow \text{Propose } {(B_{2ij^{*}})}^{*} = \hat{B}_{2ij^{*}}.  \label{eq::Bprop::line1A} \\ 
\text{If } {(E_{2i{j^{*}}})}^{*} = 1 \text{ and } E_{2ij^{*}}^{(s)} = 0 & \Rightarrow \text{Propose } {(B_{2ij^{*}})}^{*} \neq \hat{B}_{2ij^{*}}. \label{eq::Bprop::line2A} \\ 
\text{If } {(E_{2i{j^{*}}})}^{*} = 1 \text{ and } E_{2ij^{*}}^{(s)} = 1 & \Rightarrow \text{Propose } {(B_{2ij^{*}})}^{*} \not\in \lbrace \hat{B}_{2ij^{*}} \cup {B}_{2ij^{*}}^{(s)} \rbrace. \label{eq::Bprop::line3A}
\end{align}
\end{subequations}

To aid understanding, call the pool defined by $\hat{B}_{2i}$ the {reported pool} for record $i$, and any pool that is different from the reported pool an {alternate pool}. Hence, \eqref{eq::Bprop::line1A} shifts $i$ from an alternate pool to a reported pool, \eqref{eq::Bprop::line2A}
shifts from a reported pool to an alternate pool, and \eqref{eq::Bprop::line3A} shifts $i$ from one
alternate pool to another. 

In the cases of \eqref{eq::Bprop::line2A} and \eqref{eq::Bprop::line3A}, we propose ${(B_{2ij^{*}})}^{*}$ by drawing from the latent class model described in Section 2.5 of the main paper. However, we restrict ${(B_{2ij^{*}})}^{*}$ to ensure the conditions in \eqref{eq::Bprop::line2A} and \eqref{eq::Bprop::line3A} are not violated. We also disallow  ${(B_{2ij^{*}})}^{*}$ that result in MV combinations not observed in $\boldsymbol{\hat{B}}_1$. 

We define the set of \textit{legal pools} for each record $i$ to contain pools defined by $\boldsymbol{\hat{B}}_1$ and pools that fulfill the restrictions in \eqref{eq::Bprop::line1A}-\eqref{eq::Bprop::line3A}. The move restriction requires adapting the proposal step for $\boldsymbol{B}^{*}$ to only allow proposals which lead records to move to legal pools. In the proposal, we normalize each $\phi_{hj^*}$ over the set of legal pools. Specifically, define 
\begin{align}
D_{(b)} = \lbrace x \in (1:d_{j^{*}}) | x \neq {\hat{B}_{2ij^{*}}} \text{ and } B_{2ij^{*}} =x \text{ implies } B_{2i} \in \boldsymbol{\hat{B}}_1 \rbrace.
\end{align}
$D_{(b)}$ contains the set of possible proposals for $B_{2ij^{*}}$ under \eqref{eq::Bprop::line2A}. Likewise, define
\begin{align}
D_{(c)} = \lbrace x \in (1:d_{j^{*}}) | x \neq {\hat{B}_{2ij^{*}}}, x \neq {{B}_{2ij^{*}}}^{(s)} \text{ and } B_{2ij^{*}} =x \text{ implies } B_{2i} \in \boldsymbol{\hat{B}}_1  \rbrace.
\end{align}
$D_{(c)}$ then contains the set of possible proposals for $B_{2ij^{*}}$ under \eqref{eq::Bprop::line3A}. Define
\begin{align} \label{eq:phi}
    {\Phi_{ij^{*}m}}^{(s)} = \left\{
     \begin{array}{ll}
          {\phi_{h j^{*} m }}^{(s)}/ \sum_{ x \in D_{(b)} } {\phi_{h j^{*} x}}^{(s)}  & : m \in D_{(b)}, {(E_{2i{j^{*}}})}^{*} = 1 \text{ and } E_{2ij^{*}}^{(s)} = 0  \\ 
     {\phi_{h j^{*} m }}^{(s)}/ \sum_{ x \in D_{(c)} } {\phi_{h j^{*} x}}^{(s)}  & : m \in D_{(c)}, {(E_{2i{j^{*}}})}^{*} = 1 \text{ and } E_{2ij^{*}}^{(s)} = 1 \\ 
     0  & : \text{otherwise}. 
     \end{array}
   \right. 
\end{align} 
Let $\Phi_{ij^{*}}^{(s)} = \lbrace \Phi_{ij^{*}m}^{(s)} | m \in 1,\dots,d_j^{*} \rbrace$.  We then obtain the proposal for $B_{2ij^{*}}$ as follows. Under \eqref{eq::Bprop::line1A}, propose ${(B_{2ij^{*}})}^{*} = \hat{B}_{2ij^{*}}$. Under \eqref{eq::Bprop::line2A} or \eqref{eq::Bprop::line3A}, propose ${(B_{2ij^{*}})}^{*}$ by sampling from $Multinomial( 1; \lbrace 1, \dots, d_j^{*} \rbrace, \Phi_{ij^{*}}^{(s)})$. 

Requiring ${(B_{2ij^{*}})}^{*}$ to correspond to a pool in $\boldsymbol{\hat{B}}_1$ is not necessary for the sampler.  However, under the assumption that $\boldsymbol{\hat{B}}_1 =  \boldsymbol{B}_1$, we find that imposing this restriction improves  computational efficiency of the sampler. Pool moves are intended to allow records from $F_2$ to move from a potentially faulty pool into a pool where their true match lies. Restricting pool moves to combinations in $\boldsymbol{B}_1$ means that proposals more efficiently explore pools that contain possible matches.  It is possible that the MVs for some record $i$ in $F_2$ are faulty and that $B_{2i}$ actually is not spanned by $\boldsymbol{\hat{B}}_1$. In this case, and under the assumption that $\boldsymbol{\hat{B}}_1=  \boldsymbol{B}_1$, this $F_2$ record does not have a match in $F_1$. In this case, the restriction to proposing pool moves defined by $\boldsymbol{\hat{B}}_1$ would prevent the $F_2$ record from ever moving to its true pool. This should be uncommon when $F_1$ spans most of the typical possible combinations. We expect that, in most cases, the increase in computational efficiency gained by the restriction out-weighs any potential reduction in accuracy from possibly assigning the incorrect pool to a record in $F_2$ with no match in $F_1$.  If the combinations in $F_1$ do not span most of the possible pool combinations, one can remove the restriction to propose only MV combinations observed in $\boldsymbol{\hat{B}}_1$. In this case, define $ D_{(b)} = \lbrace x \in (1:d_{j^{*}}) | x \neq {\hat{B}_{2ij^{*}}} \rbrace$, and $D_{(c)} = \lbrace x \in (1:d_{j^{*}}) | x \neq {\hat{B}_{2ij^{*}}}, x \neq {{B}_{2ij^{*}}}^{(s)}  \rbrace$, and propose ${(B_{2ij^{*}})}^{*}$ by sampling from $Multinomial( 1; \lbrace 1, \dots, d_j^{*} \rbrace, \Phi_{ij^{*}}^{(s)})$, where $\Phi_{ij^{*}}^{(s)}$ is defined in \eqref{eq:phi}. 

\subsection{Proposing \texorpdfstring{${C}^{*}$}{Cstar} } \label{Supp:ssec:ProposingC}

Based on the sampled value of ${ (B_{2 i j^{*}} )}^{*}$, we shift record $i$ from the pool $k$ defined by ${B_{2i}} ^{(s)}$ into the pool $k^{*}$ defined by ${ ( B_{2 i } )}^{*}$. Pool $k$ loses a record from $F_2$ while pool $k^{*}$ gains one. We then add or remove dummy records to rebalance both affected pools. 
When pool $k^{*}$ contains at least one record from $F_{1k^*}$ matched to a dummy record, we temporarily replace one of the dummy records with the record $i$ shifted into the pool. When pool $k^{*}$ does not contain a dummy record in $F_{2k}$, we impute the missing $Y_1$ for record $i$ from
 $f(Y_1| Y_2^{(s)}, C^{(s)},\boldsymbol{B}^{*}, \Theta^{(s)})$. For pool $k$, if record $i$ was matched to a dummy record when in $F_{2k}$, we temporarily remove the dummy record from pool $k$. Otherwise, we impute the missing $Y_2$ value for record $i$ from $f(Y_2| Y_1^{(s)}, C^{(s)},\boldsymbol{B}^{*}, \Theta^{(s)})$. Let ${Y}_1^{*}$ comprise ${Y}_1$ and the imputed values of $Y_1$ after the pool move, and let ${Y}_2^{*}$ comprise ${Y}_2$ and the imputed values of $Y_2$ after the pool move. Similarly, let $(Y_{1(k)})^{*}$ and $(Y_{2(k)})^{*}$ comprise the values of $Y_1^{*}$ or $Y_2^{*}$ in pool $k$ after moving record $i$. We define $(Y_{1(k^{*})})^{*}$ and $(Y_{2(k^{*})})^{*}$ analogously. 

Once pools are balanced, two new permutations ${( C_{k})}^{*}$ and ${(C_{k^{*}})}^{*}$ need to be proposed. These permutations define the linkage structure within the proposed pooling structure defined by ${(\boldsymbol{B}_2)}^{*}$.

\subsubsection{Pool k: Case 1} \label{Supp:ssec:ProposingC:Case1}

Conditional on $C_{k}^{(s)}$, suppose record $i$ is matched to a dummy record $*$ in pool $k$. If $i$ is shifted from the pool, the dummy record is no longer necessary, and may be removed from $k$. Because the permutation $C_{k}^{(s)}$ defines the linkage structure for all remaining records in the pool, we propose $(C_{k})^{*} = C_{k}^{(s)}$, with the record $i$ removed. 

\subsubsection{Pool k: Case 2} \label{Supp:ssec:ProposingC:Case2}

Conditional on $C^{(s)}$, suppose record $i$ is not matched to a dummy record in pool $k$. We impute the missing $Y_2$ value for record $i$ from $f(Y_2| Y_1^{(s)}, C^{(s)},\boldsymbol{B}^{*}, \Theta^{(s)})$. $(Y_{2(k)})^{*}$ then comprises this imputed value and the values of $Y_{2(k)}^{(s)}$ which remain in pool $k$ after record $i$ is removed. $Y_{1}$ is unaffected by this pool move, so $(Y_{1(k)})^{*} = Y_{1(k)}^{(s)}$. If pool $k$ is small, we propose $(C_{k})^{*}$ using the Exact Step. In this case, 
\begin{align}  \label{ssec::Ch2::ComputingA::J:Smallk}
Pr( {(C_{k})}^{*} = C_{kl} \mid C_{k}^{(s)},{Y}_1^{*}, {Y}_2^{*}, \boldsymbol{B}^{*}, \Theta^{(s)})= 
\frac{ f( {({Y}_{1(k)})}^{*}, {({Y}_{2(k)})}^{*} \mid \Theta^{(s)}, C_{k\ell}, \boldsymbol{B}^{*})  }{ \sum_{l' =1}^{c_k^{*}!} f(  {({Y}_{1(k)})}^{*}, {({Y}_{2(k)})}^{*} \mid \Theta^{(s)}, C_{k\ell^{'}}, \boldsymbol{B}^{*}) }. 
\end{align}
If $k$ is large, we propose $(C_{k})^{*}$ by selecting two non-seed records and switching their positions, just as in the Switch step described in the main text. More formally, let $\mathcal{C}_{(k)}$ denote the set of all permutations which can be obtained by selecting two non-seed records and switching their positions. Let $c_k^{*}$ denote the value of $c_{k}$ after the record move. Then for a large pool, we have 
\begin{align} \label{ssec::Ch2::ComputingA::J:Largek}
Pr( {(C_{k})}^{*} = C_{kl} \mid C_{k}^{(s)},{Y}_1^{*}, {Y}_2^{*}, \boldsymbol{B}^{*}, \Theta^{(s)}) & = 
 \left\{
    \begin{array}{ll}
\frac{ ( c_k^{*} - 2 )! 2! }{ c_k^{*}! } & : \text{ if } C_{k\ell} \in \mathcal{C}_{k} \\
0 &: \text{ otherwise }.  
\end{array}
\right. 
\end{align}

\subsubsection{Pool \texorpdfstring{$k^{*}$}{kstar}: Case 1} \label{Supp:ssec:ProposingC:Case3}

Conditional on $C_{k^{*}}^{(s)}$, suppose at least one record in $F_{1(k^{*})}^{(s)}$ in pool $k^{*}$ is matched to a dummy record, denoted $*$. Conditional on ${(B_{2i})}^{*}$, we move record $i$ into pool $k^{*}$ and replace the dummy record with $i$. If there is more than one $F_2$ dummy record in the pool, we randomly select one to replace with $i$. Let $r^{'}$ denote this selected dummy record. Then $(Y_{2(k^{*})r^{'}})^{*} = Y_{2i}$ and $(Y_{2(k^{*})r})^{*} = Y_{2(k)r}^{(s)}$ for all $r = \lbrace 1,\dots, c_{(k^{*})}^{(s)} | r \neq r^{'} \rbrace$. $Y_{1}^{(s)}$ is unaffected by this pool move, so $(Y_{1(k^{*})})^{*} = Y_{1(k^{*})}^{(s)}$. There are no dummy records created for this case. We propose $(C_{k^{*}})^{*}$ analogously to Section \ref{Supp:ssec:ProposingC:Case2}. 

\subsubsection{Pool \texorpdfstring{$k^{*}$}{kstar2}: Case 2} \label{Supp:ssec:ProposingC:Case4}

Conditional on $C_{k^{*}}^{(s)}$, suppose no records in $F_{1(k^{*})}^{(s)}$ in pool $k^{*}$ are matched to a dummy record. Conditional on ${(B_{2i})}^{*}$, we move record $i$ into pool $k^{*}$ and impute the missing $Y_1$ for record $i$ from $f(Y_1| Y_2^{*}, C^{(s)},\boldsymbol{B}^{*}, \Theta^{(s)})$. Here $(Y_{2(k^{*})})^{*} = ((Y_{2(k^{*})})^{(s)},Y_{2i})$ and $(Y_{1(k^{*})})^{*}$ comprises $Y_{1(k^{*})}^{(s)}$ and the imputed $Y_1$ value. We propose $(C_{k^{*}})^{*}$ analogously to Section \ref{Supp:ssec:ProposingC:Case2}. 

\clearpage 

\section{{Sampling \texorpdfstring{$(\boldsymbol{E}_2^{*}, \boldsymbol{B}^{*}_2)$}{(E2,B2)}}: The MH Acceptance Ratio} \label{SuppA::ComputingA} 

Once values for ${({E}_{2i}})^{*}$ and ${(B_{2i})}^{*}$ have been proposed, we simultaneously accept or reject both proposals using a MH step. In this section, we derive the acceptance ratio used in this MH step. 

\subsection{Notation} 

Let $\boldsymbol{B}^{*} = ( \boldsymbol{B}_1, \boldsymbol{B}_2^{*})$, where the $i^{th}$ row of $\boldsymbol{B}_2^{*}$ is defined by the proposal ${(B_{2i})}^{*}$, and ${({B}_{2i^{'}})}^{*} = {(B_{2i^{'}})}^{(s)}$ for all $i^{'} \neq i$. Similarly, let $\boldsymbol{E}^{*} = ( \boldsymbol{E}_1, \boldsymbol{E}_2^{*})$, where the $i^{th}$ row of $\boldsymbol{E}_2^{*}$ is defined by the proposal ${(E_{2i})}^{*}$, and ${({E}_{2i^{'}})}^{*} = {(E_{2i^{'}})}^{(s)}$ for all $i^{'} \neq i$.

For each proposed pool move, the acceptance ratio $A$ takes a standard MH form, namely $A = p( \cdot) \times J (\cdot)$.
$J(\cdot)$, the ratio of transition probabilities from the proposal distribution, is  
\begin{align} \label{Ch2::eq::A::J}
J( \cdot) =  \frac{p( E_{2ij^{*}}^{(s)},B_{2ij^{*}}^{(s)},C^{(s)}| {\boldsymbol{E}}^{*},(C)^{*},{\boldsymbol{B}}^{*},Y_1^{*},Y_2^{*},\Psi^{(s)}, \boldsymbol{\hat{B}}, \Theta^{(s)},\gamma^{(s)} ) }{ p( (E_{2ij^{*}})^{*},(B_{2ij^{*}})^{*},(C)^{*}| \boldsymbol{E}^{(s)},C^{(s)},{\boldsymbol{B}}^{(s)},Y_1^{(s)},Y_2^{(s)},\Psi^{(s)}, \boldsymbol{\hat{B}}, \Theta^{(s)}, \gamma^{(s)} ) }.
\end{align}

The ratio of the probability of the new state to the current state is denoted $p(\cdot)$, with 
\begin{align} 
p( \cdot) = A_1 \times A_2 \times A_3,
\end{align} 
where 
\begin{align} \label{Ch2::eq::A::A1} 
A_1 = \frac{ f( \hat{B}_{2ij^{*}} | \boldsymbol{E}_2^{*}, \boldsymbol{B}_2^{*})}
{f( \hat{B}_{2ij^{*}} | \boldsymbol{E}_2^{(s)}, \boldsymbol{B}_2^{(s)})} 
\frac{p({B}_{2ij^{*}} = {(B_{2ij^{*}})}^{*}|\Psi^{(s)})}{p({B}_{2ij^{*}} = {B}_{2ij^{*}}^{(s)}|\Psi^{(s)}) }
\frac{ p(E_{2ij^{*}} = {(E_{2ij^{*}})}^{*}|\gamma_{2j^{*}}^{(s)})}{ p(E_{2ij^{*}} = E_{2ij^{*}}^{(s)}|\gamma_{2j^{*}}^{(s)})}  \frac{p(C = C^{*}|\boldsymbol{B}_2^{*})}{p(C = C^{(s)}|\boldsymbol{B}_{2}^{(s)}) }, 
\end{align}
\begin{align} \label{Ch2::eq::A::A2}
A_2 =  \frac{ f(Y_{1}^{*}, Y_2^{*}| \Theta^{(s)}, C^{*}, \boldsymbol{B}^{*})}{ f(Y_1^{(s)}, Y_2^{(s)}| \Theta^{(s)}, C^{(s)}, \boldsymbol{B}^{(s)})},
\end{align} 
\begin{align} \label{Ch2::eq::A::A3}
A_3 = \frac{p(\Psi^{(s)}) p(\gamma^{(s)}) p(\Theta^{(s)})}{ p(\Psi^{(s)}) p(\gamma^{(s)}) p(\Theta^{(s)})}. \end{align}

\subsection{Component \texorpdfstring{$p(\cdot)$}{p()} } \label{ssec::Ch2::ComputingA::p} 

In \eqref{Ch2::eq::A::A3}, the priors for $\gamma$ and $\Psi$ are independent of $\boldsymbol{E}_2, \boldsymbol{B}_2$ or $C$, and therefore do not contribute to the acceptance ratio. The prior $p(\Theta)$ is user-specified. For example, if the linking analysis were a regression model, a $g$ prior might be used for the coefficients. This choice for $p(\Theta)$ is conditional on the permutation $C$ through reliance on the design matrix. In our simulation studies, where $\Theta = (\boldsymbol{\beta}, \sigma_1, \boldsymbol{\eta}, \sigma_2)$, the prior  $p(\Theta) \propto (1/\sigma_1^2) (1/\sigma_2^2)$ is independent of $\boldsymbol{E}_2, \boldsymbol{B}_2$ or $C$, so \eqref{Ch2::eq::A::A3} equals 1. 

The ratio ${f( \hat{B}_{2ij^{*}} | \boldsymbol{E}_2^{*}, \boldsymbol{B}_2^{*})}/
{f( \hat{B}_{2ij^{*}} | \boldsymbol{E}_2^{(s)}, \boldsymbol{B}_2^{(s)})} $ in \eqref{Ch2::eq::A::A1} is defined by
\begin{equation} \label{Supp::eq::MeasurementError} 
   f \left( \hat{B}_{2ij}| E_{2ij}, B_{2ij} \right) = \left\{
     \begin{array}{ll}
      B_{2ij}  &  : \text{if } E_{2ij} = 0 \\
      \text{Uniform discrete on all but } B_{2ij} & : \text{if } E_{2ij} = 1. 
     \end{array}
   \right.
\end{equation}
If $E_{2ij^{*}}=0$, $\hat{B}_{2ij^{*}} = B_{2ij^{*}}$ with probability 1. If $E_{2ij^{*}}=1$, \eqref{Supp::eq::MeasurementError} assigns equal mass for all $d_{j^{*}}$ choices of $\hat{B}_{fij}$ excluding $B_{fij}$. The ratio is then 
\begin{align} \label{ssec::Ch2::ComputingA::MesError} 
\frac{ f( \hat{B}_{2ij^{*}} | \boldsymbol{E}_2^{*}, \boldsymbol{B}_2^{*})}
{f( \hat{B}_{2ij^{*}} | \boldsymbol{E}_2^{(s)}, \boldsymbol{B}_2^{(s)})}  = \left\{
	\begin{array}{ll}
       1 & : {(E_{2ij^{*}})}^{*} = E_{2ij^{*}}^{(s)} = 1 \\
       {1}/({d_{j^{*}}-1}) & :  E_{2ij^{*}}^{(s)} = 0, {(E_{2ij^{*}})}^{*} = 1  \\ 
       d_{j^{*}}-1 & : E_{2ij^{*}}^{(s)} = 1, {(E_{2ij^{*}})}^{*} = 0.  \\ 
     \end{array}
\right.
\end{align} 

The ratio $A$ is not computed when ${(E_{2ij^{*}})}^{*} = E_{2ij^{*}}^{(s)} = 0$. With such a proposal, we set $B_{2ij^{*}}^{(s+1)}= \hat{B}_{2ij^{*}}$ and the MH step is complete. 

The contribution of the model for $\boldsymbol{B}$ to \eqref{Ch2::eq::A::A1}  is the ratio of the probability of  ${p({B}_{2ij^{*}} = {(B_{2ij^{*}})}^{*}|\Psi^{(s)})}$ to $p({B}_{2ij^{*}} = {B}_{2ij^{*}}^{(s)}|\Psi^{(s)}) $. Let the latent class assignment $z_i^{(s)}=h$. Then we have 
\begin{align} \label{eq::Ch2::ARatio::BPart}
\frac{p({B}_{2ij^{*}} = {(B_{2ij^{*}})}^{*}|\Psi^{(s)})}{p({B}_{2ij^{*}} = {B}_{2ij^{*}}^{(s)}|\Psi^{(s)})} & = \frac{ \phi_{ h j^{*} {(B_{2ij^{*}})}^{*}}^{(s)} }{ \phi_{ h j^{*} B_{2ij^{*}}^{(s)}}^{(s)} }.
\end{align} 
This ratio increases the MH acceptance ratio $A$ if we propose shifting $i$ to a value ${(B_{2ij^{*}})}^{*}$ which is more likely under $\Psi^{(s)}$ than its current $B_{2ij^{*}}^{(s)}$. 

The contribution of the model for $\boldsymbol{E}_{2}$ to \eqref{Ch2::eq::A::A1}  is similarly concise. For all $j \neq j^{*}$, ${( E_{2ij} )}^{*} = E_{2ij}^{(s)}$. Hence, we have: 
\begin{align} \label{ssec::Ch2::ComputingA::p::E} 
\frac{ p(E_{2ij^{*}} = {(E_{2ij^{*}})}^{*}|\gamma_{2j^{*}}^{(s)})}{ p(E_{2ij^{*}} = E_{2ij^{*}}^{(s)}|\gamma_{2j^{*}}^{(s)})}  = \left\{
     \begin{array}{ll}
       1 & : {(E_{2fij^{*}})}^{*} = E_{2ij^{*}}^{(s)} = 1 \\
       \frac{ \gamma_{2j^{*}} }{ 1- \gamma_{2j^{*} } } & : E_{2ij^{*}}^{(s)} = 0, {(E_{2ij^{*}})}^{*} = 1   \\ 
       \frac{ 1-\gamma_{2j^{*}} }{ \gamma_{2j^{*} } }  & :  E_{2ij^{*}}^{(s)} = 1, {(E_{2ij^{*}})}^{*} = 0.  \\ 
     \end{array}
   \right.
\end{align}

The final component of \eqref{Ch2::eq::A::A1}  is the ratio $p(C^{*}|\boldsymbol{B}^{*})/p(C^{(s)}|\boldsymbol{B}^{(s)})$. Within each MH step, only record $i$ moves pools. Specially, record $i$ moves from a pool $k$ to a pool $k^{*}$. This means that the only difference between $C^{*}$ and $C^{(s)}$ are the permutations $C_{k}$ and $C_{k^{*}}$. Following our convention, let $c_{(k)}^{(s)}$ denote the length of permutation $C_k$ before the record $i$ move, and let ${(c_{(k)})}^{*}$ denote the length of permutation $C_k$ after the proposed record $i$ move. From the model in Section 2.3 of the main text, we then have
\begin{align} \label{Ch2::A::p::C} 
\frac{ p(C = C^{*}|\boldsymbol{B}^{*}) }{ p(C = C^{(s)}|\boldsymbol{B}^{(s)}) } & = \frac{ \frac{1}{ {(c_{(k)})}^{*}! }  \frac{1}{ {(c_{(k^{*})})}^{*}! } }{ \frac{1}{{c_{(k)}^{(s)}!} }  \frac{1}{ {c_{(k^{*})}^{(s)}!} } }  = \frac{ {c_{(k)}^{(s)}!} { c_{(k^{*})}^{(s)}!} }{ {{(c_{(k)})}^{*}!} {{(c_{(k^{*})})}^{*}}! },
\end{align}
where \eqref{Ch2::A::p::C} is a ratio permutation lengths. The difference $c_{(k)}^{(s)} - {(c_{(k)})}^{*} \in \lbrace 0, 1 \rbrace $, so  
\begin{align} 
\frac{ c_{(k)}^{(s)}! }{ {(c_{(k)})}^*! }  \in \lbrace 1, c_{(k)}^{(s)} \rbrace.  
\end{align}
Similarly ${(c_{(k^{*})})}^{*} - c_{(k^*)}^{(s)} \in \lbrace 0, 1 \rbrace $, so 
\begin{align} 
\frac{ c_{(k^*)}^{(s)}! }{ {({c_{(k^*)})}^*!} }  \in \lbrace 1, \frac{1}{ {(c_{(k^*)})}^{*}} \rbrace.  
\end{align}
Combining \eqref{ssec::Ch2::ComputingA::MesError} - \eqref{Ch2::A::p::C}, $p(\cdot)$ can be written as: 
\begin{equation}
\begin{aligned} \label{ssec:Ch2::ComputingA::pFinal} 
  p( \cdot) =  \frac{ f(Y_{1}^{*}, Y_2^{*}| \Theta^{(s)}, C^{*}, \boldsymbol{B}^{*})}{ f(Y_1^{(s)}, Y_2^{(s)}| \Theta^{(s)}, C^{(s)}, \boldsymbol{B}^{(s)})} \times \frac{  c_{(k)}^{(s)}! c_{(k^{*})}^{(s)}! }{{(c_{(k)})}^*! ( {(c_{(k^{*})})}^{*}!} \times \frac{ \phi_{ h j^{*} {(B_{2ij^{*}})}^{*}}^{(s)} }{ \phi_{ h j^{*} {(B_{2ij^{*}})}^{(s)}}^{(s)} } \times \\
  \left\{
     \begin{array}{ll}
       1 & : E_{2ij^{*}} ^{(s)} = { (E_{2ij^{*}} )} ^{*} = 1 \\
       \gamma_{2j^{*}}/\left( ({d_{j^{*}}-1})({ 1- \gamma_{2j^{*} } }) \right)  & : E_{2ij^{*}}^{(s)} = 0 , { (E_{2ij^{*}})} ^{*} = 1 \\ 
       ( d_{j^{*}}-1 )({ 1-\gamma_{2j^{*}} })/{ \gamma_{2j^{*} } }  & : E_{2ij^{*}}^{(s)} = 1, {(E_{2ij^{*}})}^{*} = 0.   \\ 
     \end{array}
   \right.
\end{aligned} 
\end{equation} 

The likelihood ratio in \eqref{Ch2::eq::A::A2},
simplifies to the ratio of the likelihoods of the the pools impacted by the proposed pool move. For all other pools, the ratios do not change with the proposed block move, and hence for these pools, their contribution to likelihood ratio in \eqref{Ch2::eq::A::A2} reduces to a multiplier of 1. 

In \eqref{Ch2::eq::A::A2}, we use  imputed values for $Y_{1}$ or $Y_{2}$ for records in the pool linked to dummy records.  This is done for computational convenience to avoid computation of marginal distributions of $Y_1$ and $Y_2$.  To see why this is useful, we define the density of the observed data, $f(\boldsymbol{Y}_{1}, \boldsymbol{Y}_2 \mid \Theta^{(s)}, C^{*}, \boldsymbol{B}^{*})$, 
conditional on any current draw of $(\Theta^{(s)}, C^{*}, \boldsymbol{B}^{*})$, as  follows.  For a given draw of parameters, let $\mathcal{M}_k$ denote the set of records in pool $k$ that correspond to matches to actual records in $F_1$ and $F_2$; these are the records linked across files.  Let $\mathcal{U}_{k1}$ denote the set of records in $F_1$ in pool $k$ that are linked to dummy records in $F_2$.  Let $\mathcal{U}_{k2}$ denote the set of records in $F_2$ in pool $k$ that are linked to dummy records in $F_1$. Suppressing the conditioning to save space, the density of the observed data is  
\begin{equation}
f(\boldsymbol{Y}_{1}, \boldsymbol{Y}_2 \mid --- ) = \prod_{i, i^{'} \in \mathcal{M}_k} f(y_{1i}, y_{2i^{'}} \mid ---) \prod_{i \in \mathcal{U}_{k1}} f(y_{1i} \mid ---) \prod_{i \in \mathcal{U}_{k2}} f(y_{2i} \mid ---). 
\end{equation}
In words, within any pool $k$, the contribution to the likelihood function from the linking analysis is the product of $f(y_{1i}, y_{2i})$ for all $(i, i')$ in the pool that are linked, and of $f(y_{1i})$ and $f(y_{2i'})$  for $i$ and $i'$ in the pool that are linked to dummy records. One could use this observed data density in the numerator and denominator of \eqref{Ch2::eq::A::A2}.  However, as noted in \cite{Gutman}, in a general applications the marginal distributions may be difficult to compute. We therefore follow a strategy in \cite{Gutman} and impute values for $Y_{1}$ or $Y_{2}$ for records in the pool linked to dummy records. 
The ratio in \eqref{Ch2::eq::A::A2} is then approximated by the product of the joint densities $f(y_{1i}, y_{2i})$ for all $i$ in the pool, where some $y_{1i}$ and $y_{2i}$ values may be imputations. We use this approach to compute \eqref{Ch2::eq::A::A2}, as well as in the computation of the exact step in \eqref{ssec::Ch2::ComputingA::J:Smallk}.

\subsection{Component \texorpdfstring{$J(\cdot)$}{J()} } \label{ssec::Ch2::ComputingA::J} 

The transition probability $J(\cdot)$ in \eqref{Ch2::eq::A::J} may be expressed as
\begin{equation} \label{ssec:Ch2::ComputingA::J1}
\begin{aligned}
 J(\cdot) & = \frac{ Pr(E_{2ij^{*}} = E_{2ij^{*}}^{(s)}| \dots) Pr({B}_{2ij^{*}} = B_{2ij^{*}}^{(s)}|{E}_{2ij^{*}}^{(s)}, \dots )  Pr( C = C^{(s)}| \boldsymbol{B}^{(s)}, \dots) }{  Pr(E_{2ij^{*}} = {(E_{2ij^{*}})}^{*}| \dots) Pr({B}_{2ij^{*}} = {(B_{2ij^{*}})}^{*}|{(E_{2ij^{*}}})^{*}, \dots ) Pr( C = C^{*}| \boldsymbol{B}^{*}, \dots)} 
\end{aligned}
\end{equation}
\begin{equation} \label{ssec:Ch2::ComputingA::JStart}
\begin{aligned}
& = \frac{ Pr(E_{2ij^{*}} = E_{2ij^{*}}^{(s)}|\gamma_{2j^{*}}^{(s)})}{ Pr(E_{2ij^{*}} = {(E_{2ij^{*}})}^{*}|\gamma_{2j^{*}}^{(s)})} 
 \times  \frac{ Pr({B}_{2ij^{*}} = B_{2ij^{*}}^{(s)}|\boldsymbol{E}^{(s)},\boldsymbol{E}^{*},\Psi^{(s)}, \boldsymbol{\hat{B}},\boldsymbol{B}^{*} )}{ Pr({B}_{2ij^{*}} = {(B_{2ij^{*}})}^{*}|\boldsymbol{E}^{*},\boldsymbol{E}^{(s)},\Psi^{(s)}, \boldsymbol{\hat{B}},\boldsymbol{B}^{(s)} )} \\
 & \times \frac{ Pr( C = C^{(s)}| C^{*},\boldsymbol{B}^{(s)},  {Y}_1^{(s)},{Y}_2^{(s)}, \Theta^{(s)} )}{Pr( C = C^{*}| C^{(s)}, \boldsymbol{B}^{*}, {Y}_1^{*},{Y}_2^{*}, \Theta^{(s)} )}
\end{aligned}
\end{equation}

\subsubsection{\texorpdfstring{${(E_{2ij^{*}})}^{*}$}{E2 star} }

We propose ${(E_{2ij^{*}})}^{*}|\gamma^{(s)}$ by sampling $({E_{2i{j}^{*}}})^{*}|\gamma_{2}^{(s)},j^{*} \sim Bernoulli( {\gamma_{2j^{*}}}^{(s)} )$. Hence, we have 
\begin{align} \label{ssec:Ch2::ComputingA::J::E} 
\frac{ Pr(E_{2ij^{*}} = E_{2ij^{*}}^{(s)}|\gamma_{2j^{*}}^{(s)})}{ Pr(E_{2ij^{*}} = {(E_{2ij^{*}})}^{*}|\gamma_{2j^{*}}^{(s)})}= \left\{
     \begin{array}{ll}
     1  & : {(E_{2ij^{*}})}^{*} = 1 = E_{2ij^{*}}^{(s)} \\ 
     { (1- \gamma_{2j^{*} }) }/{ \gamma_{2j^{*}} }& : E_{2ij^{*}}^{(s)} = 0, {(E_{2ij^{*}})}^{*} = 1  \\ 
      { \gamma_{2j^{*}} }/{( 1- \gamma_{2j^{*} } )}& : E_{2ij^{*}}^{(s)} = 1,  {(E_{2ij^{*}})}^{*} = 0.  \\ 
     \end{array}
   \right.
\end{align}
When multiplying \eqref{ssec:Ch2::ComputingA::J::E} by \eqref{ssec:Ch2::ComputingA::pFinal}, all terms involving $\gamma$ cancel out of $A$. 

\subsubsection{\texorpdfstring{${(B_{2ij^{*}})}^{*}$}{B2 star} }
\label{sssec:B2star}

The proposal distributions for ${(B_{2ij^{*}})}^{*}$ are described in Section \ref{Supp:ssec:ProposingB}. We denote the probability of transitioning from $B_{2ij^{*}}^{*}$ to $(B_{2ij^{*}})^{(s)}$ as $\Phi_{ij^{*}B_{fij}^{(s)}}^{(* \rightarrow s)}$. This has the same form as \eqref{eq:phi} but with the roles of $(B_{2ij^{*}})^{*}$ and $B_{2ij^{*}}^{(s)}$ reversed. The ratio of proposal distributions for $B_{2ij^{*}}$ is then 
\begin{align}
 \frac{ Pr({B}_{2ij^{*}} = B_{2ij^{*}}^{(s)}|\boldsymbol{E}^{(s)},(\boldsymbol{E})^{*},\Psi^{(s)}, \boldsymbol{\hat{B}},\boldsymbol{B}^{*} )}{ Pr({B}_{2ij^{*}} = {(B_{2ij^{*}})}^{*}|(\boldsymbol{E})^{*},\boldsymbol{E}^{(s)},\Psi^{(s)}, \boldsymbol{\hat{B}},\boldsymbol{B}^{(s)} )}  \left\{
     \begin{array}{ll}
      \Phi_{ij^{*}B_{fij}^{(s)}}^{(* \rightarrow s)}/ \Phi_{ij^{*}{(B_{fij^{*}})}^{*}}^{(s)}  & : { (E_{fij^{*}}) }^{*} = 1 = E_{fij^{*}}^{(s)} \\ 
      1/  \Phi_{ij^{*}B_{fij}^{(s)}}^{(* \rightarrow s)} & : { (E_{fij^{*}} )}^{*} = 1, E_{fij^{*}}^{(s)} = 0  \\ 
      \Phi_{ij^{*}{(B_{fij^{*}})}^{*}}^{(s)}  & : { ( E_{fij^{*}} )}^{*} = 0,  E_{fij^{*}}^{(s)} = 1.  \\ 
     \end{array}
   \right.
\end{align}

\subsubsection{\texorpdfstring{${C}^{*}$}{C star} }

The final contribution to $J(\cdot)$ in \eqref{ssec:Ch2::ComputingA::JStart} is the ratio
\begin{align} \label{ssec::Ch2::ComputingA::J::C1} 
J_{C} = \frac{ Pr( C = C^{(s)}| C^{*},\boldsymbol{B}^{(s)},{Y}_1^{(s)},{Y}_2^{(s)}, \Theta^{(s)} )}{Pr( C = C^{*}| C^{(s)}, \boldsymbol{B}^{*}, {Y}_1^{*},{Y}_2^{*}, \Theta^{(s)} )}.
\end{align} 
Within each MH step, only one record moves from a pool $k$ to a pool $k^{*}$. This means that the only difference between $C^{(s)}$ and $C^{*}$ are the permutations in pools $k$ and $k^{*}$. Following our convention, let $C_{k}^{(s)}$ denote the length $c_{(k)}^{(s)}$ permutation $C_k$ before the record $i$ move, and let ${(C_{k})}^{*}$ denote the length ${(c_{k})}^{*}$ permutation $C_k$ after the proposed record $i$ move. Then \eqref{ssec::Ch2::ComputingA::J::C1} is written as 
\begin{align} \label{ssec::Ch2::ComputingA::J::C2}
J_{C} = \frac{ Pr( C_k = C_k^{(s)} | {(C_k)}^{*}, \boldsymbol{B}^{(s)},{Y}_1^{(s)},{Y}_2^{(s)},\Theta^{(s)} ) \times Pr( C_{k^{*}} = C_{k^{*}}^{(s)}| {(C_{k^{*}})}^{*}, \boldsymbol{B}^{(s)},{Y}_1^{(s)},{Y}_2^{(s)},\Theta^{(s)} )}{ Pr( C_{k} = {(C_k)}^{*}| C_k^{(s)}, \boldsymbol{B}^{*},{Y}_1^{*},{Y}_2^{*}, \Theta^{(s)}) \times  Pr( C_{k^{*}} = {(C_{k^{*}})}^{*}| C_{k^{*}}^{(s)}, \boldsymbol{B}^{*},{Y}_1^{*},{Y}_2^{*},\Theta^{(s)} ) }.
\end{align}
Each of the four pieces of \eqref{ssec::Ch2::ComputingA::J::C2} represents the probability of proposing a given permutation conditional upon either the current or proposed state of the chain. These proposal distributions are described in Section \ref{Supp:ssec:ProposingC}. $J_{C}$ will be a product of the probabilities described in Sections \ref{Supp:ssec:ProposingC:Case1} - \ref{Supp:ssec:ProposingC:Case4}.

\subsection{Final Ratio} 

The final acceptance ratio $A$ has one of three forms, dependent on $E^{*}$ and $E^{(s)}$. 
\begin{equation}
\begin{aligned} \label{ssec:Ch2::ComputingA::Final} 
A &= p( \cdot) \times J (\cdot) \\
  & =  \frac{ f(Y_{1}^{*}, Y_2^{*}| \Theta^{(s)}, C^{*}, \boldsymbol{B}^{*})}{ f(Y_1^{(s)}, Y_2^{(s)}| \Theta^{(s)}, C^{(s)}, \boldsymbol{B}^{(s)})}  \times \frac{  c_{(k)}^{(s)}! c_{(k^{*})}^{(s)}! }{{(c_{(k)})}^*! ( {(c_{(k^{*})})}^{*}!} \times \frac{ \phi_{ h j^{*} {(B_{2ij^{*}})}^{*}}^{(s)} }{ \phi_{ h j^{*} B_{2ij^{*}}^{(s)}}^{(s)} } \times  \\
   & \times J_{C} \times
\left\{
     \begin{array}{ll}
      \Phi_{ij^{*}B_{fij}^{(s)}}^{(* \rightarrow s)}/ \Phi_{ij^{*}{(B_{fij^{*}})}^{*}}^{(s)}  & : { (E_{fij^{*}}) }^{*} = 1 = E_{fij^{*}}^{(s)} \\ 
      1/  \Phi_{ij^{*}B_{fij}^{(s)}}^{(* \rightarrow s)} & : { (E_{fij^{*}} )}^{*} = 1, E_{fij^{*}}^{(s)} = 0  \\ 
      \Phi_{ij^{*}{(B_{fij^{*}})}^{*}}^{(s)}  & : { ( E_{fij^{*}} )}^{*} = 0,  E_{fij^{*}}^{(s)} = 1.  \\ 
     \end{array}
   \right.
\end{aligned}
\end{equation}

\subsection{Accept/Reject Step} 

If, for all pools, $C^{*}$ was sampled using the Exact Step, we compute the acceptance ratio $A$ as defined in \eqref{ssec:Ch2::ComputingA::Final}. If $log(u) < log(A), u \sim Unif(0,1)$, accept $(E_{2i}^*,B_{2i}^*)$. Otherwise, reject the pool move and set $(E_{2i}^{(s+1)},B_{2i}^{(s+1)}) = (E_{2i}^{(s)},B_{2i}^{(s)})$. 

If, for any pools, $C^{*}$ was not sampled using the Exact Step, compute the acceptance ratio $A$ as defined in \eqref{ssec:Ch2::ComputingA::Final}. If $log(u) < log(A), u \sim Unif(0,1)$, accept $(E_{2i}^*,B_{2i}^*)$. Otherwise, proceed as follows. 
\begin{enumerate} 
\item For any large pool $k$ or $k^{*}$, propose a new permutation by selecting with uniform probability two within-pool records from $F_{2}$, and swapping their positions.  
\item Compute the acceptance ratio $A$ as defined in \eqref{ssec:Ch2::ComputingA::Final}
\item If $log(u) < log(A), u \sim Unif(0,1)$, accept $(E_{2i}^*,B_{2i}^*)$. Otherwise, repeat the first two steps $R>1$ times. The constant $R$ is user specified ($R=30$ in our applications). 
\item If after $R$ iterations we fail to accept, reject the pool move and set $(E_{2i}^{(s+1)},B_{2i}^{(s+1)}) = (E_{2i}^{(s)},B_{2i}^{(s)})$. 
\end{enumerate}

\clearpage

\section{DPMPM Full Conditionals} \label{SuppA::DPFCs} 

In this section, we detail the full conditionals necessary to sample the DP parameters $\Psi$. Gibbs sampling steps are adapted from \cite{IshwaranJames2001}.

\begin{enumerate}
		\item For $i \in 1, \dots , N_{B}$, update $z_{i} \in \lbrace 1, \dots , H \rbrace$ from a multinomial full conditional distribution with sample size 1 and 
		\begin{align} \label{DP:start} 
		Pr( z_{i}^{(s+1)} = h | \boldsymbol{B}^{(s+1)}, \phi^{(s)}, \pi^{(s)}, C^{(s+1)} ) = \frac{ \pi_h^{(s)} \prod_{j=1}^{J} 		\phi_{hjB_{i^{'}j}^{(s+1)}}^{(s)}  }{ \sum_{h^{'}=1}^{H} \pi_{h^{'}}^{(s)} \prod_{j=1}^{J} 		\phi_{h^{'}j {B}_{i^{'}j}^{(s+1)}}^{(s)}  } .
		\end{align} 

		\item For $h \in \lbrace 1, \dots , H-1 \rbrace$, update $V_h$ from the full conditional,
		\begin{align}
		V_h^{(s+1)} | z^{(s+1)}, \alpha^{(s)} \sim Beta\left( 1 + m_h^{(s+1)}, \alpha^{(s)} + \sum_{h^{'} = h+1 }^{H} m_h^{(s+1)} \right),
		\end{align}
		\noindent where $m_h^{(s+1)} = \sum_{i} I(z_i^{(s+1)} = h)$ denotes the number of individuals in latent class $h$. Set $V_{H} = 1$.

		\item Compute $\pi_h^{(s+1)} = V_h^{(s+1)} \prod_{g < h} (1- V_g^{(s+1)})$ for all $h \in\lbrace 1,\dots, H \rbrace $. 

		\item For $j \in \lbrace 1,\dots,J \rbrace$ and $h \in \lbrace 1,\dots, H \rbrace$, update $\phi_{hj}$ from the full conditional, 
		{
\begin{equation}
		\begin{aligned} 
		\phi_{hj}^{(s+1)}& | z^{(s+1)}, \boldsymbol{B}^{(s+1)} \sim \\
		& Dirichlet \left( a_{j1} + \sum_{ i: z_{i} = h} I( {B}_{ij}^{(s+1)} = 1 ), \dots , a_{jd_j} + \sum_{ i: z_{i} = h} I( {B}_{ij}^{(s+1)} = d_j ) \right).
		\end{aligned}
\end{equation} 
		}

		\item Update $\alpha$ from the full conditional,
		\begin{align} \label{DP:end} 
		 \alpha^{(s+1)}| \pi^{(s+1)} \sim Gamma\left( a_{\alpha} + H -1, b_{\alpha} - log( 	\pi_{H}^{(s+1)} ) \right).
		\end{align}  
	\end{enumerate}
	
	\clearpage 

\section{Simulation Study:\\ Data Generation and Simulation Design} \label{SuppA::DataGen} 

In this section, we describe the data generation and simulation design for the simulation studies presented in Section 5 of the main text. The simulation studies are designed to illustrate the performance of BLASE with varying amounts of faults, numbers of $T_1$ seeds, and hyper-parameter settings for the prior distribution of $\gamma_2$. We use two faultiness levels, high (40\%) and low (5\%), and two $T_1$ seed counts, high (60\%) and low (20\%), to define four primary simulation groups. These groups are displayed in Table \ref{table:SimSpecs1}. 

\subsection{Hyperparameter Settings}

For each simulation, we examine three prior distributions for $\gamma_{2}$ representing diffuse (D) prior beliefs, concentrated and appropriate (CA) prior beliefs, and concentrated but poorly specified (CP) beliefs. Table \ref{table:SimSpecs2} presents the simulation specifications in detail.

Broadly, with the CA prior we concentrate prior mass at the true faultiness level, e.g., in the scenario with 40\% faulty values, we specify hyperparameters that appropriately reflect a strong belief in a high degree of faultiness. The CP prior concentrates prior mass at an incorrect fault level, i.e., inappropriately reflects prior beliefs of a high degree of faultiness in the presence of low faults, and vice versa. The D prior is relatively flat with a small prior sample size.

\begin{table}[t]
\begin{center}
  \begin{threeparttable}
    \caption{Simulation group specifications}
    \label{table:SimSpecs1}
		\begin{tabular}{llll}
 			\hline
 			\hline 
 			{Sim Group}   & Overall Fault Level & Seed Level & Non-Seed Fault Level \\  			\hline
 			 HSHF    & 40\% & 60\% & 100\%  \\
 			 HSLF    & 5\% & 60\% & 12.5\%  \\
 			 LSHF    & 40\% & 20\% & 50\%  \\
 			 LSLF    & 5\% & 20\% & 6.25\%  \\
			\hline 
		\end{tabular}		       
		\begin{tablenotes}[flushleft]
      		\item NOTE: Overall Fault Level: the overall percentage of records from $F_2$ which are initialized in the incorrect pool; Seed Level: the percentage of the 5000 record pairs which are $T_1$ seeds; Non-Seed Fault Level: the percentage of non-seeds from $F_2$ which are initialized in the incorrect pools.
    	\end{tablenotes}
 	 \end{threeparttable}
	\end{center}
\end{table}

\begin{table}[t]
\begin{center}
  \begin{threeparttable}
    \caption{Simulation prior settings for $\gamma$}
    \label{table:SimSpecs2} 
		\begin{tabular}{llllll}
 			\hline
 			\hline 
 			{Sim Group}  & NSFL  & Prior & $a_{\gamma_{2\textsf{prog}}}$    & $b_{\gamma_{2\textsf{prog}}}$  & Mean (StD)\\ 
 			\hline
 			        &       & CA  & 90000  & 10000   & .9 \hspace{.4cm}(.001) \\
 			HSHF    & 100\% & D   &  2     & 10      & .167 (.1) \\
 			        &       & CP  & 12500  & 87500   & .125 (.001)\\ \hline 
		        
 			        &        & CA & 12500  & 87500   & .125 (.001) \\
 			HSLF    & 12.5\% & D  & 2      & 10      & .167 (.1)\\
 			        &       & CP  & 90000  & 10000   & .9\hspace{.4cm} (.001)\\ \hline 
 			        
 			        &       & CA  & 50000  & 50000   & .5 \hspace{.4cm}(.002) \\
 			LSHF    &50\%   &  D   & 2  & 10 & .167 (.1)  \\
 			        &       & CP   & 6250  & 93750 & .063 (.001)  \\ \hline 
 			         			 
 			        &        & CA  & 6250    & 93750 & .063 (.001) \\
 			LSLF    & 6.25\% & D   &  2      & 10    & .167 (.1) \\
 			        &        & CP  & 50000   & 50000   & .5 \hspace{.4cm}(.001) \\ \hline

			\hline 
		\end{tabular}
\begin{tablenotes}[flushleft]
      		\item NOTE: (1) NSFL = Non-Seed Fault Level: the percentage of non-seeds from $F_2$ initialized in incorrect pools; (2) Prior denotes the name of the prior setting; (3) $a_{\gamma_{2\textsf{prog}}}$, $b_{\gamma_{2\textsf{prog}}}$: the hyper-parameter values for $\gamma_{2\textsf{prog}}$; (4) Mean (Std): mean of $\gamma_{2\textsf{prog}}$ with the specified hyper-parameters, with standard deviation in parentheses.
    	\end{tablenotes}
 	 \end{threeparttable}
	\end{center}
\end{table}

\clearpage

\subsection{Simulated Data Generation} 

We use variables from the \textsf{hsbdemo} data set \citep{hsbdemo} to create simulated files $F_1$ and $F_2$. This data set focuses on school testing. Because \textsf{hsbdemo} has only 200 records, we simulate 5000 record pairs by sampling from predictive distributions estimated on the observed data. We repeat this process independently 100 times at each simulation setting, generating new sets of the 5000 records each time. In each replication, we refer to generated data with no MV faults as the perfectly blocked (PB) data, meaning $\boldsymbol{\hat{B}} = \boldsymbol{B}$. To ensure small pool sizes, we generate each replicated data set ensuring that $c_{(k)} \leq 10$ for all $k$. Specifically, the PB data sets are each generated using the following steps. 

\begin{enumerate}
\item $\textsf{cid}$: Sample $n$ values for \textsf{cid} from a Multinomial(1,$\pi$), where the probabilities $\pi$ give equal probability to every integer in $\lbrace 1, \dots, 30 \rbrace$ .
\item $\textsf{female}$: Independent of \textsf{cid}, sample $n$ values for \textsf{female} from a Binomial($n$,0.545). The success probability is the proportion of females in the \textsf{hsbdemo} data set. 
\begin{align*}
\textsf{female} \sim Binomial( n, 0.545 ). 
\end{align*} 

\item $\textsf{schtyp} | \textsf{cid} $: Using the \textsf{hsbdemo} data, fit a logistic model regressing \textsf{schtyp} on \textsf{cid}. Based on this model, estimate the probability for public and private school status for each of the \textsf{cid} values generated in Step 1, and sample $n$ values for \textsf{schtyp} based on these probabilities. 

\item $\textsf{prog}  | \textsf{schtyp}$ : Using the \textsf{hsbdemo} data, fit a multinomial regression model for \textsf{prog} on \textsf{schtyp}.  Based on this model, estimate the probability for program status for each of the $n$ \textsf{schtyp} values generated in Step 2, and sample $n$ values for \textsf{prog} based on these probabilities. 

\item  $\textsf{ses} | \textsf{prog}, \textsf{schtyp}$:  Using the \textsf{hsbdemo} data, fit a multinomial regression model for \textsf{ses} on \textsf{schtyp} and \textsf{prog}.  Based on this model, estimate the probability for program status, and sample $n$ values for \textsf{ses} based on these probabilities.

\item $\textsf{honors} | \textsf{ses}, \textsf{prog}$: Using the \textsf{hsbdemo} data, fit a logistic regression model for \textsf{honors} on \textsf{ses} and \textsf{prog}.  Based on this model, estimate the probability of being enrolled in honors, and sample $n$ values for \textsf{honors} based on these probabilities. 

\item Let $\boldsymbol{B}$ be the $n \times 6$ matrix with columns containing the generated values for \textsf{female, schtyp, ses, prog, honors} , and \textsf{cid}. 

\item $\textsf{math} | \textsf{prog}, \textsf{ses}, \textsf{female}$: Using the \textsf{hsbdemo} data, fit a normal regression model for \textsf{math} conditional on \textsf{prog,ses,female}. We obtain the model below. 
\begin{equation} \label{supp::matheq}
\begin{aligned}
\hat{\textsf{math}} & = 47.9 + 5.88 I( \textsf{prog} = academic )  - 3.84 I( \textsf{prog} = vocational )  \\ 
& + 2.93 I( \textsf{ses} = middle ) + 4.57 I( \textsf{ses} = high ) - 0.20 I(\textsf{female} = 1) + \epsilon_2 \\
\epsilon_{2i} & \sim N(0 ,6.37^2) .
\end{aligned}
\end{equation}
From \eqref{supp::matheq}, sample sample $n$ values for \textsf{math}$\mid \boldsymbol{B}$. Let $F_2$ be the $n \times 7$ matrix with columns containing the generated values for \textsf{math, female, schtyp, ses, prog, honors} , and \textsf{cid}.

\item $\textsf{read} | \textsf{math}, \textsf{prog}$: Using the \textsf{hsbdemo} data, fit a normal regression model for \textsf{read} conditional on \textsf{math} and \textsf{prog}. We obtain the following: 
\begin{equation}\label{supp:readeq}
\begin{aligned}
\hat{\textsf{read}} & = 17.1 + 0.65 \textsf{math} + 2.02 I( \textsf{prog} = academic ) - 1.20 I( \textsf{prog} = vocational ) + \epsilon_1  \\ \epsilon_{1i} & \sim N(0 , 6.25^2) .
\end{aligned}
\end{equation} 
From \eqref{supp:readeq}, sample $n$ values for \textsf{read}$\mid \textsf{math}, \boldsymbol{B}$. Let $F_1$ be the $n \times 7$ matrix with columns containing the generated values for \textsf{read, female, schtyp, ses, prog, honors} , and \textsf{cid}.

\end{enumerate} 

\begin{table}[t]
\begin{center}
  \begin{threeparttable}
    \caption{Simulated data generation patterns}
    \label{table:dependencies} 
       \label{Ch2::table::SimDependencies}
		\begin{tabular}{lllllllll}
 			\hline
 			\hline 
 		&  \textsf{read}  & \textsf{math} & \textsf{female} & \textsf{schtyp} & \textsf{ses} &  \textsf{prog} & \textsf{honors} & \textsf{cid}      \\
 		\textsf{read} & \cellcolor{gray} & \cellcolor{black} & & & & \cellcolor{black} & & \\ 
 		\textsf{math} & & \cellcolor{gray}&\cellcolor{black} & & \cellcolor{black} & \cellcolor{black} & & \\ 
 		\textsf{female} & & & \cellcolor{gray} & & & & & \\ 
 		\textsf{schtyp} & & & & \cellcolor{gray} & & & & \cellcolor{black} \\ 
 		\textsf{ses} & & & &  \cellcolor{black} &  \cellcolor{gray}&  \cellcolor{black} & & \\ 
 		\textsf{prog} & & & & \cellcolor{black} & &\cellcolor{gray} & & \\ 
 		\textsf{honors} & & & & &\cellcolor{black} & \cellcolor{black} &\cellcolor{gray} & \\
 		\textsf{cid} & & & & & & & &\cellcolor{gray} \\ 
 			\hline 
		\end{tabular}
    \begin{tablenotes}[flushleft]
      \item NOTE: Black squares in cell $(i,i^{'})$ indicate that the variable defining row $i$ was generated conditional on the variable corresponding to column $i^{'}$.
    \end{tablenotes}
  \end{threeparttable}
   \end{center}
\end{table}

\subsection{Pool Size Restriction and \texorpdfstring{$T_1$}{Type 1} seeds} 

We assign each record to a pool $k$, where each $k$ is affiliated with a unique combination of the variables in $\boldsymbol{B}$. For each $k= 1, \dots, K$, compute $c_{(k)}$. Note that because the data are perfectly pooled, $n_{1(k)}=n_{2(k)} =c_{(k)}$ for all $k$. For this simulation, we require $c_{(k)} \leq 10$ for all $k$.

To satisfy the pool size requirement, for any $k$ such that $c_{(k)} > 10$, we randomly sample $c_{(k)} - 10$ in-pool records. Selected records are declared $T_1$ seeds. Let $t^{*} = \sum_{k} max( 0, c_{(k)} -10)$, or the total number of $T_1$ seeds needed to fulfill the pool size restriction. Each simulation setting specifies a certain number $N_{T_1}$ of $T_1$ seeds. We select the remaining $N_{T_1} - t^{*}$ required $T_1$ seeds at random from the non-seed pairs.

\subsection{Pool Faults}

Of the seven categorical variables, we set \textsf{prog} as the MV and the remaining six as BVs. For the high fault (HF) scenarios, we randomly make 40\% of the record pairs have a faulty \textsf{prog}. To do so, for each selected record 
we sample a new reported value for \textsf{prog} uniformly from the available incorrect values. We apply the same procedure for the low fault (LF) scenarios for the 5\% of selected record pairs.

\subsection{Linking Analysis}

The linking analysis is a linear regression for $Y_1=$\textsf{read}, where 
\begin{align}\label{eq:Y1} 
Y_1 = \textsf{read} 
&=  \beta_0 + \beta_1 \textsf{math} + \beta_2 I(\textsf{prog} = academic) + \beta_3 I(\textsf{prog} = vocational) + \epsilon_1 
\end{align}
where $\epsilon_{1i} \stackrel{iid}{\sim} N(0,\sigma_1^2)$. In addition to the model for $Y_1$, we model a secondary regression for $Y_2=$\textsf{math}: 
\begin{equation}\label{eq:Y2}
\begin{split}
Y_2 = \textsf{math} 
& =  \eta_0 + \eta_1 I(\textsf{female} = yes) + \eta_2 I(\textsf{prog} = academic) + \\ & \eta_3 I(\textsf{prog} = vocational) + \eta_4 I(\textsf{ses} = Middle) +
\eta_5 I(\textsf{ses} = High) + \epsilon_2 \\
\end{split}
\end{equation}
where $\epsilon_{2i} \stackrel{iid}{\sim} N(0,\sigma_2^2)$. Let $\mathbf{X}_1$ denote the design matrix for \eqref{eq:Y1}, and let $\mathbf{X}_2$ denote the design matrix for \eqref{eq:Y2}. Let $\boldsymbol{\beta}  = [\beta_0, \beta_1, \beta_2, \beta_3]^{T}$, $\boldsymbol{\beta}_{1}  = [\beta_0,\beta_2,\beta_3]^{T}$, $\boldsymbol{\eta} =[\eta_0,\eta_1, \eta_2,\eta_3,\eta_4,\eta_5]^{T}$. As a prior distribution for $\Theta = (\boldsymbol{\beta}, \sigma_1, \boldsymbol{\eta}, \sigma_2)$, we use $p(\Theta) \propto (1/\sigma_1^2) (1/\sigma_2^2)$.

\subsection{Full Model}

With the linking analyses specified in the previous section, the entire BLASE model for the simulation may be expressed as follows. 

\begin{align*}
Y_1 & = \mathbf{X_1}\boldsymbol{\beta} + \epsilon_1 , \epsilon_{1i} \stackrel{iid}{\sim} N(0,\sigma_1^2)\\
Y_2 & = \mathbf{X_2}\boldsymbol{\eta} + \epsilon_2, \epsilon_{2i} \stackrel{iid}{\sim} N(0,\sigma_2^2) \\
p(\Theta)  & \propto (1/\sigma_1^2) (1/\sigma_2^2) \\ 
p(C_k = C_{kl} \mid \boldsymbol{B}) & = \frac{1}{ {c_{(k)}}!} \\
E_{2i^{'}j}| \gamma_{2j} & \sim Bernoulli(\gamma_{2j}) \\
\gamma_{2j} & \sim Beta( a_{\gamma_{2j}}, b_{\gamma_{2j}}),
\end{align*} 
\begin{equation*} 
   f \left( \hat{B}_{2i^{'}j}| E_{2i^{'}j}, B_{2i^{'}j} \right) = \left\{
     \begin{array}{ll}
      B_{2i^{'}j}  &  : \text{if } E_{2i^{'}j} = 0 \\
      \text{Uniform discrete on all but } B_{2i^{'}j} & : \text{if } E_{2i^{'}j} = 1. 
     \end{array}
   \right.
\end{equation*}

\subsection{Gibbs Full Conditionals} 

Conditional on $C^{(s+1)}$, $Y_1^{(s+1)}$, and $Y_2^{(s+1)}$, let $\mathbf{X}_1^{(s+1)}$ and $\mathbf{X}_2^{(s+1)}$ represent the design matrices of the regression of $Y_1$ and $Y_2$, respectively. Let $N_B^{(s)} = \sum_{k=1}^K n_{(k)}^{(s)}$ be the number of individuals as defined by $C^{(s+1)}$. 

\begin{enumerate}
\item Impute missing values of $Y_2$ for records linked to dummy
  records. Let $\mathbf{V}^{(s+1)}$ denote $\mathbf{X}_{1}^{(s+1)}$ without the column corresponding to $Y_2$. Dropping the superscripts $(s)$ on $\Theta$, we impute using the following conditional distribution: 
\begin{align} 
Y_2 |  Y_1^{(s)}, \Theta^{(s)}, C^{(s+1)}, \boldsymbol{B}^{(s+1)} & \sim N(\mu, \Sigma I ) \\
 \mu & = \Sigma \times ( \sigma_2^{-2} \mathbf{X}_2^{(s+1)}\boldsymbol{\eta} + \sigma_1^{-2} \beta_2 ( Y_1^{(s)}-{\mathbf{V}}^{(s+1)} \boldsymbol{\beta}_{1}) ) \\ 
 \Sigma & = (\sigma_2^{-2} + \sigma_1^{-2} \beta_2^{2})^{-1} .
\end{align}
Here $I$ is the identity matrix. Let ${Y}_2^{(s+1)}$ comprise ${Y}_2$ and the imputed values of $Y_2$.

\item Impute missing values of $Y_1$ for records linked to dummy
  records using the following conditional distribution: 
\begin{align*} 
Y_1| \mathbf{X}_1^{(s+1)}, C^{(s+1)}, \Theta^{(s)} \sim N( {\bf X_1^{(s+1)} } \boldsymbol{\beta}^{(s)}, {(\sigma_1^2)}^{(s)}I ).
\end{align*} 
Here $I$ is the identity matrix. Let ${Y}_1^{(s+1)}$ comprise ${Y}_1$ and the imputed values of $Y_1$.

\item Sample $\sigma_1^{(s+1)}$ from the full conditional
\begin{align}
\sigma_1^2 | Y_1^{(s+1)}, \mathbf{X}_{1}^{(s+1)},\boldsymbol{\beta}^{(s)} & \sim InvGamma\left( \frac{N_B^{(s+1)} - p}{2}, SSE_1/2 \right) \\
SSE_1 & = (Y_1^{(s+1)} - {\bf X_1^{(s+1)}} \boldsymbol{\beta}^{(s)})^{T} (Y_1^{(s+1)} - {\bf X_1^{(s+1)}} \boldsymbol{\beta^{(s)}} ).
\end{align}

\item Sample $\boldsymbol{\beta}^{(s+1)}$ from the full conditional 
\begin{align}
\boldsymbol{\beta} | \sigma_1^{(s+1)}, Y_1^{(s+1)},Y_2^{(s+1)}, C^{(s+1)}, \boldsymbol{B}^{(s+1)}& \sim N \left( \boldsymbol{\hat{\beta}}, ( {\mathbf{X}_1^{(s+1)}}^{T} \mathbf{X}_1^{(s+1)})^{-1} {(\sigma_1^2)}^{(s+1)} \right) \\ 
\boldsymbol{\hat{\beta}} & = ({\bf X_1^{(s+1)}}^{T} {\bf X_1^{(s+1)}})^{-1} {\bf X_1^{(s+1)}}^{T} Y_1^{(s+1)}.
\end{align}

\item Sample $\sigma_2^{(s+1)}$ from the full conditional
\begin{align}
\sigma_2^2 |\mathbf{X}_{2}^{(s+1)},\boldsymbol{\eta}^{(s)} & \sim InvGamma\left( \frac{N_B^{(s+1)} - q}{2}, SSE_2/2 \right) \\
SSE_2 & = (Y_2^{(s+1)} - {\bf X_2^{(s+1)}} \boldsymbol{\eta}^{(s)})^{T} (Y_2^{(s+1)} - {\bf X_2^{(s+1)}} \boldsymbol{\eta^{(s)}} ).
\end{align}

\item Sample $\boldsymbol{\eta}^{(s+1)}$ from the full conditional 
\begin{align}
\boldsymbol{\eta} | \sigma_2^{(s+1)}, Y_2^{(s+1)}, C^{(s+1)}, \boldsymbol{B}^{(s+1)}& \sim N \left( \boldsymbol{\hat{\eta}}, ( {\mathbf{X}_2^{(s+1)}}^{T} \mathbf{X}_2^{(s+1)})^{-1} {(\sigma_2^2)}^{(s+1)} \right) \\ 
\boldsymbol{\hat{\eta}} & = ({\bf X_2^{(s+1)}}^{T} {\bf X_2^{(s+1)}})^{-1} {\bf X_2^{(s+1)}}^{T} Y_2^{(s+1)}.
\end{align}

\end{enumerate} 

\section{Simulation Study: Extended Results} \label{SuppA::Results} 

In this section we present further simulation results from the BLASE simulation discussed in Section 5 of the main paper. These results are summarized in Table \ref{Ch2::table::SimResultsY1} in this supplement, which is the same as Table 2 in the main paper. For each of the four simulation groups, we provide figures for (1) PMR, (2) RMSE and (3) parameter estimation for the $Y_1$ regression. Supplementary Table \ref{Ch2::table::SimResultsY2} summarizes the parameter estimation from the $Y_2$ regression, but because the behavior mimics that observed in the $Y_1$ regression estimates, we do not include plots for the $Y_2$ estimates. 

\begin{table}[t]
\begin{center}
  \begin{threeparttable}
    \caption{Summary of $Y_1$ regression results of simulation studies}
    \label{Ch2::table::SimResultsY1}
		\begin{tabular}{cccccccc}
  \hline
\hline
  &  & \multicolumn{4}{c}{$Y_1$ Regression Linking Analysis Parameters} &  & \\ \cline{3-6} 
 			{}    &     & {Intercept}    & Math  & \textsf{prog} = Acad & \textsf{prog} = Voc  & {dPMR } & {dRMSE}       \\[+5pt]
 			\hline
			        & CA & \textbf{29.5} & \textbf{15.2} & \textbf{-14.7} & \textbf{42.8}  & \textbf{11.4} & \textbf{3.3}\\
			HSHF    & D  & \textbf{25.6} & \textbf{13.3}  & \textbf{-14.0} & \textbf{41.0}   & \textbf{10.1}& \textbf{2.5}\\
 			        & CP & \textbf{11.9}& \textbf{6.3}  & -3.1 & \textbf{55.0} & \textbf{5.5}& \textbf{.50} \\
 			         \\[-4pt]

 			        & CA & \textbf{.9} & \textbf{.5} & \textbf{2.1} & \textbf{4.4}   & \textbf{-1.2} & .006\\
 			HSLF    & D   & .16 & \textbf{.15} & \textbf{1.7} & \textbf{3.2}  & \textbf{-1.6} & \textbf{-.02} \\
 			        & CP & \textbf{-21.4} & \textbf{-11.0} & \textbf{-37.6} & \textbf{-69.1}  & \textbf{-10.6}  & \textbf{-.96} \\
 			        \\[-4pt]
 			         
 			     & CA & {.3} & .06 & \textbf{-21.2}& 1.2&  \textbf{.95}& \textbf{.003}\\
 			LSHF   & D  & \textbf{5.9} &\textbf{2.8} & \textbf{-9.9} & \textbf{20.3}  & \textbf{3.3} & \textbf{.008}\\
 			        & CP & \textbf{9.9}& \textbf{4.8}  & \textbf{-4.4} & \textbf{27.0}  & \textbf{4.3}  & \textbf{.011}  \\
 			        \\[-4pt]
 			          			         			 
 			        & CA & \textbf{-12.6}& \textbf{-6.3}  & .6 &  \textbf{-9.3} & \textbf{-1.9}  &  \textbf{-1.3} \\
 		LSLF	         & \multicolumn{1}{l}{D}   & \textbf{-9.9} & \textbf{-5.0} & \textbf{-6.2} & \textbf{-13.0}& \textbf{-1.3}  & \textbf{-.9}\\
 			        & CP & \textbf{-40.0}  & \textbf{-20.2} & \textbf{-45.7}  & \textbf{-82.3}  & \textbf{-9.9}&  \textbf{-3.4} \\
			 \\[-4pt]
		t-values & & 27.7&  56.3 & 8.9 & -5.3 & $R^2$ & .46 \\ 
		\hline 
		\end{tabular}
    	\begin{tablenotes}[flushleft]
      		\item NOTE: Positive values indicate BLASE is more accurate than the file linking model that incorrectly treats the faulty MV as a BV, and negative values indicate the opposite. Letting GM stand for the latter modeling strategy, and using subscripts to indicate the method used, the entries under each predictor are averages across 100 replications of $100\left( \left| \hat{\theta}_{GM} -\theta \right| - \left| \hat{\theta}_{BLASE} - \theta\right|\right)/\theta$; the entries under dPMR equal 100$(PMR_{BLASE} - PMR_{GM})$; and, the entries under dRMSE are averages across 100 replications of $( RMSE_{GM} - RMSE_{BLASE} )/ RMSE_{PB}$. Bold values indicate differences more extreme than would be expected under Monte Carlo error (p-values $<.05$ in appropriate t-tests) if both modeling strategies had the same properties. Final row is the average t-values and $R^2$ values of the regression linking analysis across the 100 PB data sets.
    	\end{tablenotes}
 	 \end{threeparttable}
	\end{center}
\end{table}

\begin{table}[t]
\begin{center}
  \begin{threeparttable}
    \caption{Summary of $Y_2$ regression results of simulation studies}
    \label{Ch2::table::SimResultsY2}
		\begin{tabular}{cccccccc}
 			\hline
 			\hline 
 			  &  & \multicolumn{6}{c}{$Y_2$ Regression Linking Analysis Parameters}  \\ \cline{3-8} 
 			{}    &     & {Int}  & \textsf{prog} = Acad & \textsf{prog} = Voc & \textsf{ses} = middle  & \textsf{ses} = high & {female} \\ 
 			        & CA & \textbf{-.10}   & \textbf{42.6} & \textbf{16.5}  & \textbf{.68} & \textbf{16.11} & \textbf{5.9}  \\
 			HSHF    & D  & \textbf{-.11}    & \textbf{42.6}& \textbf{15.7} & .48    & \textbf{-12.1}& \textbf{5.4}  \\
 			        & CP & \textbf{-.13} & \textbf{37.2}&   \textbf{11.2} & .20  & \textbf{-9.5} & \textbf{4.8}  \\
 			        \\[-4pt]			        
	        
 			        & CA & \textbf{-.053} & \textbf{3.3} & \textbf{-.80}  & .20   & -.71  & -.35  \\
 			HSLF    & D  & \textbf{-.072} & \textbf{4.6} & \textbf{-1.3}  & 0.02  & -1.3  &  .35\\
 			        & CP & \textbf{-2.2}& \textbf{-60.3} & \textbf{-15.2} & \textbf{-9.7} &  \textbf{.03} & \textbf{-.006} \\
 			        \\[-4pt]
 			        
 			        & CA & \textbf{-1.3} & \textbf{-21.3} & \textbf{8.5} & \textbf{5.3}& \textbf{.003} & \textbf{.11} \\
 			LSHF    & D  & \textbf{-1.2} & \textbf{-29.2} & \textbf{15.9}& \textbf{8.0}    &  \textbf{.04} & \textbf{.02}  \\
 			        & CP & \textbf{-1.3}  & \textbf{-13.5} & \textbf{17.0} & \textbf{8.9} & \textbf{.03} & \textbf{.02} \\
 			        \\[-4pt]
 			         			 
 			        & CA & \textbf{-.1} & \textbf{6.3} & \textbf{2.4} & 0.05  & \textbf{23.1} & \textbf{-6.1} \\
 			LSLF    & D  & \textbf{-.78} &  1.0 & -.33 & \textbf{-1.3} & \textbf{.02} & \textbf{.001}  \\
 			        & CP & \textbf{-.70} & \textbf{-72.3} & \textbf{-16.3} & \textbf{-3.6} & \textbf{.03}  & \textbf{-.001}  \\
 			      \\[-4pt]
		t-values & & 218.9 & 7.9 & -21.2  & 6.5   & 2.6  & -6.4  \\
		\hline 
		\end{tabular}
		\begin{tablenotes}[flushleft]
      		\item NOTE: Positive values indicate BLASE is more accurate than the file linking model that incorrectly treats the faulty MV as a BV, and negative values indicate the opposite. Letting GM stand for the latter modeling strategy, and using subscripts to indicate the method used, the entries under each predictor are averages across 100 replications of $100\left( \left| \hat{\eta}_{GM} -\eta \right| - \left| \hat{\eta}_{BLASE} - \eta\right|\right)/\theta$. Bold values indicate differences more extreme than would be expected under Monte Carlo error (p-values $<.05$ in appropriate t-tests) if both modeling strategies had the same properties. Final row is the average t-values of the regression linking analysis across the 100 PB data sets. The average $R^2$ value is .14.
    	\end{tablenotes}
 	 \end{threeparttable}
	\end{center}
\end{table}

\clearpage

\subsection{High Seed High Fault}

We begin with a scenario with high levels of faulty values, which is where one would expect accounting for errors in the MVs to matter the most. Recall that the focus of these simulation studies is to examine potential gains when error in the MVs is addressed in the model. Consider Figure \ref{Ch2::fig::BLASE::HEHS::Coefs}. The posterior means based on BLASE with the appropriately concentrated or diffuse priors tend to be similar to those based on the analysis with the PB data.  Not surprisingly, using the poorly concentrated prior distribution worsens the performance of BLASE. Figure \ref{Ch2::fig::BLASE::HEHS::Coefs}, as well as Table \ref{Ch2::table::SimResultsY1}, also suggest that accounting for potential errors in the MVs can improve inferences relative to treating the faulty MV as an error-free BV.  This is corroborated in Figure \ref{Ch2::fig::BLASE::HSHF::Match}, which shows that allowing the file matching algorithm to change faulty MVs from their reported values can increase match rates.  While some increase in PMR in BLASE is expected due to chance, Figure \ref{Ch2::fig::BLASE::HSHF::Match} shows a sizable increase in PMR at all three prior settings. The increase in PMR for BLASE in the HSHF scenario is accompanied by a reduction in RMSE (Figure \ref{Ch2::fig::BLASE::HEHS::MSE}) and more accurate estimates of three of the four linking parameters (Figure \ref{Ch2::fig::BLASE::HEHS::Coefs}).

\begin{figure}[t]
\begin{minipage}{.5\linewidth}
\centering
\subfloat[Intercept]{\label{HEHS:a}\includegraphics[scale=.35]{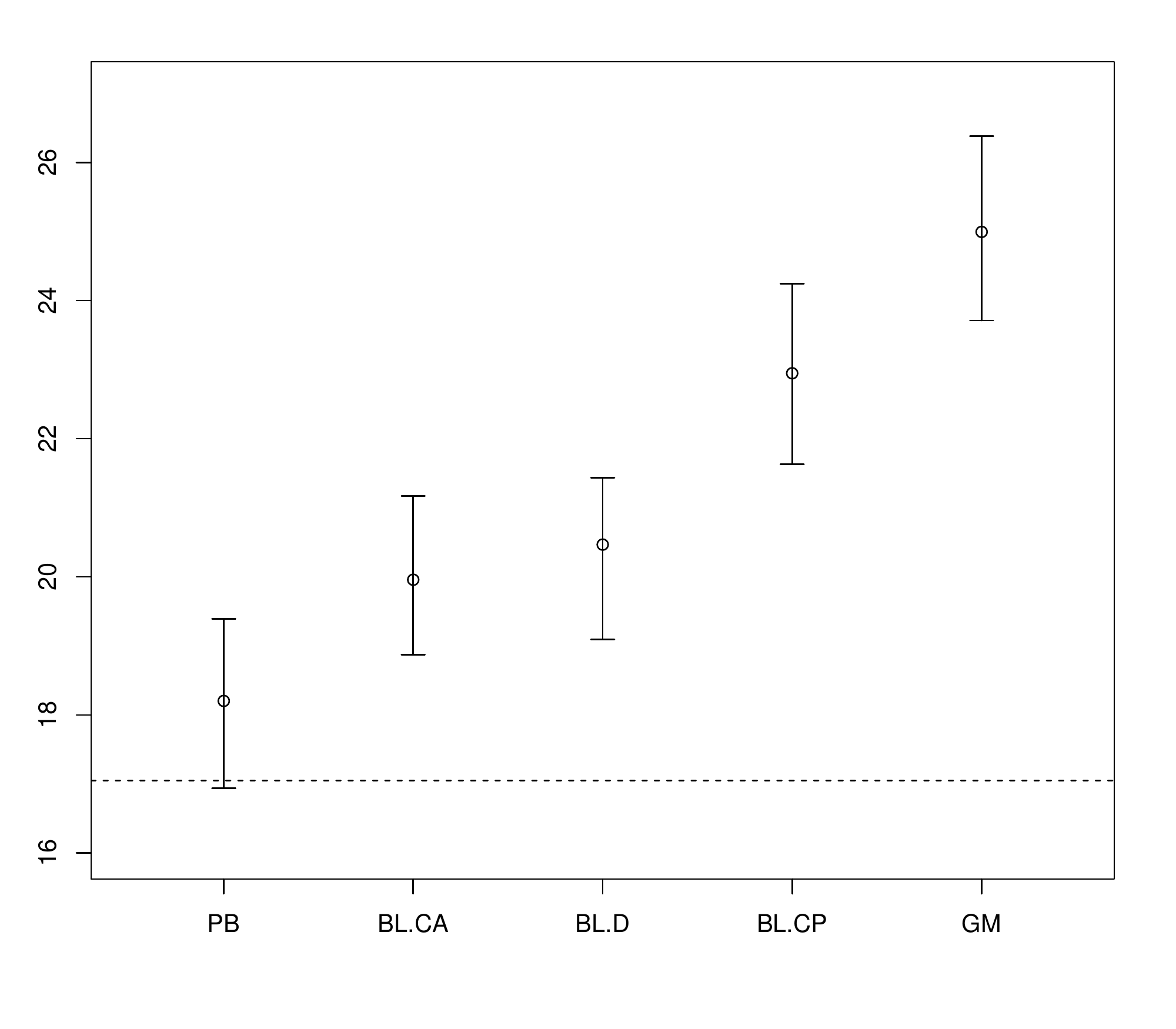}}
\end{minipage}%
\begin{minipage}{.5\linewidth}
\centering
\subfloat[Math]{\label{HEHS:b}\includegraphics[scale=.35]{./Figures/BLASE_HEHS/Update_HEHS_All_Math_CIs.pdf}}
\end{minipage}\par\medskip
\centering
\begin{minipage}{.5\linewidth}
\centering
\subfloat[\textsf{prog} = Academic]{\label{HEHS:c}\includegraphics[scale=.35]{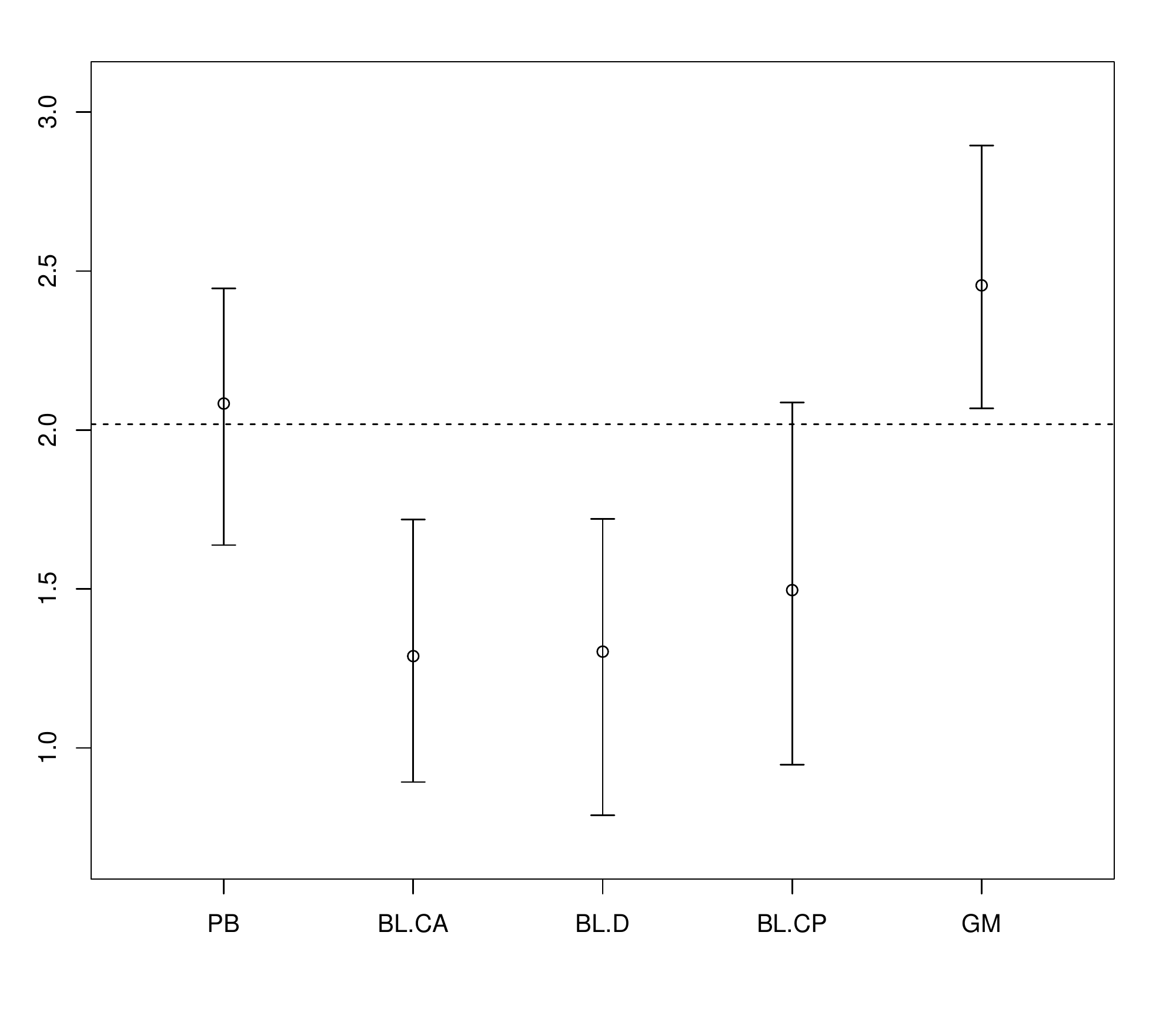}}
\end{minipage}%
\begin{minipage}{.5\linewidth}
\centering
\subfloat[\textsf{prog} = Vocational]{\label{HEHS:d}\includegraphics[scale=.35]{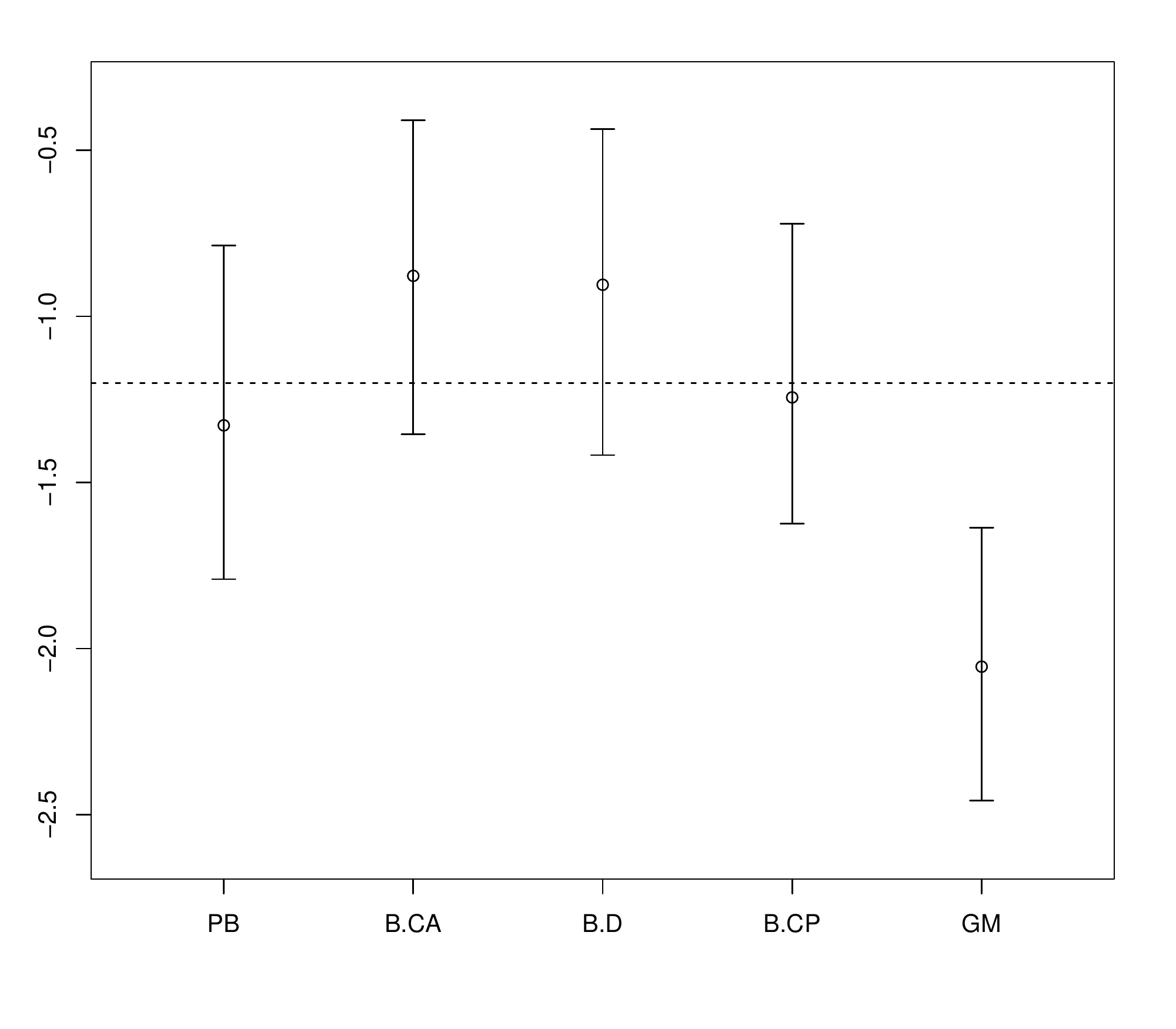}}
\end{minipage}\par\medskip
\caption{Results from the HSHF simulation. Top left panel (a) displays 95\% interval for the 100 posterior means for intercept. Top right panel (b) displays 95\% interval for the 100 posterior means for \textsf{math} coefficient. Lower left panel (c) displays 95\% interval for the 100 posterior means for \textsf{prog} = Academic coefficient. Lower right panel (d) displays 95\% interval for the 100 posterior means for \textsf{prog} = Vocational coefficient. In all panels, PB stands for using the GAZM on the perfectly blocked data, BL.CA for using BLASE with the concentrated and appropriate prior, BL.D for using BLASE with the diffuse prior, BL.CP for using BLASE with the concentrated but poorly specified prior, and GM for using the GAZM on the faulty data. Dashed horizontal line indicates true value of coefficient.}
\label{Ch2::fig::BLASE::HEHS::Coefs}
\end{figure}

\begin{figure}[ht]
    \centering
    \subfloat[HSHF: Non-seed PMRs.]{{\includegraphics[width=.8\textwidth]{./Figures/BLASE_HEHS/Update_HEHS_Match.pdf} }\label{Ch2::fig::BLASE::HSHF::Match}}%
    \qquad
     \subfloat[HSHF: Out-of-sample RMSE.]{{\includegraphics[width=.8\textwidth]{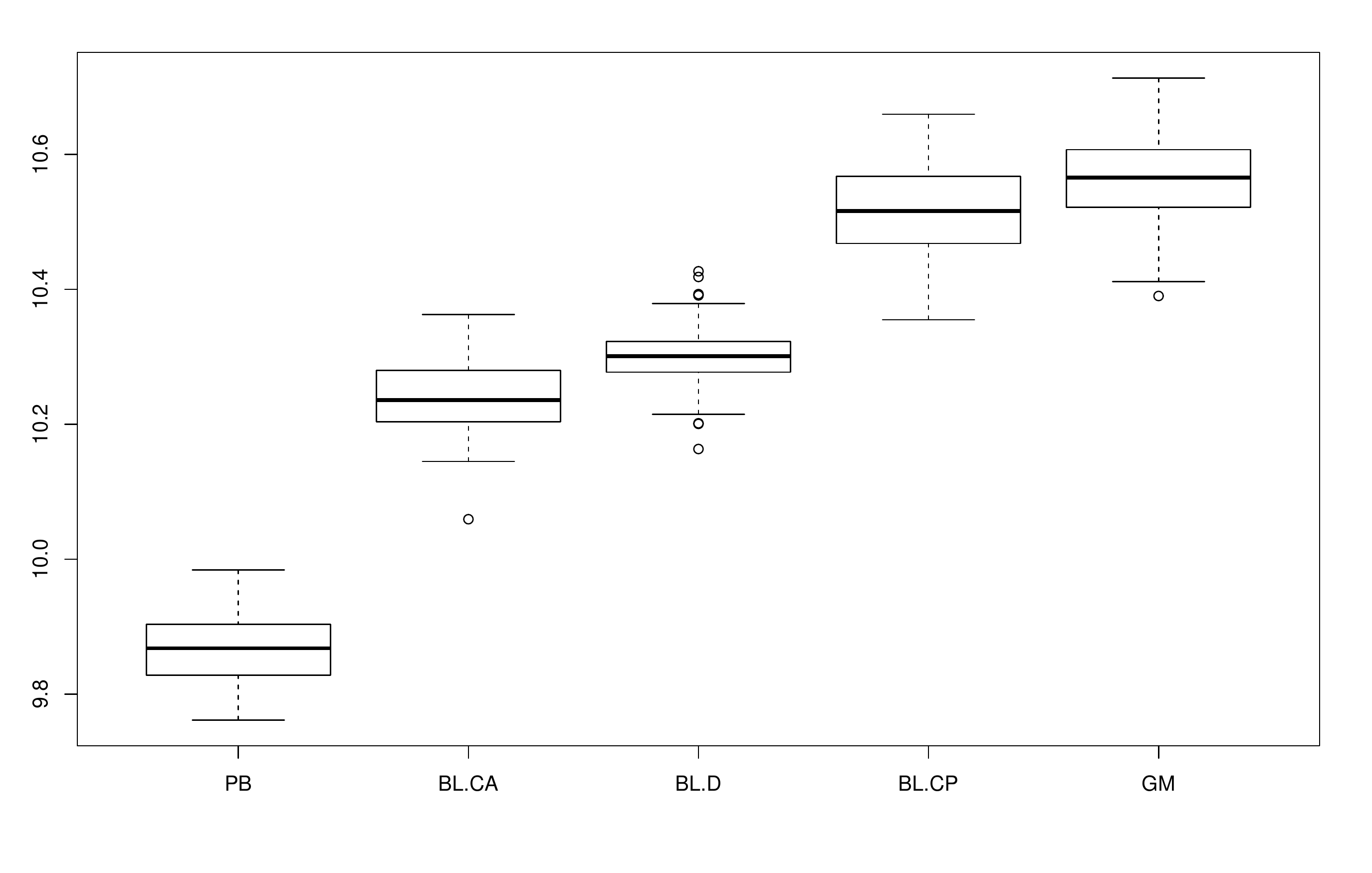}} \label{Ch2::fig::BLASE::HEHS::MSE}}%
    \caption{Results from the HSHF simulation. Top panel (a) displays average posterior match rates from 100 simulations. Bottom panel (b) displays average RMSE from 100 simulations. In both panels, PB stands for using the GAZM on the perfectly blocked data, BL.CA for using BLASE with the concentrated and appropriate prior, BL.D for using BLASE with the diffuse prior, BL.CP for using BLASE with the concentrated but poorly specified prior, and GM for using the GAZM on the faulty data.}
    \label{fig:HEHS}
\end{figure}

\subsection{High Seed Low Fault}

Out of the four simulation scenarios, the two Low Fault (LF) simulation groups are the ones in which we expect accounting for errors in the MVs to matter the least. Indeed, the HSLF scenario is the closest of the four groups to be ideal for file matching; there are a high number of seeds, and low faults. Accordingly, Figure \ref{Ch2::fig::BLASE::LEHS::Coefs} indicates that the BLASE estimates for the academic and vocational coefficients are comparable to the PB estimates, and are similar to the PB estimates for the remaining coefficients. The PMR for BLASE is, predictably, higher in the HSLF scenario than in the HSHF setting. In this setting, the results indicate similar performance in BLASE versus the model that treats the faulty MV as a BV. Specifically, the RMSE and match rates for BLASE are comparable to the model treating the faulty MV as an error-free BV. This actually is encouraging for BLASE, as it appears not to destroy the quality of inferences that could be obtained from using the PB.

\begin{figure}[t]
\begin{minipage}{.5\linewidth}
\centering
\subfloat[Intercept]{\label{LEHS:a}\includegraphics[scale=.35]{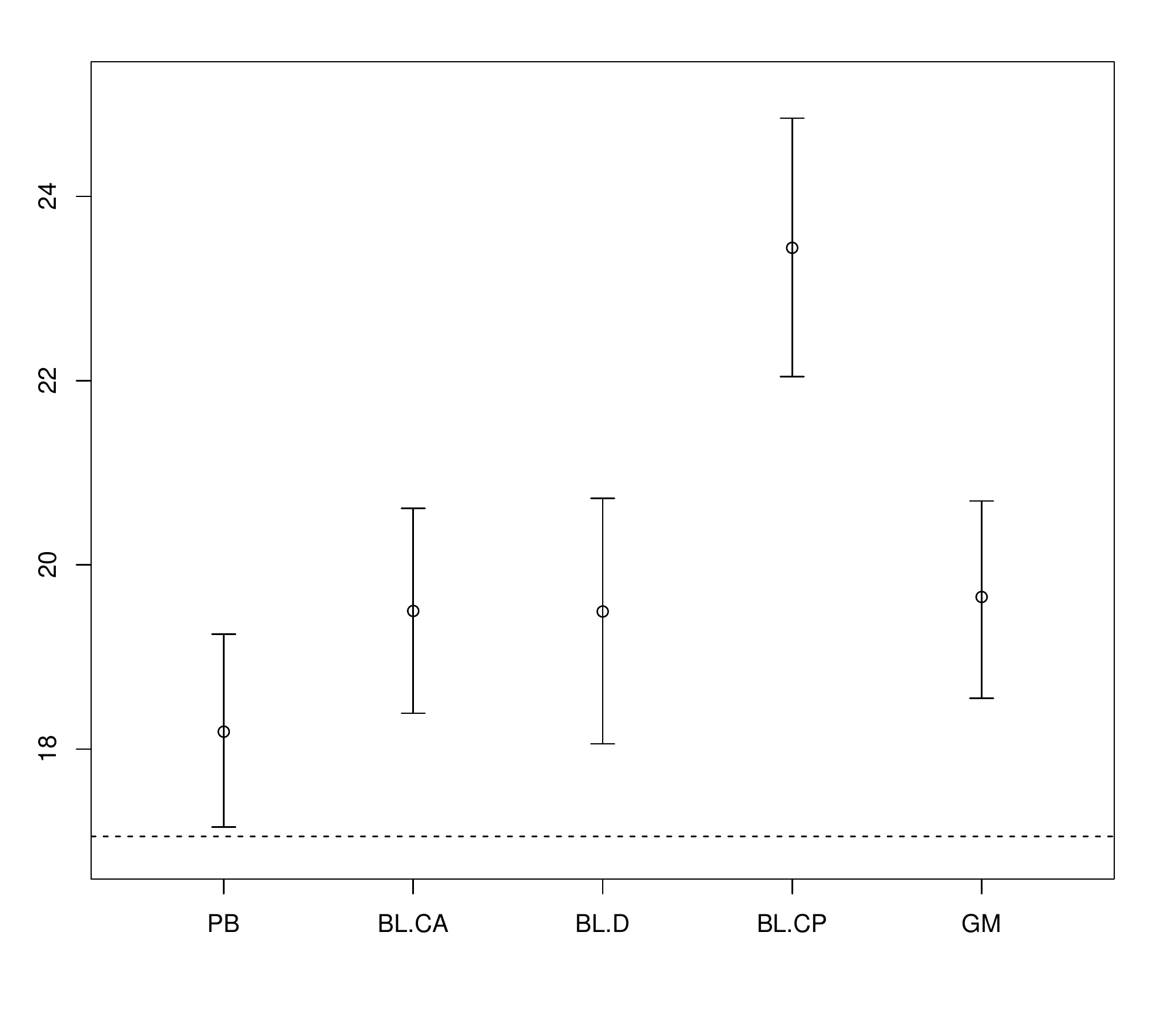}}
\end{minipage}%
\begin{minipage}{.5\linewidth}
\centering
\subfloat[Math]{\label{LEHS:b}\includegraphics[scale=.35]{./Figures/BLASE_LEHS/Update_LE_HS_All_Math_CIs.pdf}}
\end{minipage}\par\medskip
\centering
\begin{minipage}{.5\linewidth}
\centering
\subfloat[Academic]{\label{LEHS:c}\includegraphics[scale=.35]{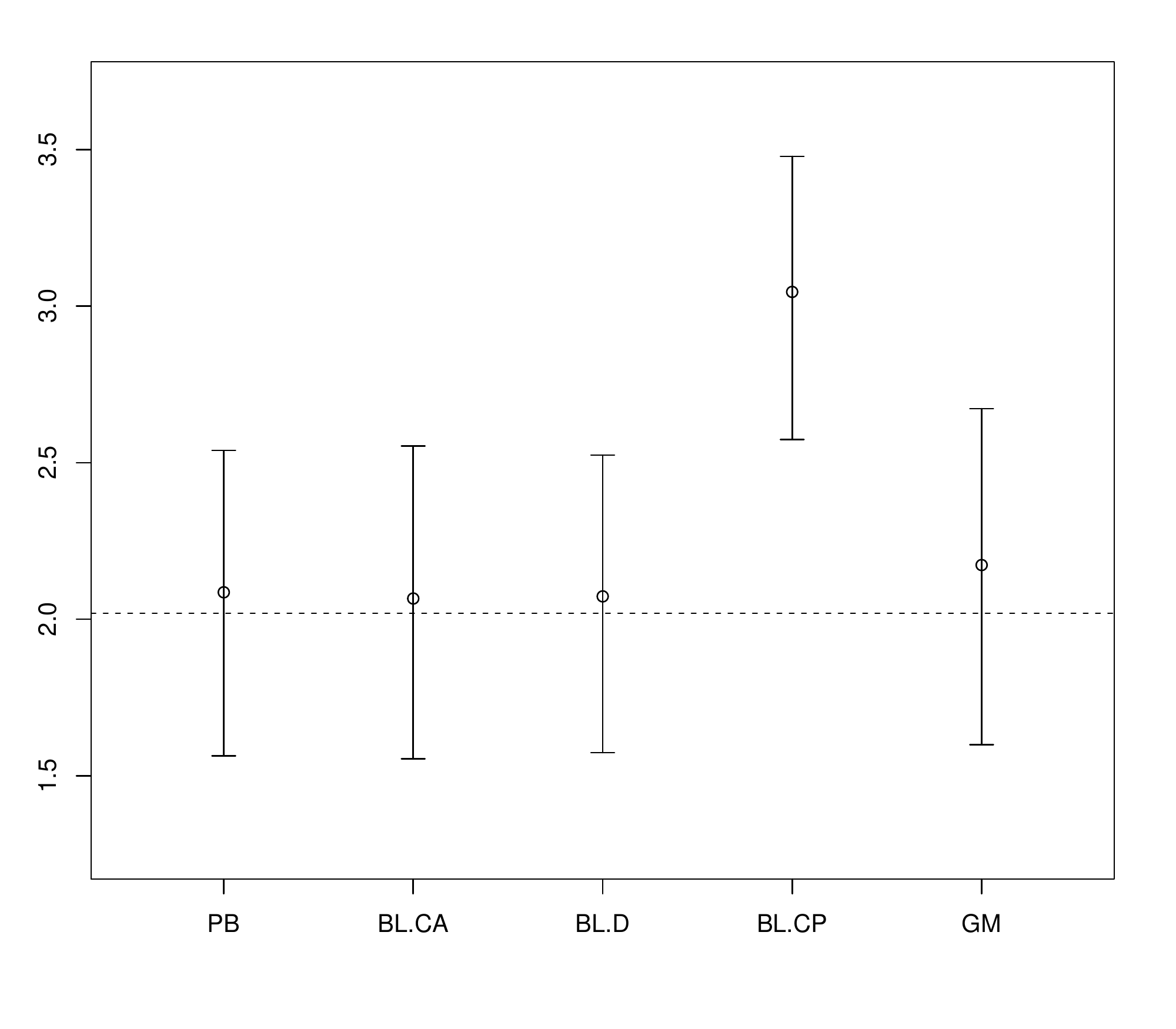}}
\end{minipage}%
\begin{minipage}{.5\linewidth}
\centering
\subfloat[Vocational]{\label{LEHS:d}\includegraphics[scale=.35]{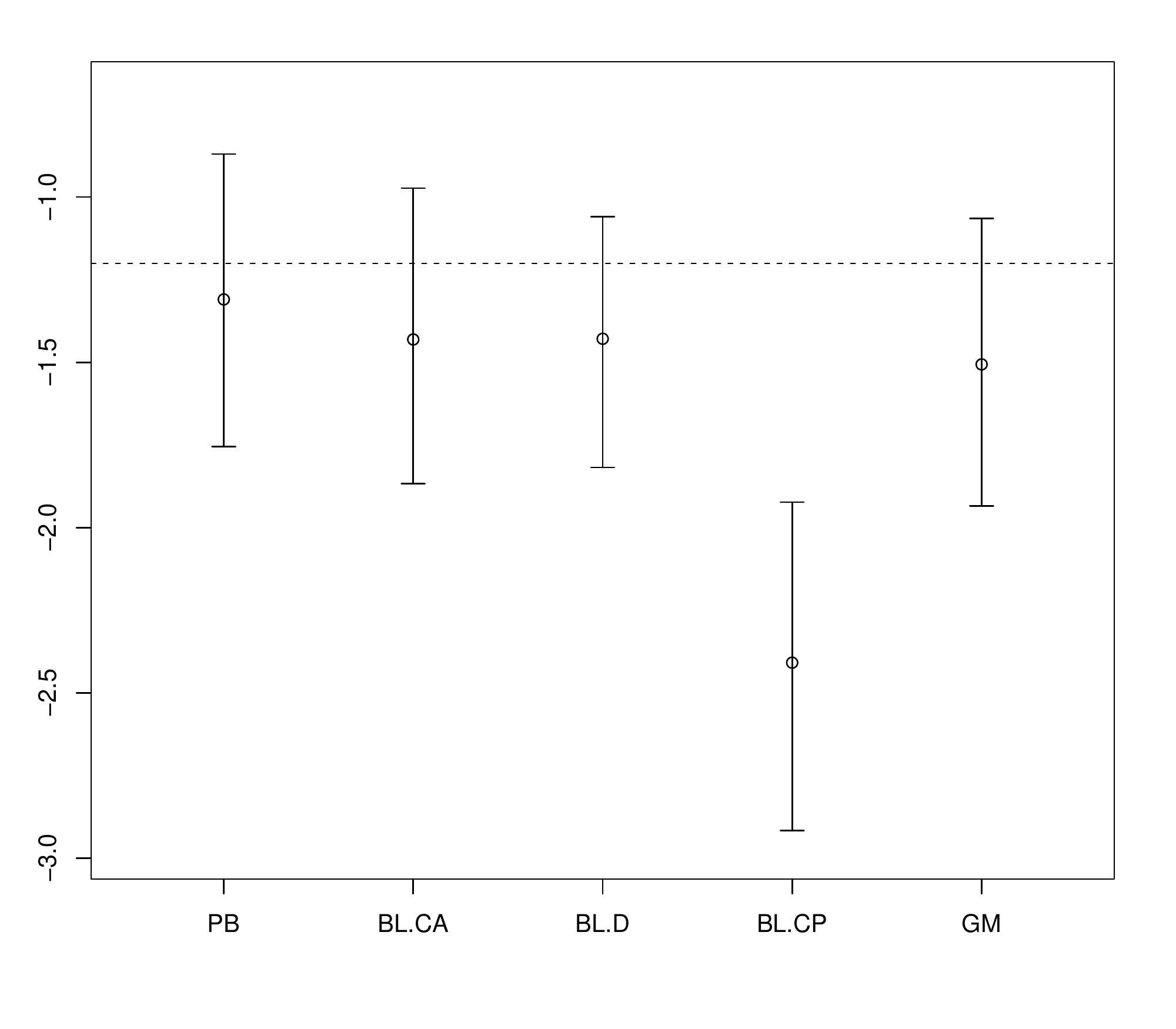}}
\end{minipage}\par\medskip

\caption{Results from the HSLF simulation. Top left panel (a) displays 95\% interval for the 100 posterior means for intercept. Top right panel (b) displays 95\% interval for the 100 posterior means for \textsf{math} coefficient. Lower left panel (c) displays 95\% interval for the 100 posterior means for \textsf{prog} = Academic coefficient. Lower right panel (d) displays 95\% interval for the 100 posterior means for \textsf{prog} = Vocational coefficient. In all panels, PB stands for using the GAZM on the perfectly blocked data, BL.CA for using BLASE with the concentrated and appropriate prior, BL.D for using BLASE with the diffuse prior, BL.CP for using BLASE with the concentrated but poorly specified prior, and GM for using the GAZM on the faulty data. Dashed horizontal line indicates true value of coefficient.}
\label{Ch2::fig::BLASE::LEHS::Coefs}
\end{figure}

\begin{figure}[ht]
    \centering
    \subfloat[HSLF: Non-seed PMRs.]{{\includegraphics[width=.8\textwidth]{./Figures/BLASE_LEHS/Update_LEHS_Match.pdf} }\label{Ch2::fig::BLASE::LEHS::Match}}%
    \qquad
     \subfloat[HSLF: Out-of-sample RMSE.]{{\includegraphics[width=.8\textwidth]{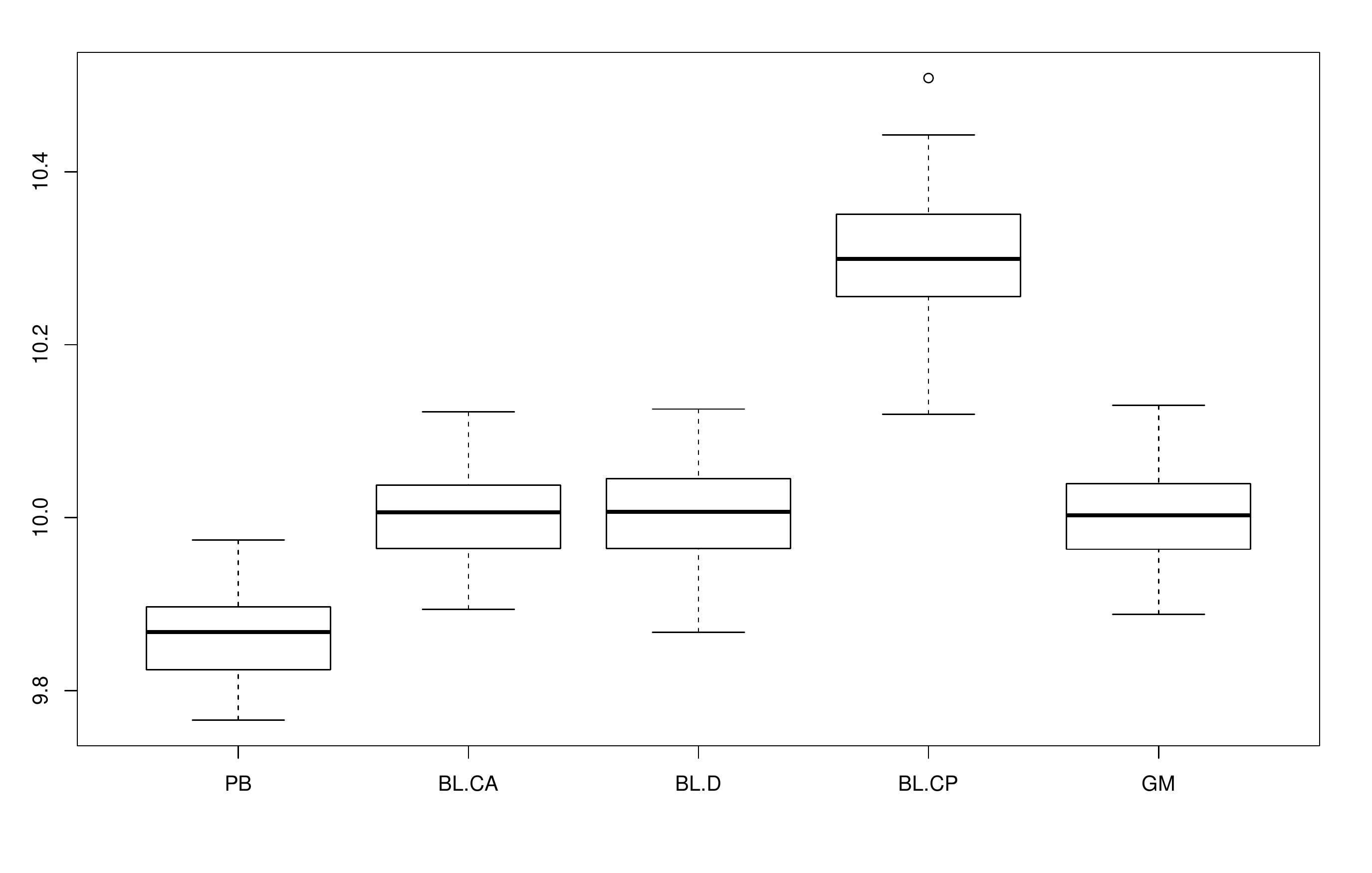}} \label{Ch2::fig::BLASE::LEHS::MSE}}%
    \caption{Results from the HSLF simulation. Top panel (a) displays average posterior match rates from 100 simulations. Bottom panel (b) displays average RMSE from 100 simulations. In both panels, PB stands for using the GAZM on the perfectly blocked data, BL.CA for using BLASE with the concentrated and appropriate prior, BL.D for using BLASE with the diffuse prior, BL.CP for using BLASE with the concentrated but poorly specified prior, and GM for using the GAZM on the faulty data.}
    \label{fig:HSLFSupp}
\end{figure}

\clearpage 

\subsection{Low Seed High Fault}

The role of $T_1$ seeds in the posterior performance of BLASE and the GAZM is highlighted in the contrast of the results of the HSHF and LSHF simulation groups. In the HSHF simulation, the coefficient estimates obtained by BLASE were similar to those obtained using PB. In contrast, with low seeds and high faults, BLASE is unable to accurately estimate the coefficients, and has a low PMR relative to PB. We note that in the LSHF scenario, PB is unable to estimate the math coefficient. However, the behavior of BLASE under the CP and diffuse priors is comparable in estimation to the model that assumes the faulty MV is a BV, and yields a slightly higher match rate. There are low seeds to inform the block moves, but BLASE does result in some records moving from incorrect pools into correct ones, yielding the slight increase in PMR. The RMSE is comparable whether one treats all variables as BVs or allows \textsf{prog} to be an MV.


\begin{figure}[t]
\begin{minipage}{.5\linewidth}
\centering
\subfloat[Intercept]{\label{HELS:a}\includegraphics[scale=.35]{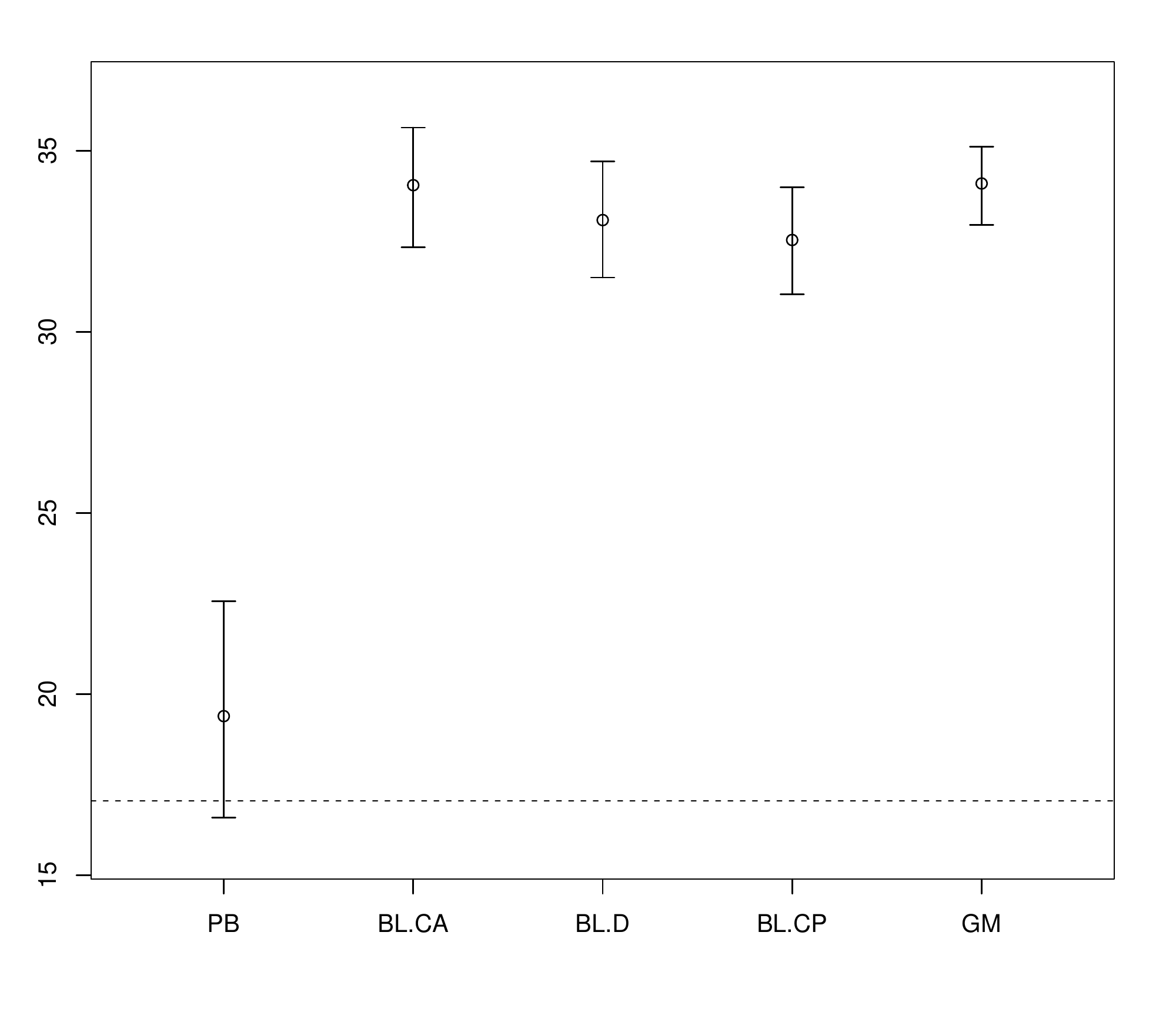}}
\end{minipage}%
\begin{minipage}{.5\linewidth}
\centering
\subfloat[Math]{\label{HELS:b}\includegraphics[scale=.35]{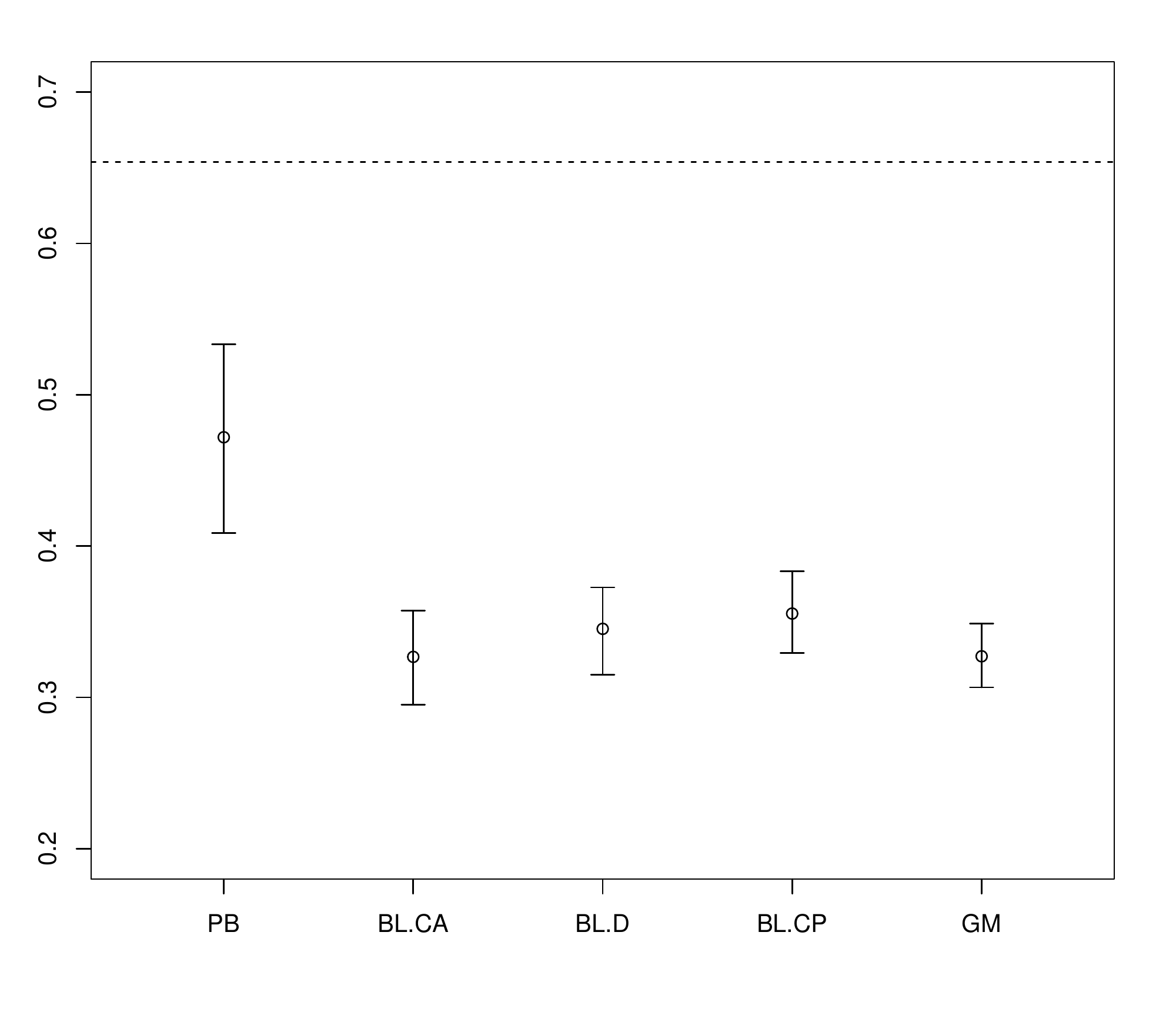}}
\end{minipage}\par\medskip
\centering
\begin{minipage}{.5\linewidth}
\centering
\subfloat[Academic]{\label{HELS:c}\includegraphics[scale=.35]{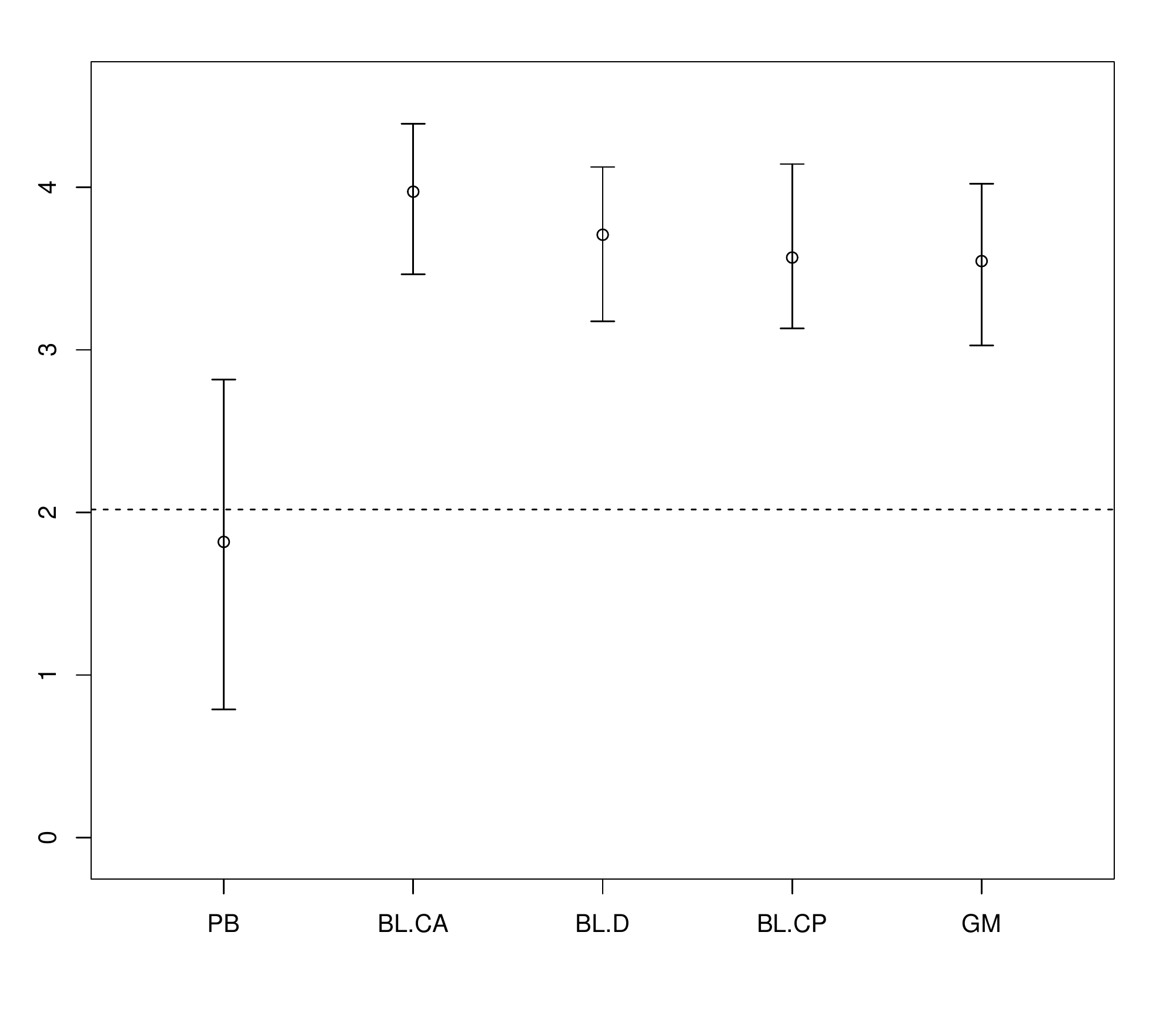}}
\end{minipage}%
\begin{minipage}{.5\linewidth}
\centering
\subfloat[Vocational]{\label{HELS:d}\includegraphics[scale=.35]{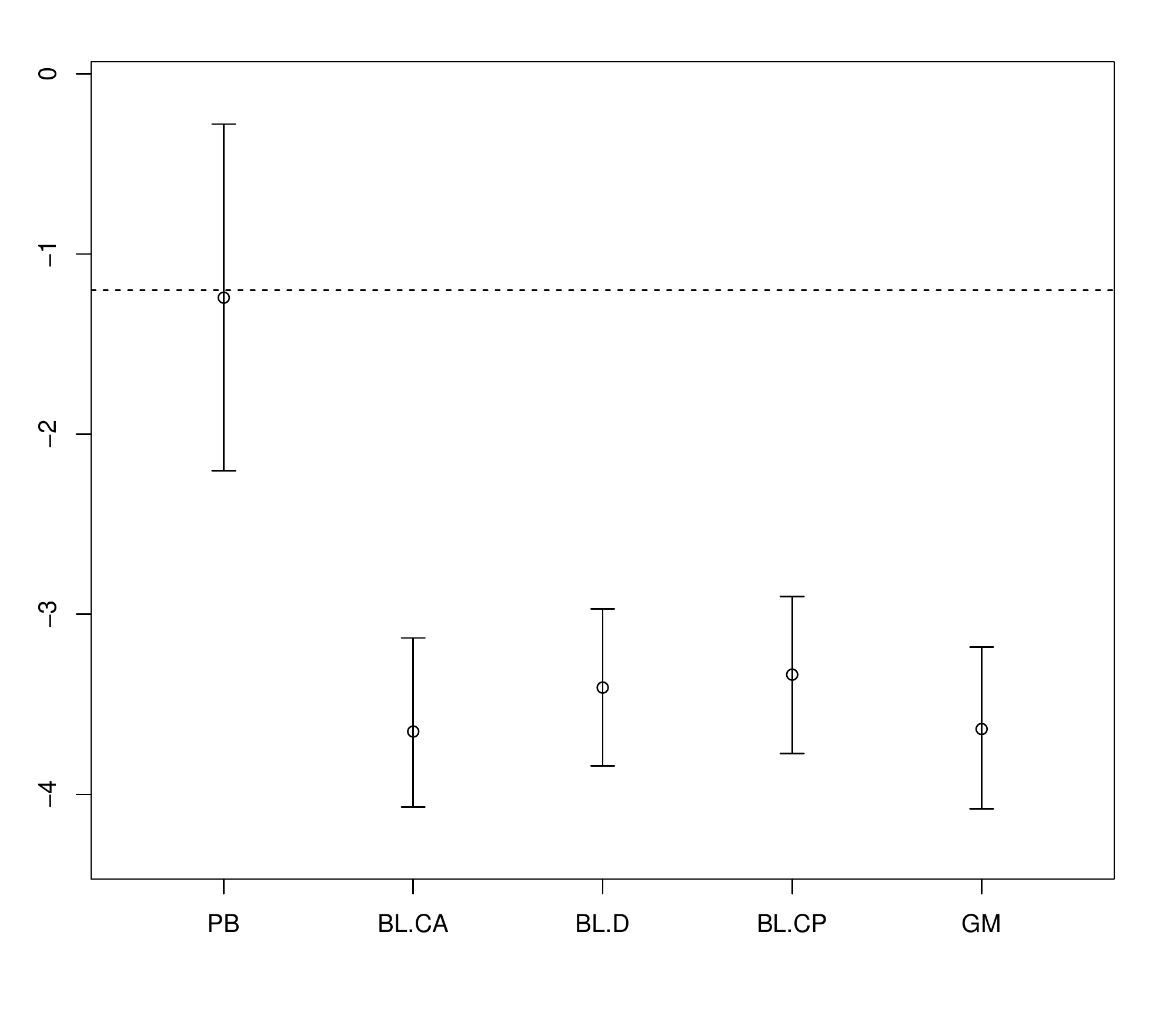}}
\end{minipage}\par\medskip

\caption{Results from the LSHF simulation. Top left panel (a) displays 95\% interval for the 100 posterior means for intercept. Top right panel (b) displays 95\% interval for the 100 posterior means for \textsf{math} coefficient. Lower left panel (c) displays 95\% interval for the 100 posterior means for \textsf{prog} = Academic coefficient. Lower right panel (d) displays 95\% interval for the 100 posterior means for \textsf{prog} = Vocational coefficient. In all panels, PB stands for using the GAZM on the perfectly blocked data, BL.CA for using BLASE with the concentrated and appropriate prior, BL.D for using BLASE with the diffuse prior, BL.CP for using BLASE with the concentrated but poorly specified prior, and GM for using the GAZM on the faulty data. Dashed horizontal line indicates true value of coefficient.}
\label{Ch2::fig::BLASE::HELS::Coefs}
\end{figure}

\begin{figure}[ht]
    \centering
    \subfloat[LSHF: Non-seed PMRs.]{{\includegraphics[width=.8\textwidth]{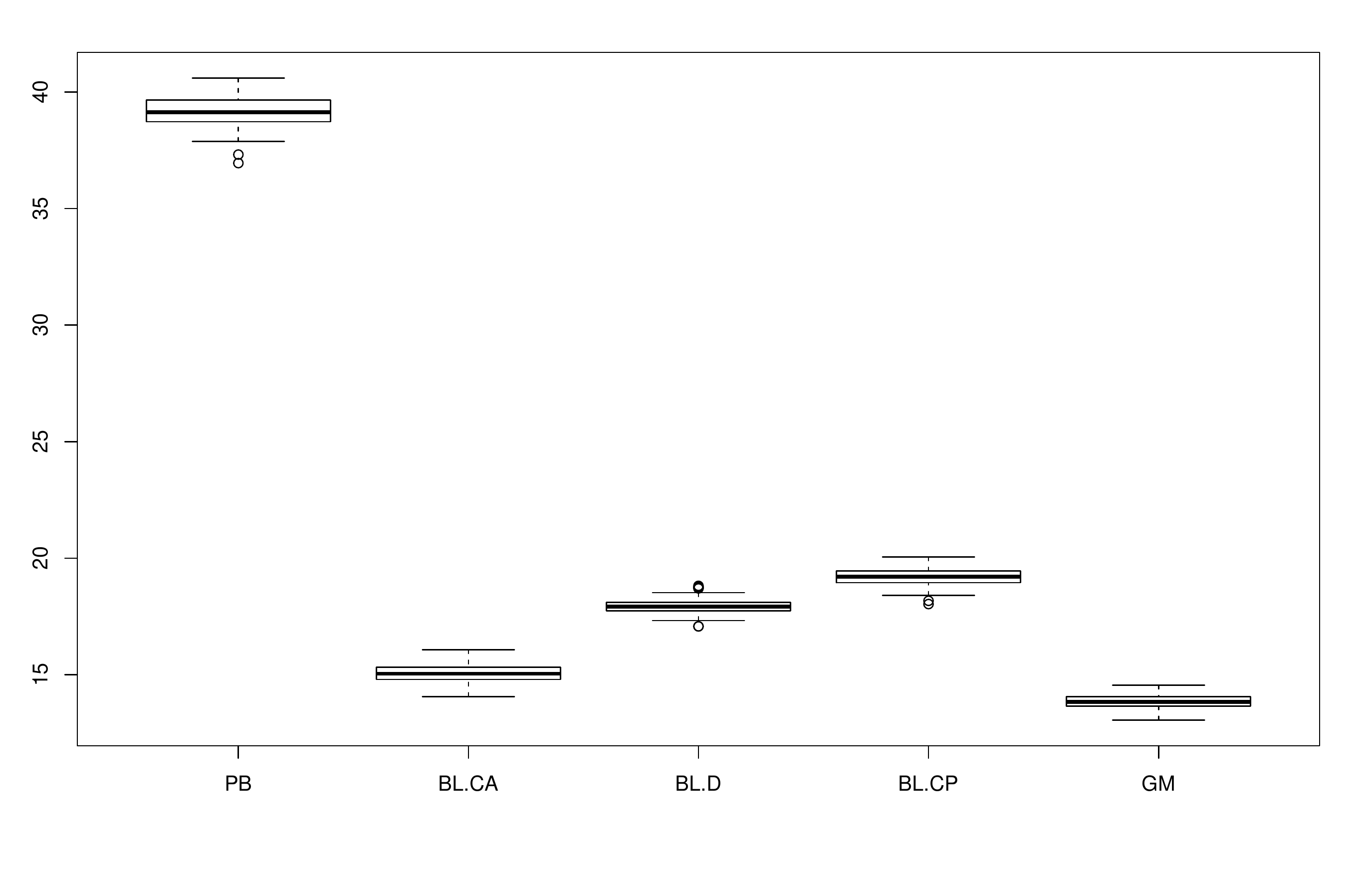} }\label{Ch2::fig::BLASE::HELS::Match}}%
    \qquad
     \subfloat[LSHF: Out-of-sample RMSE.]{{\includegraphics[width=.8\textwidth]{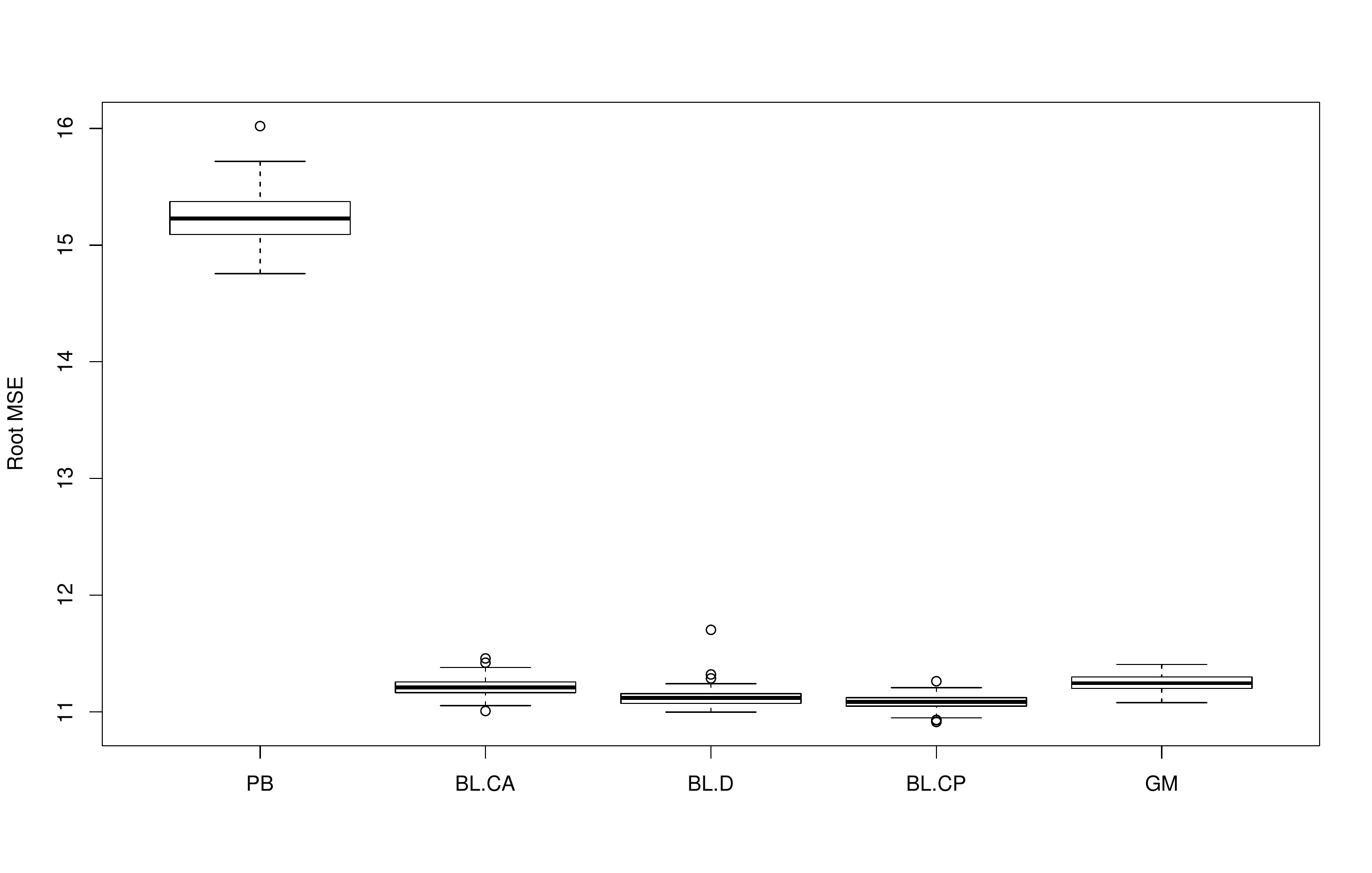}} \label{Ch2::fig::BLASE::HELS::MSE}}%
    \caption{Results from the LSHF simulation. Top panel (a) displays average posterior match rates from 100 simulations. Bottom panel (b) displays average RMSE from 100 simulations. In both panels, PB stands for using the GAZM on the perfectly blocked data, BL.CA for using BLASE with the concentrated and appropriate prior, BL.D for using BLASE with the diffuse prior, BL.CP for using BLASE with the concentrated but poorly specified prior, and GM for using the GAZM on the faulty data.}
    \label{fig:HELS}
\end{figure}

\subsection{Low Seed Low Fault}

The final simulation group, LSLF, is the situation in which we would expect to see the least impact from applying BLASE. In this scenario, BLASE is not as effective as treating all the variables as BVs. Without adequate numbers of $T_1$ seeds, BLASE does not have enough information to determine accurate moves.

As seen in Figure \ref{Ch2::fig::BLASE::HELS::Coefs} and \ref{Ch2::fig::BLASE::LELS::Coefs},in scenarios with low numbers of $T_1$ seeds, none of the file matching algorithms provide accurate estimates of the coefficients.

\begin{figure}
\begin{minipage}{.5\linewidth}
\centering
\subfloat[Intercept]{\label{LELS:a}\includegraphics[scale=.35]{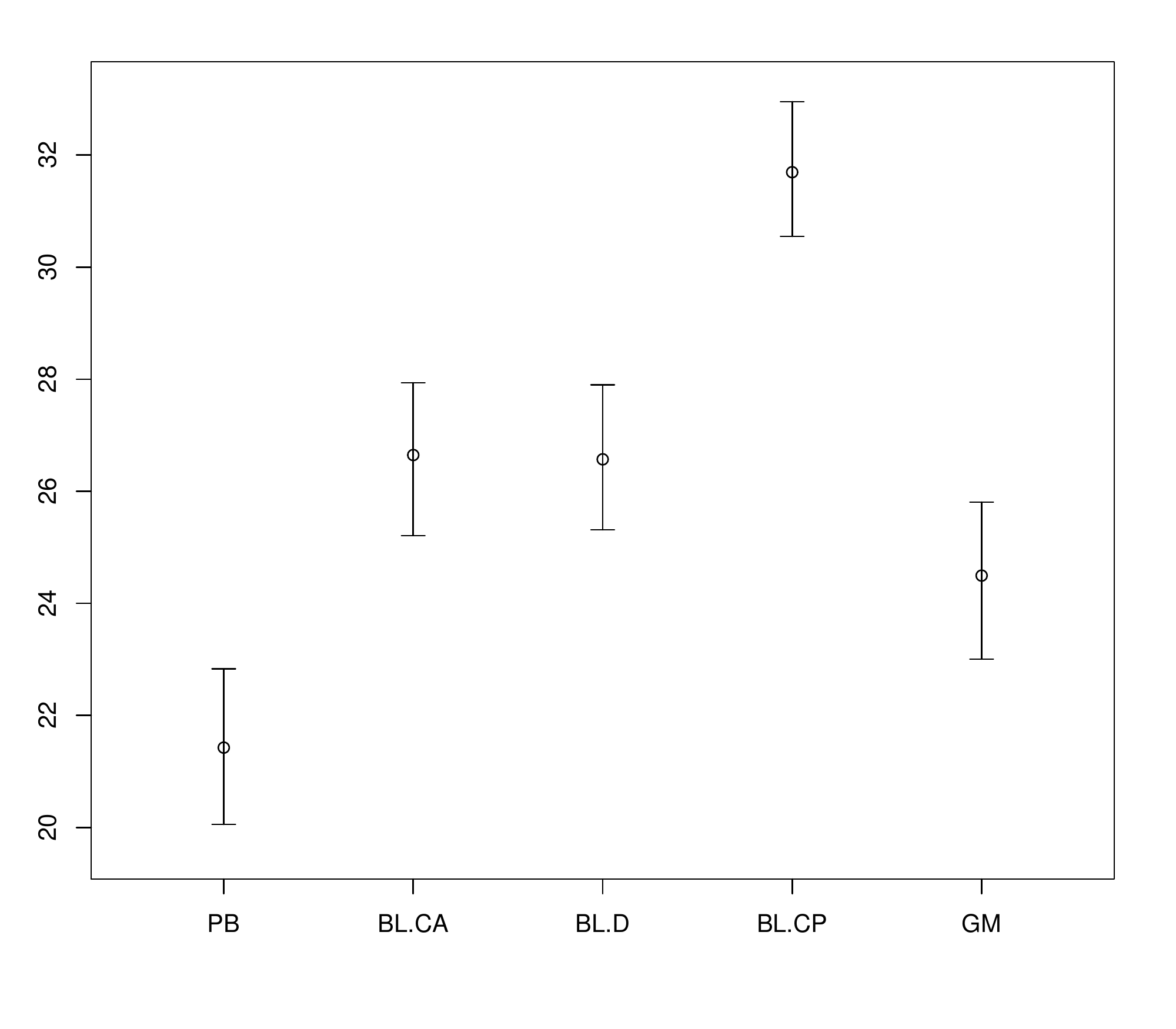}}
\end{minipage}%
\begin{minipage}{.5\linewidth}
\centering
\subfloat[Math]{\label{LELS:b}\includegraphics[scale=.35]{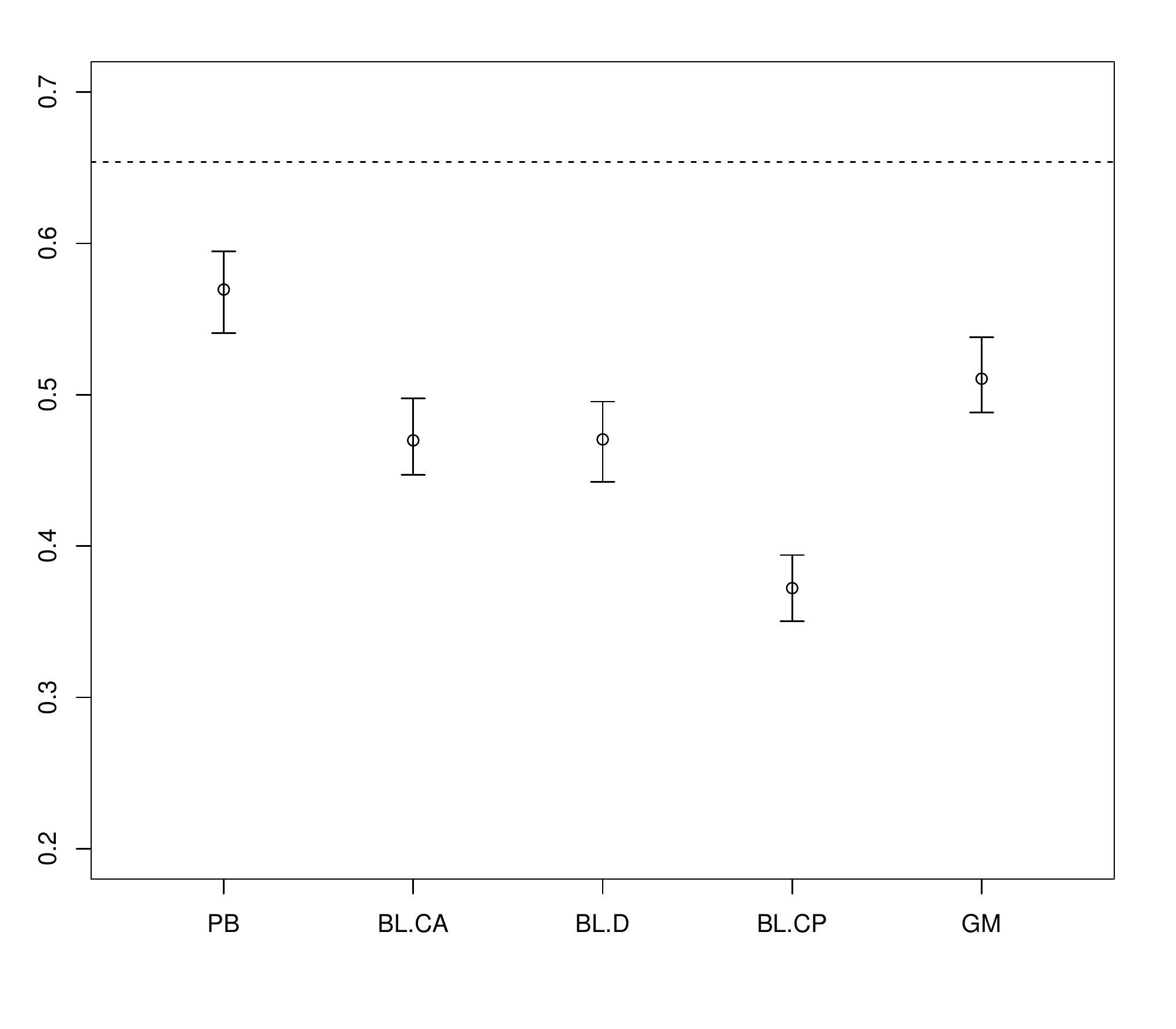}}
\end{minipage}\par\medskip
\centering
\begin{minipage}{.5\linewidth}
\centering
\subfloat[Academic]{\label{LELS:c}\includegraphics[scale=.35]{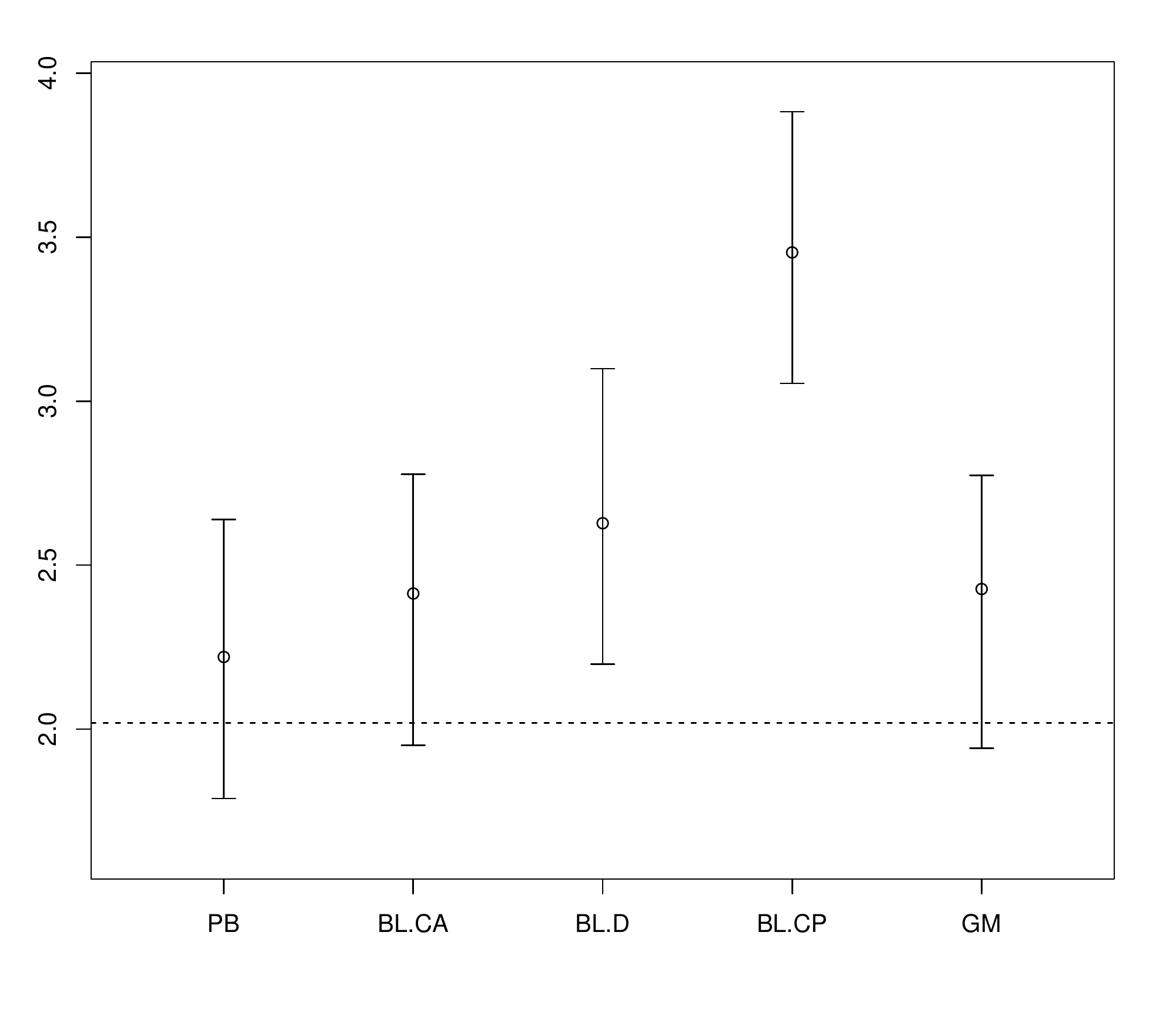}}
\end{minipage}%
\begin{minipage}{.5\linewidth}
\centering
\subfloat[Vocational]{\label{LELS:d}\includegraphics[scale=.35]{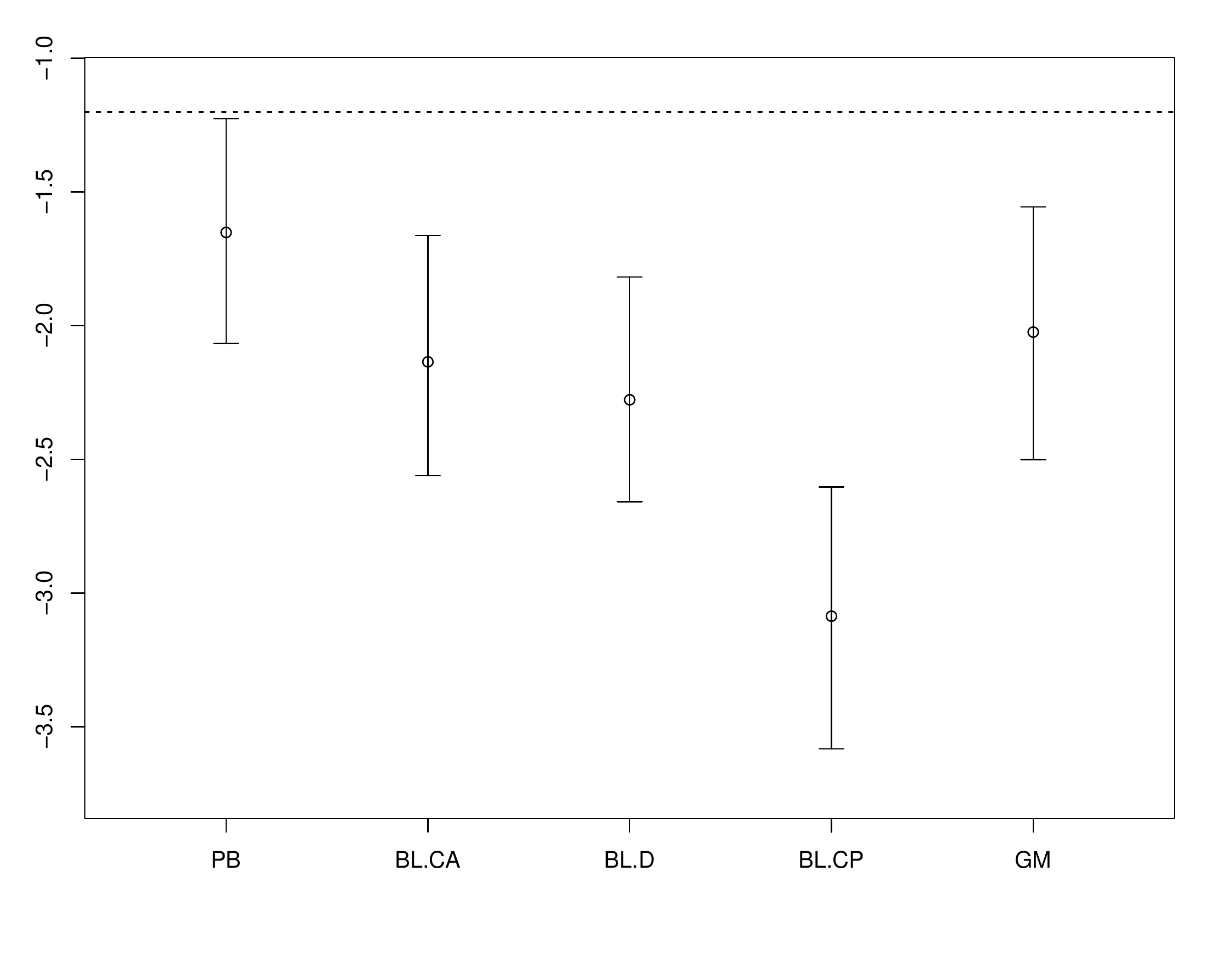}}
\end{minipage}\par\medskip

\caption{Results from the LSLF simulation. Top left panel (a) displays 95\% interval for the 100 posterior means for intercept. Top right panel (b) displays 95\% interval for the 100 posterior means for \textsf{math} coefficient. Lower left panel (c) displays 95\% interval for the 100 posterior means for \textsf{prog} = Academic coefficient. Lower right panel (d) displays 95\% interval for the 100 posterior means for \textsf{prog} = Vocational coefficient. In all panels, PB stands for using the GAZM on the perfectly blocked data, BL.CA for using BLASE with the concentrated and appropriate prior, BL.D for using BLASE with the diffuse prior, BL.CP for using BLASE with the concentrated but poorly specified prior, and GM for using the GAZM on the faulty data. Dashed horizontal line indicates true value of coefficient.}
\label{Ch2::fig::BLASE::LELS::Coefs}
\end{figure}

\begin{figure}[ht]
    \centering
    \subfloat[LSLF: Non-seed PMRs.]{{\includegraphics[width=.8\textwidth]{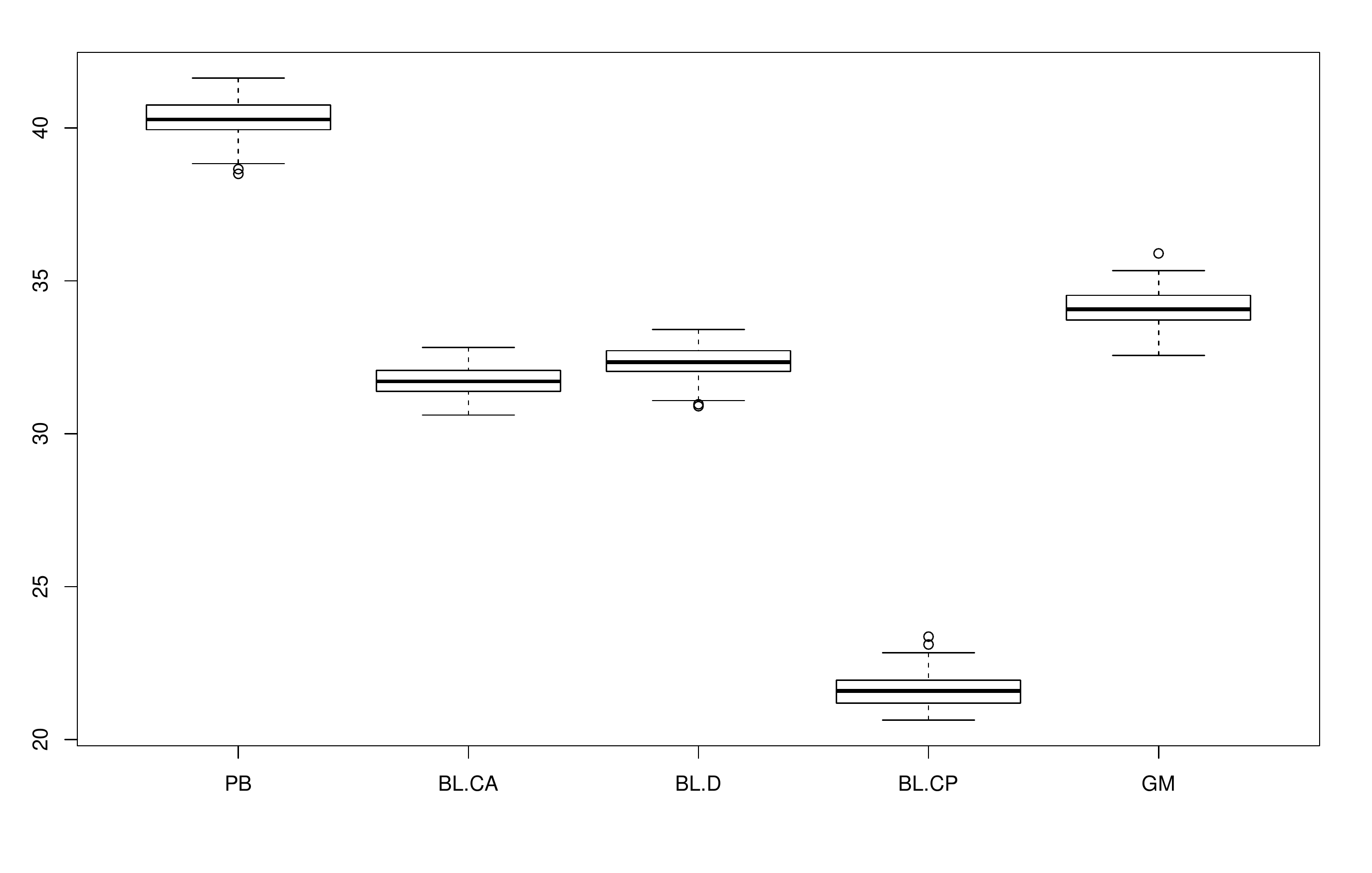} }\label{Ch2::fig::BLASE::LELS::Match}}%
    \qquad
     \subfloat[LSLF: Out-of-sample RMSE.]{{\includegraphics[width=.8\textwidth]{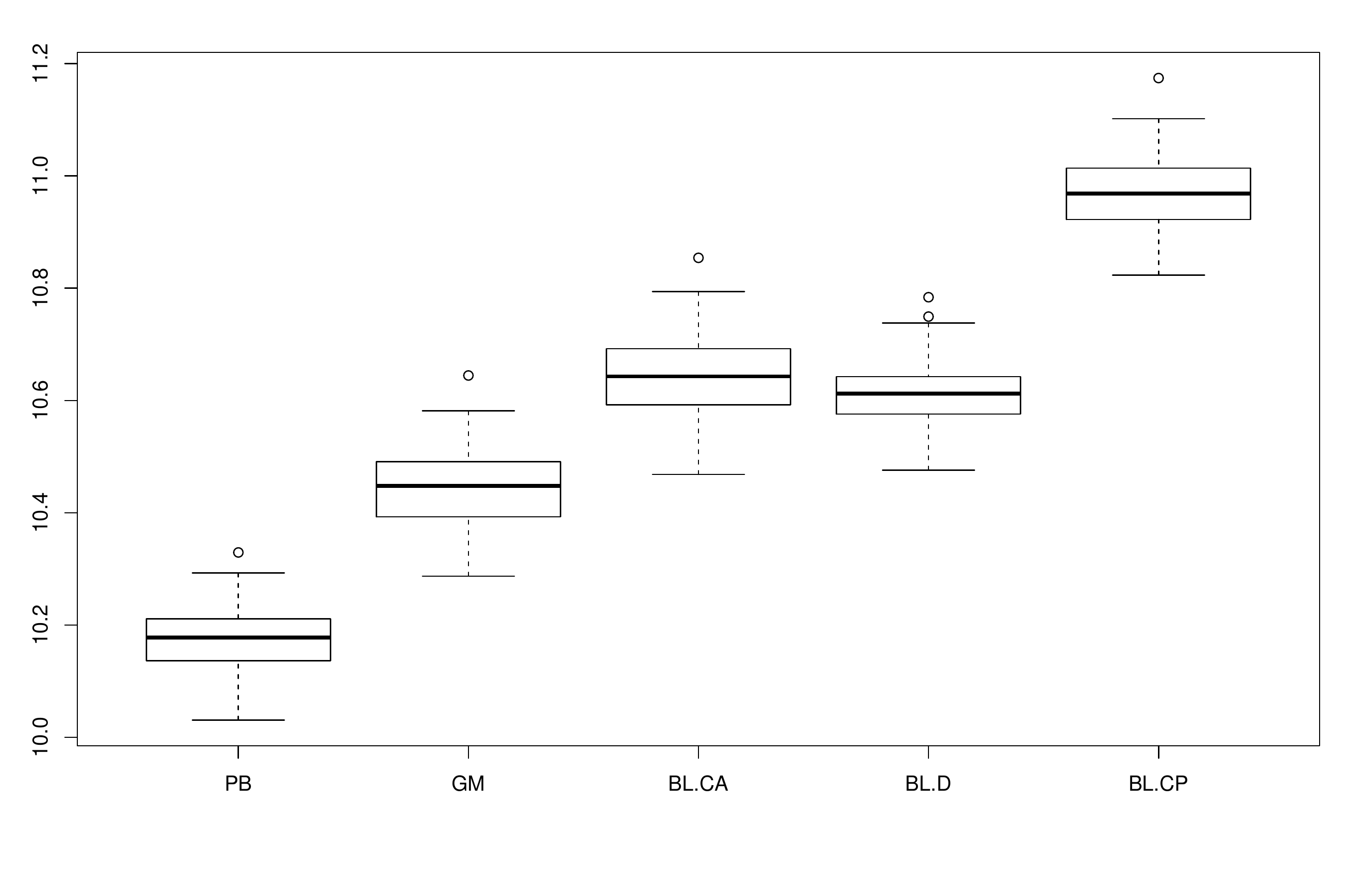}} \label{Ch2::fig::BLASE::LELS::MSE}}%
    \caption{Results from the LSLF simulation. Top panel (a) displays average posterior match rates from 100 simulations. Bottom panel (b) displays average RMSE from 100 simulations. In both panels, PB stands for using the GAZM on the perfectly blocked data, BL.CA for using BLASE with the concentrated and appropriate prior, BL.D for using BLASE with the diffuse prior, BL.CP for using BLASE with the concentrated but poorly specified prior, and GM for using the GAZM on the faulty data.}
    \label{fig:LSLF}
\end{figure}

\clearpage 

\section{NCERDC: Supplemental Plot} \label{Supp::NCERDC} 

\begin{figure}[!ht]
\begin{center}
\includegraphics[scale=0.35]{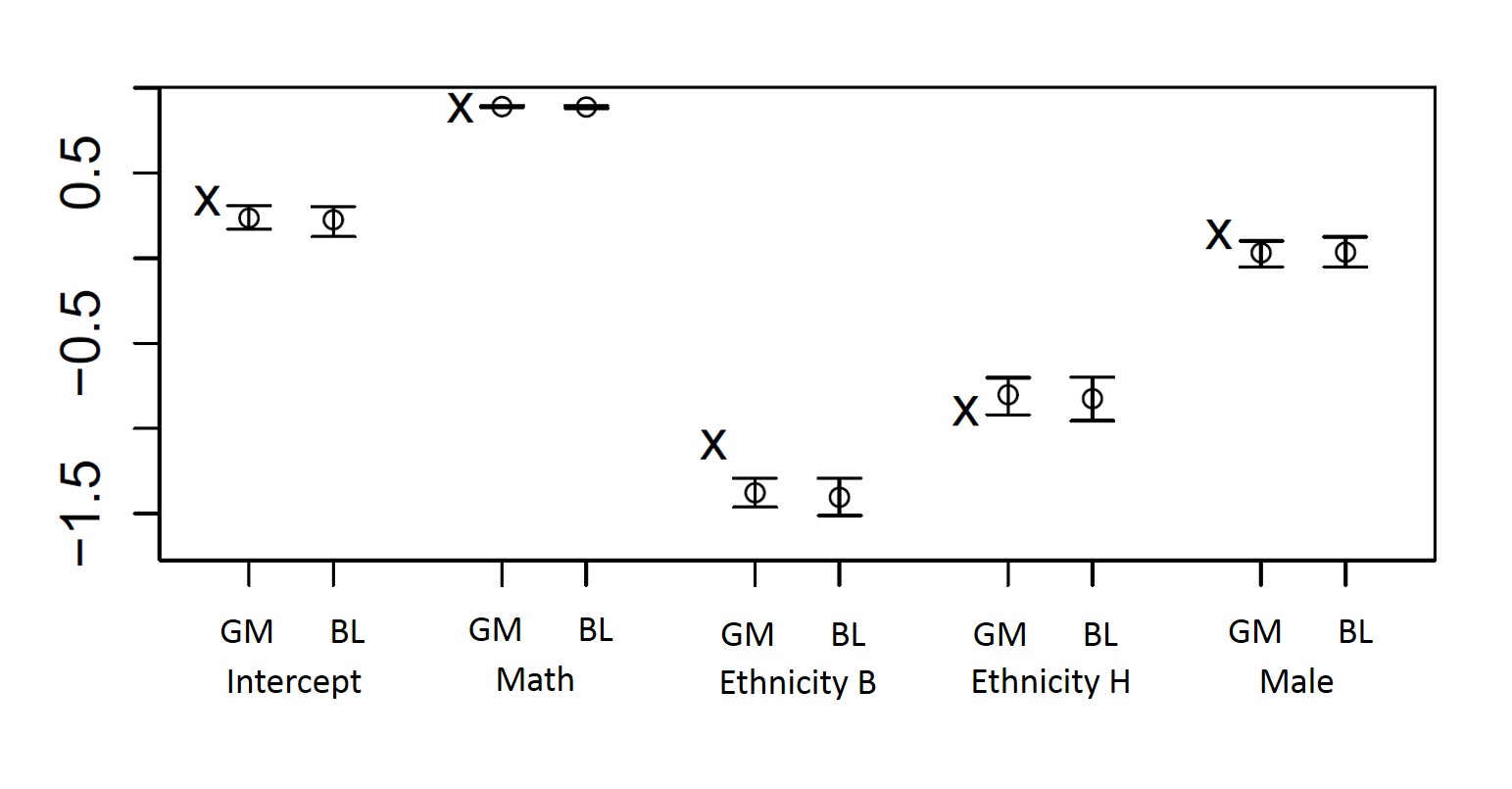}
\caption{Results from the simulation with school as the MV. 95\% posterior intervals for the coefficients of the linking analysis. ``X" represent OLS estimates from the correctly linked data. BL for using BLASE with the diffuse prior, and GM for using the GAZM on the faulty data. Ethnicity B stands for Black, and Ethnicity H stands for Hispanic. }
\label{fig::NCERDC}
\end{center}
\end{figure}

This section contains a supplemental plot from the simulation in Section 5 of the main paper. Some of this material is repeated from the main text for completeness. Recall that we work specifically with a subset of 77,998 record pairs which correspond to the three largest ethnic groups in the data set. Only about 9.2\% of the true matches disagreeing on at least one in-common variable, with school being the most commonly different. This arises mainly because some students move schools over the year.  We fit BLASE (using the diffuse prior) using school as the MV, as well as the  model that treats all seven categorical variables as BVs.  Not surprisingly given the low error rates and high fraction of $T_1$ seeds, there is little difference in their performances (Figure \ref {fig::NCERDC}). Additionally, in these data the school attended is not an important predictor of the other variables. As a result, BLASE has little information to encourage moves to the correct pools.   

\end{document}